\documentclass{aa}  

\usepackage{graphicx}
\graphicspath{{FIGs/}}
\usepackage{txfonts}

\usepackage[utf8]{inputenc}
\usepackage{textcomp}
\usepackage{txfonts}
\usepackage{float}
\usepackage{natbib}
\usepackage{caption}

\newcommand{\doce}{\rm $^{12}$CO}       
\newcommand{\trece}{\rm $^{13}$CO}

\newcommand{\kms}{\mbox{km~s$^{-1}$}}

\newcommand{\s}{\mbox{$''$}}
\newcommand{\mloss}{\mbox{$\dot{M}$}}
\newcommand{\my}{\mbox{$M_{\odot}$~yr$^{-1}$}}
\newcommand{\ls}{\mbox{$L_{\odot}$}}
\newcommand{\ms}{\mbox{$M_{\odot}$}}

\newcommand{\secp}{\mbox{\rlap{.}$''$}}
\newcommand{\secs}{\mbox{\rlap{.}$^{\rm s}$}}

\newcommand{\h}{$^{\rm h}$}
\newcommand{\m}{$^{\rm m}$} 
         
\newcommand{\afgl}{AFGL\,2233}
\newcommand{\tas}{\mbox{T\rlap{$_A$}$^*$}}
\newcommand{\trot}{$T_\mathrm{rot}$}

\begin{document}

   \title{Chemistry in the High expansion-velocity C-rich evolved star AFGL\,2233
   \thanks{This work is based on observations carried out under project numbers V016 and V074 with the IRAM Plateau de Bure Interferometer, and 050-19 with the IRAM 30m telescope. IRAM is supported by INSU/CNRS (France), MPG (Germany) and IGN (Spain).}
   \thanks{$Herschel$ is an ESA space observatory with science instruments
provided by European-led Principal Investigator consortia and with important participation from NASA. $HIFI$ is the Herschel HeterodyneInstrument for the Far Infrared.}   
   }
   
   \subtitle{Isotopic ratios, peculiarities and evolutionary status.}

   \author{G. Quintana-Lacaci
          \inst{1}
          \and
          M. Ag\'undez\inst{1}
          \and
          L. Velilla-Prieto\inst{1} 
          \and
          J. Alcolea\inst{2}
          \and
          A. Castro-Carrizo\inst{3}
          \and
          J.P. Fonfr\'ia\inst{1}
          \and
          J. Cernicharo\inst{1}
          }

   \institute{Department of Molecular Astrophysics, Instituto de Física Fundamental (IFF-CSIC), C/ Serrano 123, 28006 Madrid, Spain
   \and
   Observatorio Astron\'omico Nacional (IGN), Alfonso XII No. 3, 28014 Madrid, Spain
   \and
   Institut de Radioastronomie Millimetrique, 300 rue de la Piscine, 38406 Saint Martin d$'$Heres,
France
   }
   

   \date{Received September 15, 1996; accepted March 16, 1997}

 
  \abstract
   {High expansion velocity carbon stars (HVCs) are a rare class of evolved stars whose circumstellar envelope (CSE) combine a C-rich chemistry with unusually high expansion velocities that are typical in O-rich massive evolved stars. These objects, in particular AFGL\,2233, have been
   proposed to be high-mass evolved objects with an exhausted hot-bottom burning. 
 Studying their chemistry is essential to understand the nature of these objects.   
   }
   {We aim to characterize the chemical composition and isotopic ratios of the circumstellar envelope of AFGL\,2233. 
   We also investigate the origin of several chemical peculiarities observed in this source, including the presence of nitrogen- and oxygen-bearing species in a C-rich environment.}
   {We carried out a complete line survey at 3\,mm and 1\,mm using the IRAM 30m telescope, complemented by Herschel/HIFI FIR observations and interferometric maps of species such as SiO, C$_2$H or HCN. The molecular emission was analyzed using rotational diagrams and radiative transfer modeling under the LVG approximation. Column densities and fractional abundances were derived for over 30 molecular species, including isotopologues. The results were compared with those of other well-studied evolved stars.}
   {The revised Gaia DR3 distance of 1.236 kpc implies a luminosity of $\sim 2\times10^{4}\ls$   
   consistent with an initial mass of $4.5$–$9\,M_\odot$. The molecular inventory confirms a C-rich chemistry, but also reveals unusually high abundances of NH$_3$, H$_2$O, and SiN. 
   The derived isotopic ratios show significant variations across different species, with $^{12}$C/$^{13}$C ranging from 7 to 55, { which reflects opacity effects in some cases.} 
   The C$_2$H/C$_4$H ratio is abnormally high compare with C-rich AGB stars.
   The presence of SiN and the high NH$_3$ abundance can either point to a N-enrichment or the influence of a companion.}
   {AFGL\,2233 is likely a high-mass AGB or super-AGB star that experienced hot-bottom burning in the past. Its current C-rich chemistry, combined with signs of N-enrichment and the presence of O-bearing species, suggests a complex evolutionary history. The observed chemical anomalies may result from a combination of nucleosynthesis and other chemical processing effects such as shock-induced chemistry and binary influence. 
   }


   \maketitle
%

\section{Introduction} \label{sec:intro}

The role of the evolved stars is essential for the chemical evolution of the Galaxy \citep{AGBs2003}. The elements synthesized in their interiors are transported to the photosphere during the late stages of stellar evolution, and subsequently expelled into the interstellar medium (ISM) through intense mass-loss episodes. This process, which is particularly efficient during the Asymptotic Giant Branch (AGB) and Red Supergiant (RSG) phases, leads to the formation of extended circumstellar envelopes (CSEs) where a rich molecular chemistry develops. The chemical composition of these envelopes is { mainly} determined by the nucleosynthesis processes that have taken place in the stellar interior, as the Hot Bottom Burning (HBB),  and by the efficiency of mixing mechanisms such as the third dredge-up.

In the case of intermediate-mass stars ($M_\mathrm{init} \sim 1.5$–$4\,M_\odot$), the third dredge-up (TDU) leads to a C-rich photosphere (C/O > 1), while in more massive stars ($M_\mathrm{init} \gtrsim 4$–$8\,M_\odot$), the activation of HBB prevents the formation of carbon stars by converting $^{12}$C into $^{14}$N in the stellar core. As a result, C-rich envelopes are typically associated with low- and  intermediate-mass AGB stars. In addition, due to their luminosities and due to the radiation pressure acting as the main driver of mass ejection, the expansion velocities are lower in C-rich stars \citep[$v_\mathrm{exp} \sim 10$–$20\,\kms$, e.g.][]{massalkhi-Crich} than in massive O-rich stars \citep[$v_\mathrm{exp} \sim 30$–$40\,\kms$, ][]{chemyhg}. However, a small group of evolved stars challenges this paradigm: the so-called High expansion Velocity Carbon stars (HVCs), which combine a C-rich chemistry with unusually high expansion velocities ($v_\mathrm{exp} \gtrsim 25\,\kms$). These objects are rare and poorly understood, and their nature remains uncertain.

AFGL\,2233 is one of these HVC stars. Its CO line profiles reveal a terminal expansion velocity of $\sim32\,\kms$, significantly higher than the average for C-rich AGB stars \citep{Loup93}. Its infrared spectrum shows the characteristic 11\,$\mu$m SiC feature \citet{ISOSiC}, and previous studies have reported strong emission from molecules such as HCN \citep{Likkel96} and high abundances of C$_2$H \citep{c2h2}. These properties, together with its long pulsation period ($P \sim 687$ days), initially suggested that AFGL\,2233 could be a massive object with a C-rich envelope, possibly formed after the exhaustion of HBB \citep{HVCs}.

\citet{HVCs} assumed a distance to AFGL\,2233 4.16 kpc (from Gaia DR2), leading to a derived luminosity of $\log(L/L_\odot) \sim 5.4$ and a total envelope mass of $\sim1.7\,M_\odot$, values consistent with a massive origin. 
However, the recent release of Gaia DR3 has provided a revised parallax for this source, yielding a Bayesian distance of 1.236\,kpc (see Sect.\,\ref{sect:dist}). 

In this paper, we first revisited the luminosity, density and temperature profiles with this new distance estimate. 
After this update, we present a detailed chemical analysis of AFGL\,2233 based on a complete line survey at 3\,mm and 1\,mm obtained with the IRAM 30m telescope, complemented by HIFI observations from the Herschel Space Observatory and interferometric maps of different species. 
This study provides new constraints on the physical and chemical properties of AFGL\,2233.
Our goals are to characterize the molecular content and isotopic composition of the envelope, to investigate the origin of the observed chemical anomalies, and to understand the evolutionary status of AFGL\,2233 in the context of massive AGB and super-AGB stars.

\section{Observations}



\subsection{HIFI Data}

AFGL\,2233 was observed in April 2013 using the Heterodyne Instrument for the Far Infrared \citep[HIFI,][]{HIFI} onboard the HSO \citep{HSO}.
We observed bands 1b, 2a, 5a, 4b, 7b to cover transitions from species such as H$_2$O, CO, SiO, NH$_3$ or H$^{13}$CN.
The values of the system temperature were T$_{sys} \sim 90 - 390$\,K for bands 1 to 4, and 830\,K and 1240\,K for bands 5a and 7b respectively (see table\,\ref{hifiobs}). 


Observational procedure was similar to that used in the HIFISTARS programme (P.I. V. Bujarrabal). This procedure is fully described in \citet{AlcoleaVYCMa} ; here we briefly summarize its main characteristics. The observations were  performed using the two orthogonal linearly polarized receivers available at each band, named H and V (horizontal and vertical) after their mutually perpendicular orientations. The receivers worked in double side-band mode (DSB), providing an instantaneous frequency coverage of 4 plus 4\,GHz for bands 1 to 5, and 2.6 plus 2.6\,GHz for band 7. The observations were all performed in the dual-beam switching (DBS) mode, alternating  between the ON source position and two OFF reference positions located 3 arc-minutes away  at either side of the science target \citep[see][for additional details]{HIFIobs}. This DBS subtraction procedure worked well except for band 7, where strong ripples (generated by electrical standing waves) are often found in the averaged spectra, especially in the case of the V-receiver. The data shown here were taken using the wide-band spectrometer (WBS),  with an effective spectral resolution slightly varying across the band, with a mean value of 1.1\,MHz.  The data were retrieved from the Herschel Science Archive and were reprocessed using a modified version of the standard HIFI pipeline using HIPE\footnote{HIPE is a joint development by the Herschel Science Ground Segment Consortium, consisting of ESA, the NASA Herschel Science Center, and the HIFI, PACS, and SPIRE consortia. Visit 
https://www.cosmos.esa.int/web/herschel/hipe-download for additional information.}, providing as final result individual ON-OFF elementary integrations without performing the final time-averaging. Later on, these spectra were exported to CLASS\footnote{CLASS is part of the GILDAS software package, developed and maintained by IRAM, LAOG/Univ. de Grenoble, LAB/Obs. de Bordeaux, and LERMA/Obs. de Paris. For more details, see http://www.iram.fr/IRAMFR/GILDAS} using the hiClass tool within HIPE for futher processing. Time-averaging was also performed in CLASS, as well as baseline removal, and combination of the results from the H and V receivers. 

In general, the data presented no problems and did not need a lot of flagging, except for band 7, for which a semiautomated procedure was designed in CLASS to detect and remove the sub-scans in which the ripples were more severe. The application of this procedure normally results in the rejection of 30\% to 50\% of the non-averaged spectra, which produces a final spectrum slightly noisier, but with a more reliable baseline.

The original data were oversampled to a uniform channel spacing of 0.5\,MHz, but we smoothed all spectra down to a velocity resolution of about 1\kms. The data were re-calibrated into (Rayleigh-Jeans equivalent) main-beam temperatures ($T_\mathrm{mb}$) adopting the latest available values for the telescope and main beam efficiencies (Roelfsema et al. 2012). In all cases we assumed a side-band gain ratio of one. 


\begin{table*}
\caption{Main parameters of the HIFI observations.} 
\centering                          
\begin{tabular}{l c c c c c c c c}        
\hline\hline                 
Herschel  & Obs. date & Durat. & \multicolumn{2}{c}{Sky frequency coverage}           &  T$_{\rm sys}$ &  HIFI & HPBW &  Cal. \\
OBSID     & yr:mo:day &  (s)   & LSB (GHz)  & USB (GHz) & (K)        &  band & ($\s$) &  uncer \\
\hline 
1342269367& 2013:04:04& 1830&556.44--560.57 &568.42--572.56    &88   & 1b  &39.6 &  15\% \\
1342269369& 2013:04:04& 3421&645.45--649.59 &657.44--661.58    &122  & 2a  &34.3 &  15\% \\
1342269370& 2013:04:04& 1350&678.36--682.50 &690.35--694.48     &132  & 2a  &32.7 &  15\% \\
1342269377& 2013:04:04& 3720&1149.61--1153.75 &1161.60--1165.73 &831  & 5a  &19.4 &  20\% \\
1342269400& 2013:04:04& 2258&1099.04--1103.17 &1111.02--1115.16 &394  & 5a  &20.3 &  20\% \\
1342269402& 2013:04:04& 2361&1833.41--1835.98 &1840.66--1843.23 &1245 & 7b  &12.3 &  30\% \\
\hline                                   
\end{tabular}
\label{hifiobs}
\end{table*}

\subsection{IRAM 30m data}

We used the IRAM 30m radiotelescope to obtain a complete line
survey at 3\,mm and 1\,mm for the HVC AFGL\,2233.
We observed AFGL\,2233 
at the position coordinates (J2000) 
18\h42\m24$\secs$680	-02$\rm^o$17${'}$25\secp200
with an $\rm{v_{LSR}}$ = 2.80\,km/s.
The observations were obtained during September and November 2019. We used the EMIR receiver,  simultaneously using the receivers E090 and E230 with a bandwidth of 4\,GHz. 
We used 7 different setups to cover the atmospheric windows at 
3\,mm and 7 setups at 1\,mm covering both polarizations. 
For each setup, we observed 
every setup twice, introducing a frequency shift of 50 MHz between the two observations for easing the identification of spectral features arising from the image band. 
In total each setup was observed for 3h (= $2\times\,1.5$h).
We used the wobbler switching mode to minimize the ripples in the baselines.
The system temperatures
during the observations were in the range 70--280\,K for the E090 receiver (3\,mm receiver band)
and between 200 and 425\,K
for the E230 receiver (1.3\,mm band). The weather conditions during the observations 
were good with an amount of precipitable water vapor ranging between 
2 and 6 mm.  

The backends used were WILMA (spectral resolution 2\,MHz) and the FFT with a
spectral resolution of 0.195\,MHz.
Since the profiles of AFGL\,2233 are known to be relatively wide the spectra
could be smoothed to increase the SNR.

The pointing correction was checked frequently and, therefore, we expect 
pointing errors of $\sim$3''. 
Focus was checked at the beginning of the observing runs, and after changes in the temperature of the telescope
structure (i.e. during dusk and dawn).
The telescope beam size at 3\,mm is 21--29$''$ and 9--13$''$ at 1\,mm. 
The Atmospheric Transmission Model (ATM) is adopted
at the IRAM 30 m \citep{Cerni85,Pardo01}.
The data presented are calibrated in antenna 
temperature ($T^*_A$). The calibration error is expected to be 10\% at 3\,mm and 
30\% at 1\,mm.
The data were processed using the GILDAS package
\footnote{See {\tt http://www.iram.fr/IRAMFR/GILDAS} for more informa-tion about the GILDAS software.}. 
Both polarization were averaged, since no changes were observed in the line intensities along the line survey.
Baselines were subtracted using only first order polynomials.


\subsection{IRAM PdBI data}

Observations towards AFGL\,2233 were performed with the former Plateau de Bure Interferometer
(with 6 antennas, early version of the NOEMA interferometer). Three short tracks were obtained
in 6Cq configuration in November 2011 (project name V016), and one in 6Bq configuration in March
2012 (project name V074). Single sideband receivers were tuned at 88.630\,GHz to observe the
frequency interval from 87.83 to 91.43 GHz, with different spectral resolutions, from 1 to 8\kms .
Four intervals were obtained with the highest spectral resolution: [89.15 - 89.23], [88.60 - 88.67], [87.5 - 87.6] and [86.8 - 86.9] GHz.
All observations were performed in good conditions, with 1749+096 and 1827+062 as phase calibrators.
Data calibration was straightforward for all the tracks, resulting in uncertainties below 5\% and 20 degrees for amplitudes and phases.
The absolute flux calibration was made by using the main reference at the observatory, MWC\,349, 
for which a flux of 1.1\,Jy at 88.630\,GHz was adopted.
Uncertainties in the absolute flux calibration in 2011-2012 were below 10\% at 3\,mm.
The calibration and data analysis were performed in the standard way by using the GILDAS software package.


We were able to detect
the emission from species as H$^{13}$CN $1-0$, SiO $5-4$, CCH $1-0$, HC$_5$N $33-32$, H$^{13}$CCCN $10-9$, HCN $1-0$, C$_3$N $9-8$. The
obtained total integrated emission spectrum  
is presented in Fig.\,\ref{pdbi}, and the channel maps are shown in Appendix \ref{sect:maps}.

%

   \begin{figure}
   \centering
   \includegraphics[width=\hsize]{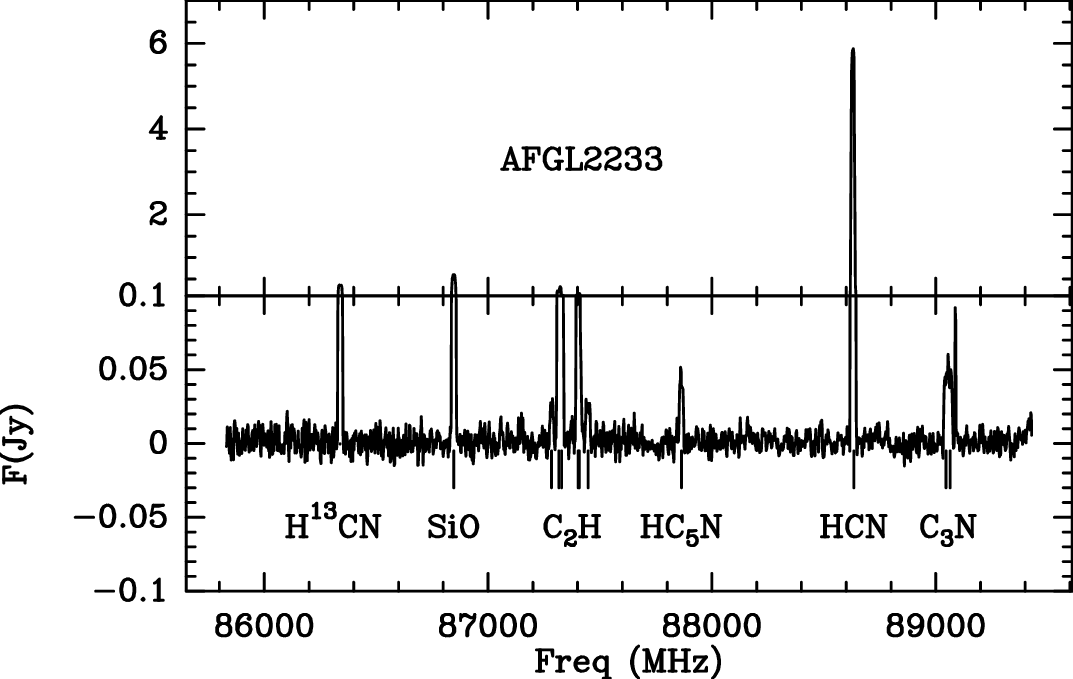}
      \caption{Spectra of the integrated flux obtained with IRAM PdBI towards AFGL\,2233.
              }
         \label{pdbi}
   \end{figure}

\section{2019 results revisited. Gaia DR3.}
\label{sect:dist}

Since the publication of \citet{HVCs} a new Gaia data release has been published \citep{gaia3}. The new measurement of the paralax of \afgl\ 
is better constrained, being the error significantly lower than that provided in GAIA DR2 \citep{Gaia2-1}. 

In \citet{HVCs} we assumed a distance of 4.16\,Kpc \citep[from][]{Yuasa99}. { This estimate was derived based on radial velocities and its result was uncertain. Also Gaia DR2 data had a very large error, similar to the derived distance itself, but the derived distance estimate was similar to that from \citet{Yuasa99}. Due to this, \citet{Yuasa99} estimate was adopted by \citet{HVCs}.
However, the new Bayesian \citep[GSP-Phot-Aeneas,][]{GSP-Phot} 
distance of GAIA DR3 is much better constrained. This latter estimate is 1.236\,Kpc (with a confidence interval ranging from 1.165 to 1.431 pc).}
While the distance determination is still unsure for Red giant 
stars, we have revisited the results of the cited paper with the new distance.
It is worth noting, in any case, that the column densities derived { in this work} are independent of the distance.


The methodology applied is exactly the same presented in \citet{HVCs}. 

\subsection{Spectral energy distribution (SED).}
\label{SED}

\begin{figure}
   \centering
   \includegraphics[width=\hsize]{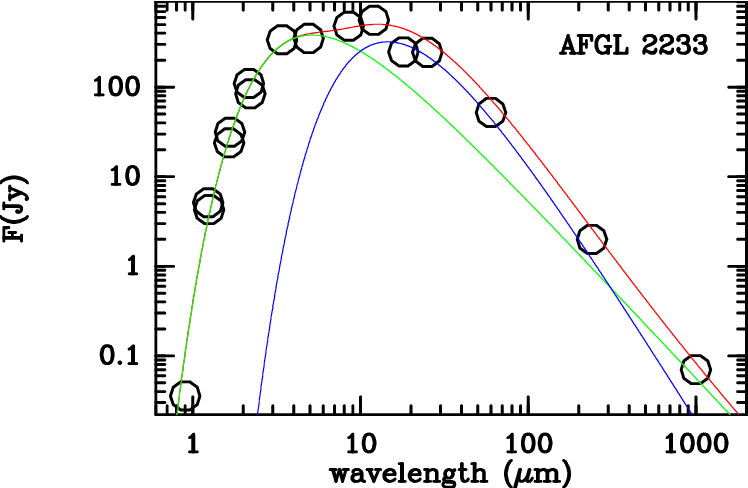}
      \caption{SED fitting obtained with the new distance estimate for AFGL\,2233. { Green and blue curves correspond to the two components, and the red curve to the sum of both components.}
              }
         \label{bb}
   \end{figure}

As in \citet{HVCs} we found that, to be able to properly fit the SED of \afgl, we needed two independent black bodies. We found that the best fit -- following a $\chi^2$ methodology  --  is obtained for a $T_\star = 900\,K$ and $r_\star = 3.1\times10^{14}$cm and $T_{gas} = 300$\,K and $r_{gas} = 10^{15}$cm  (see Fig.\ref{bb}). 

The luminosity derived for this new distance is log($L/\ls$)=4.3 \citep[instead of log($L/\ls$)=5.4,][]{HVCs}. This suggests, according to the evolutionary models by \citet{Meynet03} or \citet{tracks} that AFGL\,2233 { would have an initial mass,} $M_\mathrm{init}$ between 4.5\ms\ and 9\ms\ depending on different factors such as metallicity or rotation.{ We conclude} that this object probably is a RSG or a super-AGB. 

\subsection{Molecular emission model.}
\label{model}

Following the same method as in \citet{HVCs} -- modeling the CO emission adopting an LVG aproximation, assuming a constant mass loss and expansion velocity, and a temperature following $T(r) = T_\mathrm{16}(r/10^{16})^{\alpha_t} + T_\mathrm{min}$ within a certain region (or shell)  --, and assuming the new distance, we fitted both the single-dish CO $J=1-0$ and $J=2-1$ observations obtained at the IRAM 30m radiotelescope and the azimuthal averaged profiles of the interferometric maps obtained with IRAM PdBI for the same transitions. The results are listed in the Table\,\ref{result:1}.

 \begin{table*}
\caption{Best fit parameters for AFGL\,2233. }             
\label{result:1}      
\centering                          
\begin{tabular}{l c c c c c c c}        
\hline\hline                 
Shell & $R_\mathrm{in}$ (cm) & $R_\mathrm{out}$ (cm) &\mloss (\my) & $V_\mathrm{exp}$ ($\kms$) & $T_{16}$ (K) & $\alpha_t$ & $T_\mathrm{min}$ (K) \\
\hline  
1&  1$\times 10^{14}$&  5$\times 10^{16}$& 1.8$\times 10^{-5}$&32 &50  &0.6&2.73\\
\hline 
2&  5$\times 10^{16}$&  9$\times 10^{16}$& 2.6$\times 10^{-5}$&32 &150 &0.6&2.73\\
\hline 
3&  9$\times 10^{16}$&  3$\times 10^{17}$& 2.1$\times 10^{-5}$&32 &500 &0.6&2.73\\
\hline
4&  3$\times 10^{17}$& 7$\times 10^{17}$&  3.2$\times 10^{-5}$&32 &450 &0.6&2.73\\
\hline    

\end{tabular}
\end{table*}

   \begin{figure}
   \centering
   \includegraphics[width=\hsize]{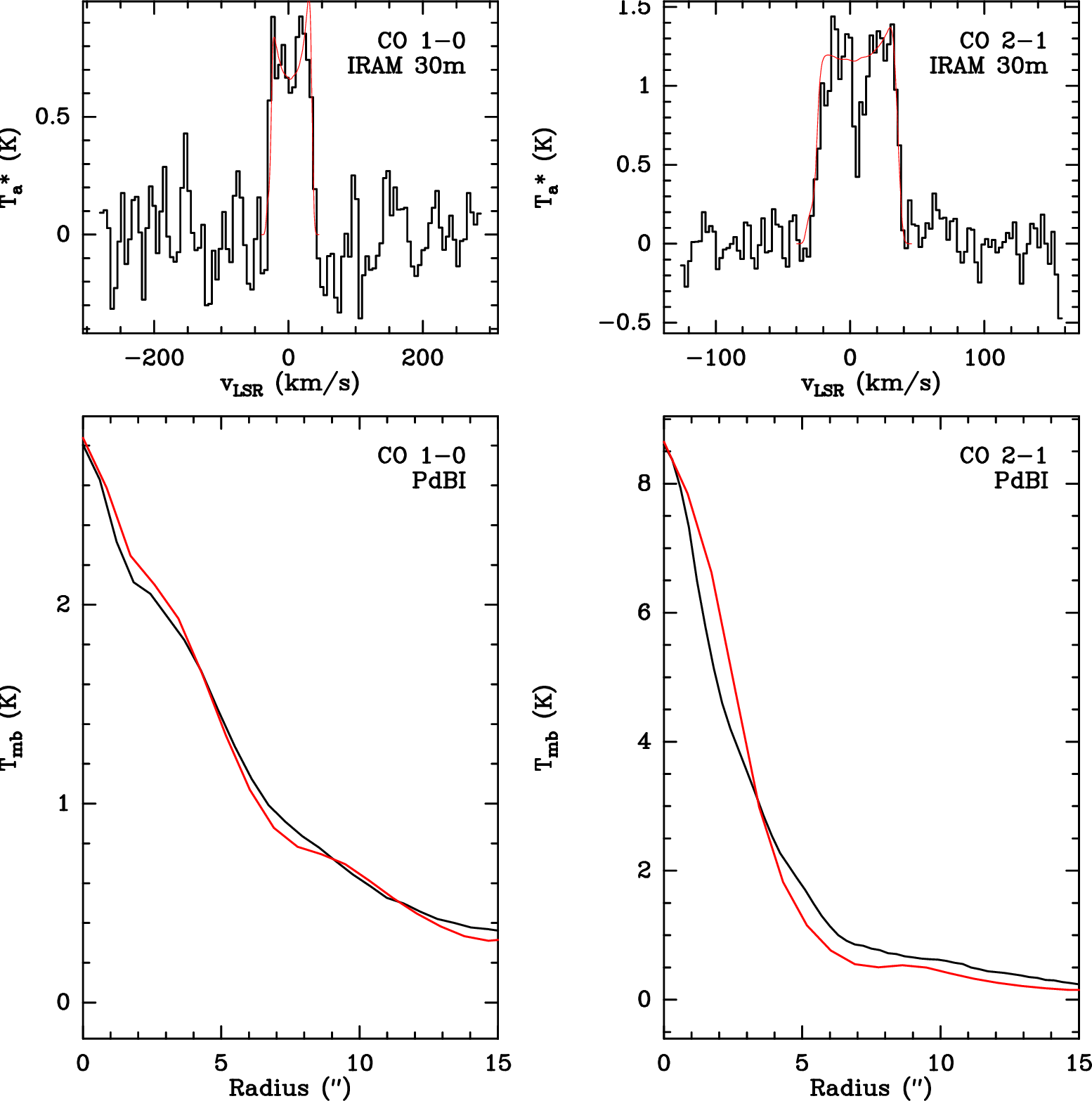}
   \caption[]{\emph{Top-left:} CO $J=1-0$ line profile obtained with the IRAM 30m 	telescope for AFGL\,2233.
   \emph{Top-right:} CO $J=2-1$ line profile obtained with the IRAM 30m telescope for AFGL\,2233. 
   \emph{Bottom-left:} Azimuthally averaged profile of the CO $J=1-0$ molecular emission of the central velocity channel obtained towards AFGL\,2233 with PdBI.
   \emph{Bottom-right:}  Azimuthally averaged profile of the CO $J=2-1$ molecular emission of the central velocity channel obtained towards AFGL\,2233 with PdBI.
   { The black lines correspond to the observations and the red lines to the results of the model. }
   }
              \label{fit2233}%
\end{figure}


The total { circumstellar mass derived from the fitting of these observations, for the new distance estimate,} is 0.19\ms (f. This value, for an outer radius of $7\times10^{17}$cm results in a column density $N_\mathrm{H_2} = 7.4\times10^{19}$cm$^{-2}$ for CO.

\section{Results} 

In this section we will present the results obtained from the observations presented above. 

\subsection{Structure of the ejecta}
\label{shape}

\begin{figure}
   \centering
   \includegraphics[width=\hsize]{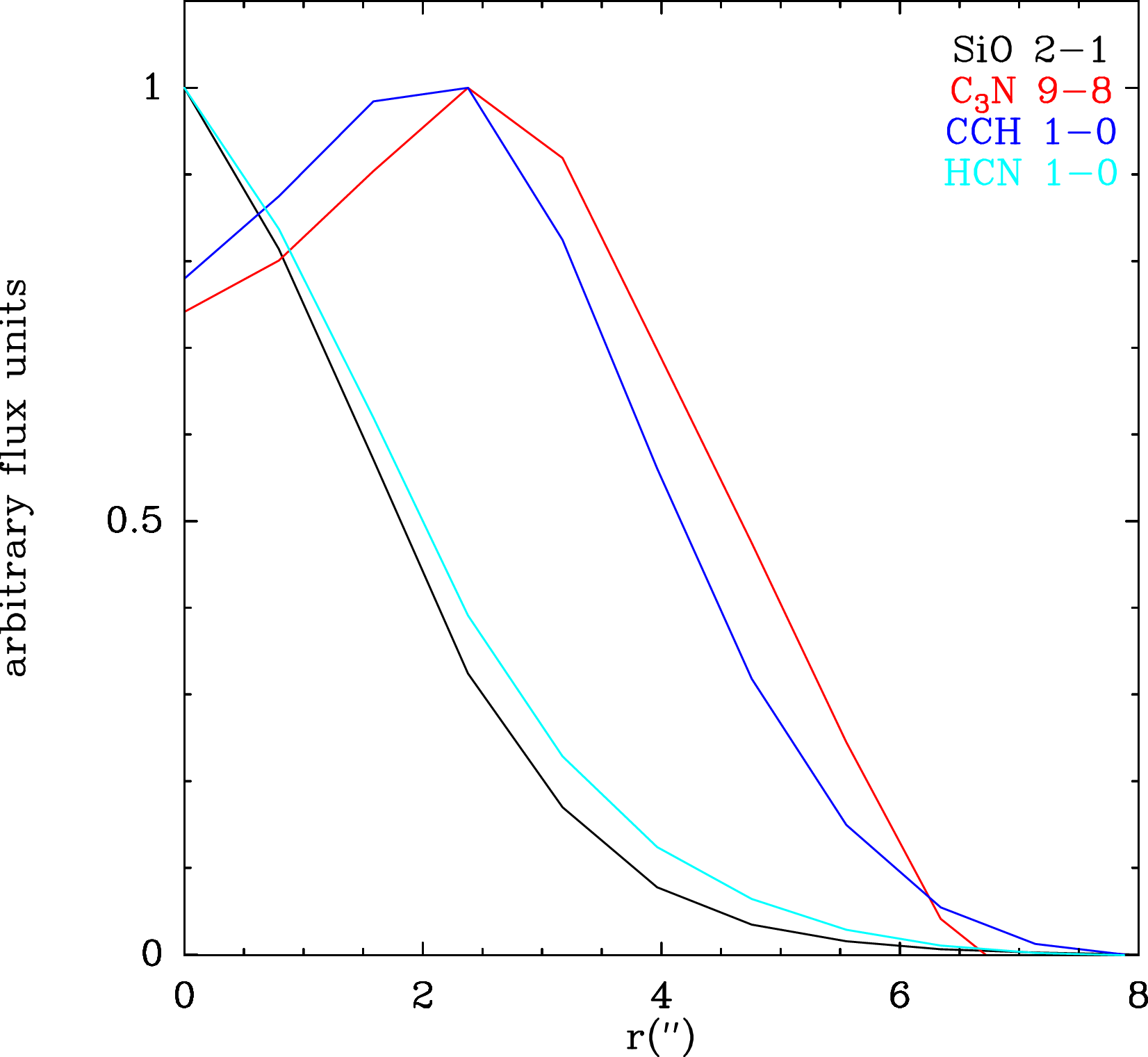}
      \caption{Azimuthal averaged intensity of the different transitions observed with PdBI for the central velocity channel, normalized to its maxium intensity.
              }
         \label{azave}
   \end{figure}

In addition to the CO $J=1-0$ and $J=2-1$ interferometric maps obtained by \citet{HVCs} we have maps of SiO $J=2-1$, C$_2$H $N=1-0$, HCN $J=1-0$, C$_3$N $N=9-8$, H$^{13}$CN $J=1-0$ and HC$_5$N $J=9-8$ (see Figs\,\ref{SiO}--\ref{HC5N}). These maps allowed us to have an estimate of the extent of the different line emissions. All these maps confirm the spherical structure observed in the CO maps. The emission from C$_2$H { and C$_3$N} shows clear hints of a hollow shell, however the size of the inner hole cannot be resolved. The size estimated for the different species is presented in Table\,\ref{sizes}. This estimate was obtained from azimuthal averaged maps of the central channels of the different molecular lines (see Fig.\,\ref{azave}).

%
%

\begin{table}
\caption{Extent of the emission observed for different species and H$_2$ column 
density within these radii.}             
\centering                          
\begin{tabular}{l c c c}        
\hline\hline                 
Molecule & $R_\mathrm{out}$(cm)& $N_\mathrm{H_2}$ (cm$^{-2}$) & Comment\\
\hline 
CO 			&	7\,10$^{17}$		&7.4\,10$^{19}$		&\\
HCN 			&	7.8\,10$^{16}$	&5.0\,10$^{20}$		&\\
SiO 			&	7.0\,10$^{16}$ 	&5.5\,10$^{20}$		&\\
C$_3$N 		&	1.1\,10$^{17}$	&<3.7\,10$^{20}$		&Hollow shell structure.\\
C$_2$H    	&	1.1\,10$^{17}$	&<3.7\,10$^{20}$		&"  \\

\hline                                   
\end{tabular}
\label{sizes}
\end{table}

To estimate the H$_2$ column density associated with the molecules found in the inner regions, relative to the full extent of CO, we used the interferometric maps presented here.

We estimated the total mass within the values of $R_\mathrm{out}$ obtained, and given this size we estimated the value of 
$N_\mathrm{H_2}$ asssociated with each species. 
{ In particular, we integrated the density profile obtained for CO and constrained that integration to the radii where each species are
located from the interferometric maps (i.e. $N_\mathrm{H_2} (mol) = \int_{Rin}^{R_\mathrm{out}^{mol}}\,n(r)\,dr$).}
These estimates are presented in Table\,\ref{sizes}.

{ The H$_2$ column density associated with the C$_2$H emitting region} is an upper limit as long as the inner hole for this species is not resolved. As the HPBW is $4\secp25\times2\secp78$, we would set as an upper limit to the size of this hole the same as the HPBW (i.e.$\sim$3\secp4 = $3.2\,\times\,10^{16}$cm). This would lead to an upper limit $N_\mathrm{H_2}$ < $1.4\times\,10^{20}$ { for C$_2$H}. In a similar way, the emission from C$_3$N has also been observed to have an inner hole, which is not resolved in our case \citep{agundez17}.
{ In any case, it is worth noting that the impact of the inner radius in the estimate of column density is low. This is particularly relevant for species presnting inner holes as C$_2$H or C$_3$N. In those particular cases, varying $R_\mathrm{in}$ from $10^{14}$cm to $10^{16}$cm only modifies $N_\mathrm{H_2}$ by a 8\%.}

\subsection{Line identification}

The IRAM-30m data showed a rich spectrum. We identified { more than 150} spectral features arising from 31 molecular species (see Table\,\ref{species}).

The complete spectra obtained in the 3\,mm and 1\,mm bands is shown in Sect.\ref{surveyimages}. The EMIR receiver used for the observations provides a sideband rejection ratio exceeding 13\,dB, which ensures that only the most intense spectral features exhibit counterparts from the image band. In our particular case, 
by comparing the results from the two observations of each setup, which differ by 50\,MHz in their tuning, we conclude that there is no image-band contribution in the spectra.

Spectral line identification was conducted using the internal database of the MADEX radiative transfer code \citep{MADEX}.

Table A.1 list all the molecular transitions detected in the 3 mm and 1 mm spectral scans. For each transition, we report the velocity-integrated intensity, peak intensity, and the rms noise level. As a result of the large expansion velocities present in AFGL\,2233, the hyperfine structure of species such as CN is not spectrally resolved. Consequently, the reported integrated intensities for these species correspond to the sum over all hyperfine components.


%

 \begin{table}
\caption{Molecular species detected and number of molecular transitions observed in the differents sets of data presented.}             
\centering                          
\begin{tabular}{l c | c c}        
\hline\hline                 
Species & lines &Species &lines \\
\hline 
CO			&	5 	&	\trece				& 2\\
HCN			&	2	&	HNC 					& 2 \\
HC$_3$N		&	12	&	HC$^{13}$CCN			& 2	\\
H$^{13}$CN	&	3	&	H$^{13}$CCCN			& 2\\
HCC$^{13}$CN	&	3	&	HC$_{5}$N			& 15 \\
HC$_{7}$N	&	1	&	C$_3$N				& 6 \\
CN			&	5	&	$^{13}$CN			& 5	\\
C$_{4}$H		&	20	&	H$_2$O				& 2	\\
CCS			&	4	&	NH$_3$				& 1	\\
CCH			&	16	&	C$^{13}$CH			& 1	\\
SiO			&	4	&	$^{29}$SiO			& 3 \\
SiS			&	7	&	$^{30}$SiO			& 3 \\
SiN			&	2	&	$^{13}$CS			& 2	\\
CH$_{3}$CN	&	4	&	C$^{34}$S			& 2	\\
CS			&	2	&	SiC$_{2}$			& 23	\\
C$^{33}$S	&	2	&	\\
\hline                                   
\label{species}
\end{tabular}
\end{table}

 \begin{table*}
\caption{Column densities and rotational temperatures. }             
\centering                          
\begin{tabular}{l c c c c c c c c}        
\hline\hline                 
Molecule & $N^{C}$ (cm$^{-2}$) & $T_\mathrm{rot}^{C}$ (K) & $N^{H}$ (cm$^{-2}$) & $T_\mathrm{rot}^{H}$ (K) &  $\langle X \rangle$&N. Lines fit & N$_\mathrm{H_2}$& Comment\\
\hline 
CO				&2.4(6)\,10$^{17}$&16(3)	  	   &1.4(4)\,10$^{14}$   &413(124)    & 3.6\,10$^{-4}$   &5	 &CO	\\      
$^{13}$CO		&4.4(2)\,10$^{15}$&7.8(2)       &--                 &--           & 6.8\,10$^{-6}$   &2	 &CO	\\	    
\hline
HCN				&2.80(3)\,10$^{14}$& 51(2)	   &	--		           &--      	    & 5.6\,10$^{-7}$  &2 &HCN& Intense $J=1-0$\\ 
H$^{13}$CN		&8.79(1)\,10$^{13}$&8.7(0)	   &1.6(1)\,10$^{13}$  &20(1)   	    & 2.1\,10$^{-7}$  &3	 &HCN\\    
\hline
HC$_3$N			&2.8(6)\,10$^{14}$&22(4)        &6(3)\,10$^{12}$&75(18)	        & 7.6\,10$^{-7}$  &12 &C$_3$N\\         
H$^{13}$CCCN		&4.9(8)\,10$^{12}$&12(1)            &--		    &--			    & 1.4\,10$^{-8}$	 &3  &C$_3$N\\	
HC$_5$N			&3.2(3)\,10$^{13}$&28(1)	       &1.5(4)\,10$^{12}$  &149(53)	    & 8.9\,10$^{-8}$  &15	 &C$_3$N\\ 
\hline
SiO				&1.89(6)\,10$^{14}$&14.3(6)     &3.0(1)\,10$^{13}$  &40(1)        & 4.0\,10$^{-7}$  &4	 &SiO	\\ 
$^{29}$SiO		&1.38(2)\,10$^{12}$&13.3(3)     &3.0(5)\,10$^{12}$  &38(5)        & 3.1\,10$^{-8}$  &3	 &SiO	\\ 
$^{30}$SiO		&1.04(1)\,10$^{13}$&14.6(3)     &4.8(8)\,10$^{12}$  &23(2)        & 2.8\,10$^{-8}$  &3	 &SiO   \\ 
SiS				&1.14(6)\,10$^{14}$&50(4)		  & -- &	--  		                 & 2.1\,10$^{-7}$  &7	 &SiO & High excitation levels \\  
CS				&1.13(4)\,10$^{14}$&14.8(0)  &--                &--		    & 2.1\,10$^{-6}$  &2	 &SiO	\\ 
$^{13}$CS		&2.06(7)\,10$^{13}$&13.5(0.8)	  &--                &--			& 3.8\,10$^{-8}$  &2	 &SiO	\\ 
C$^{34}$S		&6.87(3)\,10$^{13}$&16.3(1)	  &--                &--			    & 1.3\,10$^{-7}$   &2	 &SiO	\\ 
\hline
SiC$_2$			&2.7(2)\,10$^{14}$&23(2)       &1.5(8)\,10$^{13}$&204(218)   	    & 7.7\,10$^{-7}$   &23	 &C$_2$H	\\ 
HNC				&5.10(4)\,10$^{13}$&7.2(1)      &	--	    & --			        & >1.4\,10$^{-7}$  &2	 &C$_2$H \\  
CCS 				&2.5(9)\,$^{13}$&15(4)		  &--			&--				    &>6.7\,10$^{-8}$	 &4	 &C$_2$H	\\  
CCH				&4.0(1)\,10$^{15}$&16(1)        &--                &--			&>1.1\,10$^{-5}$  &10	 &C$_2$H	\\  
C$_4$H$^a$		&6.6(2)\,10$^{14}$&46(2)  	  &--                &--			    &>1.8\,10$^{-7}$  &8	 &C$_2$H	\\  
C$_4$H$^b$		&7.4(4)\,10$^{14}$&50(3)  	  &--				  &--		    &>2.0\,10$^{-7}$  &8	 &C$_2$H	\\  
CN				&1.7(3)\,10$^{13}$&4.2(0.4)	  &--  		      &--			    &>1.3\,10$^{-5}$  &3	 &C$_2$H&\\       
CCCN				&5.6(6)\,10$^{13}$&29(4)   	  &--                &--			    &>1.7\,10$^{-7}$   &3  &C$_2$H&		\\   
$^{13}$CCCN		&8\,10$^{12}$&29					&-- 				&--			    &>$2.2\,10^{-8}$    &1	&C$_2$H&LTE.$T_\mathrm{rot}$(C$_3$N) assumed 	\\        


\hline                                   
\label{density}
\end{tabular}
\end{table*}

\begin{table*}
\caption{Ratio betwen the abundance obtained for AFGL\,2233 (abundance in second column) and other sources ($X(source)/X(AFGL2343)$.
\emph{a:\citet{2014apn6.confE..88S}; b:\citep{debeck2020}; c:\citet{survey10420}; d:\citet{zhang2009}; e:\citet{Bujarrabal94a};
f:\citet{agundez17}}.
O-B94, C-B94 and S-B94 correspond to the average values for O-rich, C-rich and S-type stars respectively from \citet{Bujarrabal94a}.
}             
\centering                          
\begin{tabular}{l c c c c c c c c c c c}        
\hline\hline                 
Molecule 		& AFGL2233	& OH231$^a$ 	& IK\,Tau$^b$ 	& IRC+10420$^c$ & CIT\,6$^d$ & IRC+10216$^b$ & W\,Aql$^b$ & O-B94$^e$ & C-B94$^e$ & S-B94$^e$\\
\hline 
CO    			&3.6\,10$^{-4}$	&0.42	&0.31	&1.49	&--		&2.25	&--		&0.84	&2.25	&1.69\\
$^{13}$CO  		&6.5\,10$^{-6}$	&7.72	&2.16	&11.6	&--		&--		&--		&1.70	&3.71	&--\\
\hline 
HCN    			&5.6\,10$^{-7}$	&0.04	&1.18	&1.96	&2.14	&17.9	&5.54	&0.66	&13.2	&7.14\\
H$^{13}$CN  		&2.1\,10$^{-7}$	&--		&0.39	&0.67	&0.96	&--		&--		&--		&--		&--\\
\hline
HC$_3$N   		&7.6\,10$^{-7}$	&0.01	&--		&--		&1.71	&0.26	&1.11	&0.30	&2.90	&--\\
H$^{13}$CCCN  	&1.4\,10$^{-8}$	&--		&--		&--		&--		&--		&--		&--		&--		&--\\
HC$_5$N   		&8.9\,10$^{-8}$	&--		&--		&--		&145.5	&0.90	&--		&--		&--		&--\\
\hline
SiO    			&4.0\,10$^{-7}$	&0.10	&20.1	&3.27	&1.76	&0.15	&0.73	&24.7	&1.79	&40.3\\
$^{29}$SiO 	 	&3.1\,10$^{-8}$	&--		&58.8	&9.14	&2.58	&--		&--		&--		&--		&--\\
$^{30}$SiO  		&2.8\,10$^{-8}$	&--		&47.1	&9.78	&2.90	&--		&--		&--		&--		&--\\
SiS    			&2.1\,10$^{-7}$	&0.14	&22.2	&0.34	&16.4	&0.96	&7.24	&3.47	&10.6	&--\\
CS    			&2.1\,10$^{-6}$	&0.05	&0.39	&0.06	&0.97	&0.10	&0.58	&0.05	&0.83	&0.29\\
$^{13}$CS  		&3.8\,10$^{-8}$	&--		&2.24	&--		&4.00	&--		&--		&--		&--		&--\\
C$^{34}$S  		&1.3\,10$^{-7}$	&--		&0.48	&--		&2.40	&--		&--		&--		&--		&--\\
\hline
SiC$_2$   		&7.6\,10$^{-7}$	&--		&--		&--		&3.14	&--		&0.66	&--		&--		&--\\
HNC    			&1.4\,10$^{-7}$	&0.29	&0.06	&0.70	&1.67	&0.58	&0.22	&0.59	&5.22	&--\\
CCS    			&6.7\,10$^{-8}$	&--		&--		&--		&--		&--		&--		&--		&--		&--\\
CCH    			&1.1\,10$^{-5}$	&--		&--		&--		&0.50	&0.19	&0.93	&--		&--		&--\\
C$_4$H$^a$ 	 	&1.8\,10$^{-7}$	&--		&--		&--		&22.4$^f$	&--		&--		&--		&--		&--\\
C$_4$H$^b$	  	&2.0\,10$^{-7}$	&--		&--		&--		&--		&--		&--		&--		&--		&--\\
CN    			&1.4\,10$^{-5}$	&--		&0.01	&0.10	&1.92	&0.07	&0.42	&--		&--		&--\\
CCCN   		 	&1.7\,10$^{-7}$	&--		&--		&--		&11.2	&2.36	&--		&--		&--		&--\\
\hline                                  
\label{comparison}
\end{tabular}
\end{table*}

 \begin{table}
\caption{Isotopic ratios derived comparing the derived molecular abundances.}             
\centering                          
\begin{tabular}{l c c}        
\hline\hline                 
Ratio & Value & From\\
\hline 
$^{12}$C/$^{13}$C	&55			&CO\\
$^{12}$C/$^{13}$C	&2.5--7.9	&HCN\\
$^{12}$C/$^{13}$C	&55 			&CS\\
$^{12}$C/$^{13}$C	&57			&HC$_3$N\\
$^{12}$C/$^{13}$C	&7			&C$_3$N\\
$^{32}$S/$^{34}$S	&16			&CS\\
$^{28}$Si/$^{29}$Si	&13			&SiO	\\
$^{28}$Si/$^{30}$Si	&14			&SiO	\\
\hline                                   
\label{ratios}
\end{tabular}
\end{table}

\subsection{Abundance and temperature estimate}

In general we have used the method of the rotational diagrams \citep{rtd}. In most of the cases a single component was enough to fit the data obtained. However, in certain cases, in which the presence of two components was clearly observed in the rotational diagrams, we fitted both components which allowed us to obtain a column density and rotational temperature for each of them. These results are presented in Table \ref{density}. In the cases in which we fitted two components the density and temperature are presented as $N^{C}$ \& $T_\mathrm{rot}^{C}$ and $N^{H}$ \& $T_\mathrm{rot}^{H}$ for the cold and hot components respectively. When only a single component was fitted, it was labeled as the cold component. 

To obtain an estimate of the total abundance of each species we used that $X_\mathrm{mol} = N_\mathrm{mol}/N_\mathrm{H_2}$. Since we have an estimate of $N_\mathrm{H_2}$ for the outer species and the innermost ones, and given that $N_\mathrm{mol} = N_\mathrm{mol}^C + N_\mathrm{mol}^H$ we could have an estimate of the fractional abundances of the different species. These results are also shown in Table \ref{density}.

In particular we adopted that CO and \trece\ would have similar $N_\mathrm{H_2}$, Si-bearing species and CS would have a value of $N_\mathrm{H_2}$ similar to SiO, HCN and H$^{13}$CN also would present similar values of $N_\mathrm{H_2}$, and the carbon chains would have a similar $N_\mathrm{H_2}$ as C$_2$H and C$_3$N. The emission observed from HC$_5$N 9--8 is relatively weak; its real extent can be assumed to be smaller. Therefore, we assumed that its extent is similar to that of C$_2$H (or C$_3$N). Previous results, as those presented by \citet{agundez17}, show that these are reasonable assumptions.

The results for the different species are commented in { the following sub-sections.} 

\subsubsection{CO \& $^{13}$CO.}
\label{CO}


In the case of CO we count with high excitation lines, and thus we were able to fit both a hot and a cold component.
The hot component has a column density of $1.4 \times 10^{14}$\mbox{cm$^{-2}$} and a \trot\ of 413\,K, while the cold one has a column density of $2.46 \times 10^{17}$\mbox{cm$^{-2}$} and \trot = 16\,K.  In the case of \trece\ only two transitions where available from our data and a single component was fitted ($N = 4.4\,10^{15}$\mbox{cm$^{-2}$}, \trot = 7.8\,K). 
{ The temperature obtained for \trece\ is significantly lower than that obtained for the cold component of \doce. However, if we just only take into account for the \doce\ fitting up $J=1-0\ \&\ 2-1$, the rotational temperature derived is $\sim 10$\,K while the column density remains the same. This suggests that the value of $T_\mathrm{rot}$ derived is artificially low, but the column density is reliable.}

The column density derived via our LVG modeling (see Sect.\,\ref{model}) is $3.4 \times 10^{20}$\mbox{cm$^{-2}$} and thus the fractional abundances with respect to H$_2$ obtained are $7 \times 10^{-4}$ and $1.3 \times 10^{-5}$ respectively for CO and \trece.

The value of the CO fractional abundance ($3 \times 10^{-4}$) was already an input in our LVG model (Sect.\,\ref{model}) and deriving it here -- from the rotational diagrams and the density structure obtained above -- works as a test of the validity of the approximations made.

The optical depths obtained in the rotational diagrams for the different CO transitions range between 0.15 and 2 for the low-excitation lines (much lower values are obtained for the high-excitation lines). These values are taken into account and corrected in our rotational diagram fitting \citep[see][]{survey10420}. 



\subsubsection{HCN, HC$_3$N, CN and HNC}
\label{HCN}

The particularly high intensity of the HCN $1-0$ line has been already reported by \citet{Likkel96}. These authors 
suggested that this high intensity could be due to 
an abnormally high HCN abundance.
The latter case is associated by these authors to a N-enrichment related with the massive nature of this object. 

We have tried to model the HCN 
line profiles as well as the HCN $J=1-0$ interferometric maps using a LVG approach. However, the intensity of this transition seemed too intense, suggesting indeed the presence of another contribution we could not model.
{ The fitting revealed that our synthetic profiles lacked on emission for the $J=1-0$ emission in the inner regions of the CSE ($r<4''$).
A possible interpretation of this fact is that IR pumping has an important effect in the excitation of HCN $J=1-0$.
Also, the spatial resolution of the interferometric HCN $J=1-0$ maps is poor to clearly identify any emitting region that could have 
importance in this fitting like high density regions. }
Thus, we ignored this transition for the density and temperature determination; we used only $3-2$ and $13-12$ transitions within the rotational diagram approach. 

The value of $E_\mathrm{upp}$ of these two transitions are very different (12.8 and 387.0\,K respectively) and so they probably trace different regions of the envelope as it can be seen in other species here presented. The properties derived for HCN would be a mix of these two different regions. 

The opacities obtained for the two transitions are low (0.15 and $4.3 \times 10^{-3}$ respectively) so no deviations are expected due to opacity effects. Mixing two regions is expected to derive higher temperatures for the external (cold) region as well as lower values of the density in the rotational diagrams. Thus, until further lines are observed to fill the gap between the two points presented in the HCN rotational diagram, the values of $T_\mathrm{rot}=51$\,K and $N = 2.8 \times 10^{14}$cm$^{-2}$ 
{ shall be regarded as intermediate values between that of the hot and the cold regions.}
{ If we use this column density and temperature to generate a synthetic HCN $J=1-0$ profile, the result is significantly lower than the profile observed.}

From the HCN $1-0$ interferometric map we constrained the extent of this emission and the associated column density (Sect.\,\ref{shape}). 
{ The line intensity observed in this interferometric map is compatible with that obtained with the IRAM 30m telescope. Also, it is worth noting that, even thought the emission of this line is abnormaly high, we can use this map to contrain the extent of the HCN emission.} 
These values lead to an abundance value $>1.1 \times 10^{-6}$. This value is low for C-rich AGBs but at the same time high if compared with that of standard O-rich AGBs \citep[][hereafter B94]{Bujarrabal94a}, but similar to that obtained for the yellow hypergiant star IRC+10420 \citep{survey10420}. 

\begin{figure}
   \centering
      \includegraphics[width=\hsize]{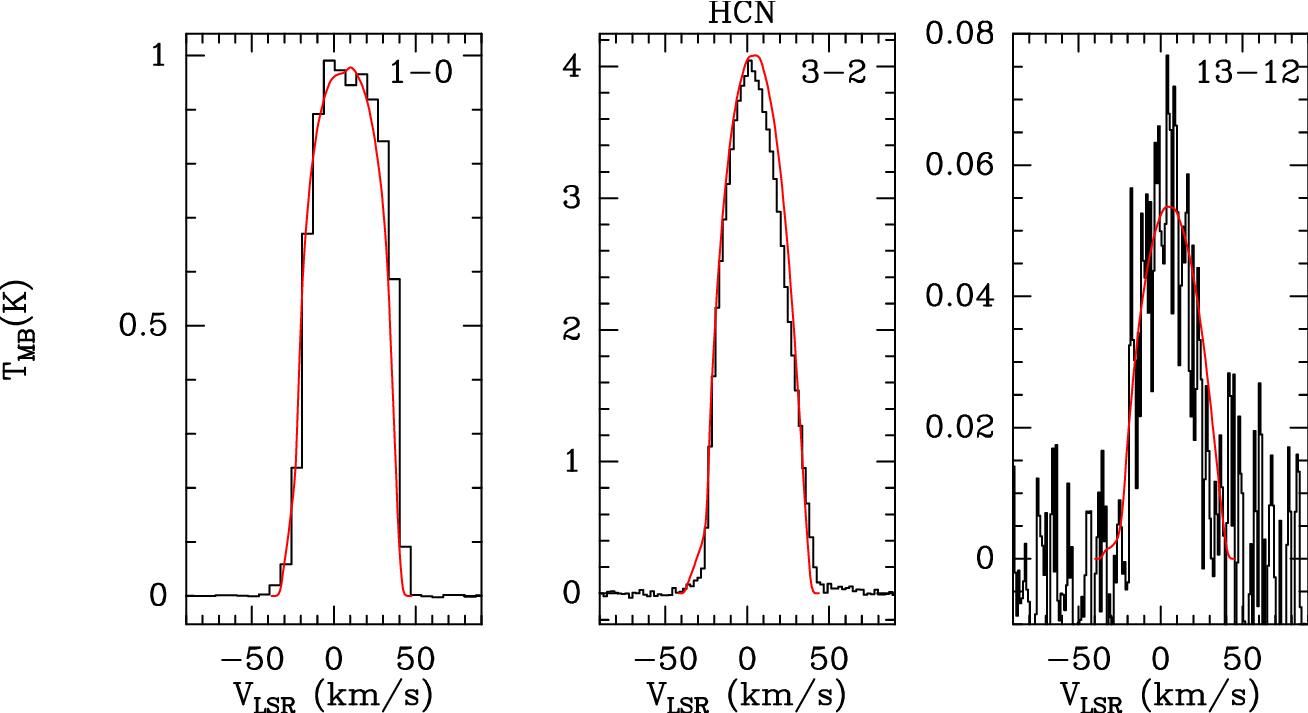}
      \caption{Fitting of the HCN transitions observed, taking into account the IR-pumping.
              }
         \label{HCN_vib}
   \end{figure}

{ In order to understand the impact of the IR-pumping in the relative strenght of the HCN $J=1-0$ line, we imposed HCN $v$>0  collisional rates to be 10$^{4}$ times lower than those of the known $v$=0 values \citep{HCN_cols}. With this aproximation, and adopting the density and temperature values obtained from CO (see Table \ref{result:1}), we were able to fit the three observed transitions (see Fig.\,\ref{HCN_vib}) by imposing a dust temperature of 900\,K for $r<5 \times 10^{15}$cm and 100\,K beyond that radius until $7.8\time 10^{16}$cm, which is the HCN $J=1-0$ extent. This dust temperature distribution was meant to maximize the effect of the IR-pumping on the $J=13-12$ transition in the innermost regions and  that of the $J=1-0$ in the outermost regions of the HCN emitting region, while keeping the balance for the $J=3-2$ transition, which is affected by both regions. In order to fit the data we also had to scale the profiles, multiplying them by 0.92, 1.3 and 0.9 respectively. These values are within the calibration errors of the different telescopes and bands.
The HCN abundance derived from this fitting is $10^{-5}$, which is similar to those values obtained for C-rich AGB stars \citep{Schoier13}.}

The abundance derived for CN is $>1.4 \times 10^{-5}$. This value is high for both C-rich and O-rich stars (except for CIT\,6). On the contrary, HNC abundance ($\sim 1.4 \times 10^{-7}$) falls between both chemical types, while HC$_3$N abundance is similar to that of C-rich AGB stars. 


{ The adjustement of the rotational diagrams of H$^{13}$CN required two components. However, since we have observed that the IR-pumping plays an essential role in the excitation of the HCN lines, it might also impact on the H$^{13}$CN excitation (i.e. the LTE assumption is not valid in this case), specially on the $J=1-0$ emission. Since, as it happens for HCN, we do not count with collisional rates for H$^{13}$CN, the derived LTE abundances are uncertain, and thus any isotopic ratio derived comparing both abundances would not represent the reallity.}

Six transitions of the different isomers H$^{13}$CCCN, HC$^{13}$CCN and HCC$^{13}$CN have been observed in total. 
However, due to sensitivity, only two lines were used to estimate the column density and rotational temperature. 

\subsubsection{C$_3$N \& $^{13}$CCCN.}


For AFGL 2233 the intensity of the C$_3$N lines is stronger than that of the C$_4$H lines, however for IRC +10216 (a well-studied nearby C-rich AGB star) the intensities of the C$_3$N lines are similar to those of the C$_4$H lines. \citet{Guelin78} found that for IRC +10216 the abundance of C$_4$H is 4 times higher than that of C$_3$N. However, in our case the abundances are similar.
If we compare case by case, 
in AFLG\,2233 $X_\mathrm{C_4H}$ is low compared with C-rich stars (IRC\,+10216 and CIT-6) and $X_\mathrm{C_3N}$ is similar to that of IRC\,+10216 but low if compared with CIT\,6.

To obtain the estimate of the column density of $^{13}$CCCN, since we only have detected a single line, we imposed 
the same temperature as for C$_3$N and, assumed LTE. The abundance obtained is $> 2.1\times 10^{-8}$. The 
$^{12}$C/$^{13}$C ratio derived for these species is $\sim 8$, very low compared with the rest of the C-bearing molecules for which this ratio could be derived (see Table\,\ref{ratios}), { except for HCN for which the derived value is uncertain (see Sect.\,\ref{HCN}). }
This suggests that the assumptions taken to derive the density of $^{13}$CCCN might not be accurate.

\subsubsection{Si-bearing species \& CS}


In general, the abundances derived for SiO and its isotopologues are low compared with O-rich stars 
as IK\,Tau or the values obtained by B94 for this type of objects (see Table.\,\ref{comparison}). 
It is somehow similar to that of IRC+10420, however this object is known to have SiO constrained to 
a detached shell; this emission is suspected to arise from SiO freed from heated dust grains \citep{ccsio}. 
On the other hand, these abundances are similar to those of CIT\,6 and the values obtained for C-rich stars 
by B94.

The abundance derived for SiS is similar to that obtained by B94 for O-rich stars. 



CS abundance is in general high compared with most of the estimates presented in Table\,\ref{comparison}, 
but closer to that of C-rich stars. 
{ The opacities derived from the two transitions observed are 0.27 and 0.40 for $J=2-1\ \&\ J=5-4$ respectively.}
Interestingly, the $^{13}$CS abundance is similar ($\sim 2$ and $\sim$ 4 times lower)
in AFGL\,2233 compared with IK\,Tau and CIT\,6 respectively. This shows a clear difference in the carbon 
isotopic ratio between these objects. In the case of AFGL\,2233, { regarding CS,} $^{12}$C/$^{13}$C is 55, while for IK\,Tau 
is 9.5 and 13.3 for CIT\,6. These values have 
evolutionary implications as shown by \citet{cisotopic} (see below for a further discussion on the isotopic ratios).

   \begin{figure*}
   \centering
      \includegraphics[width=\hsize]{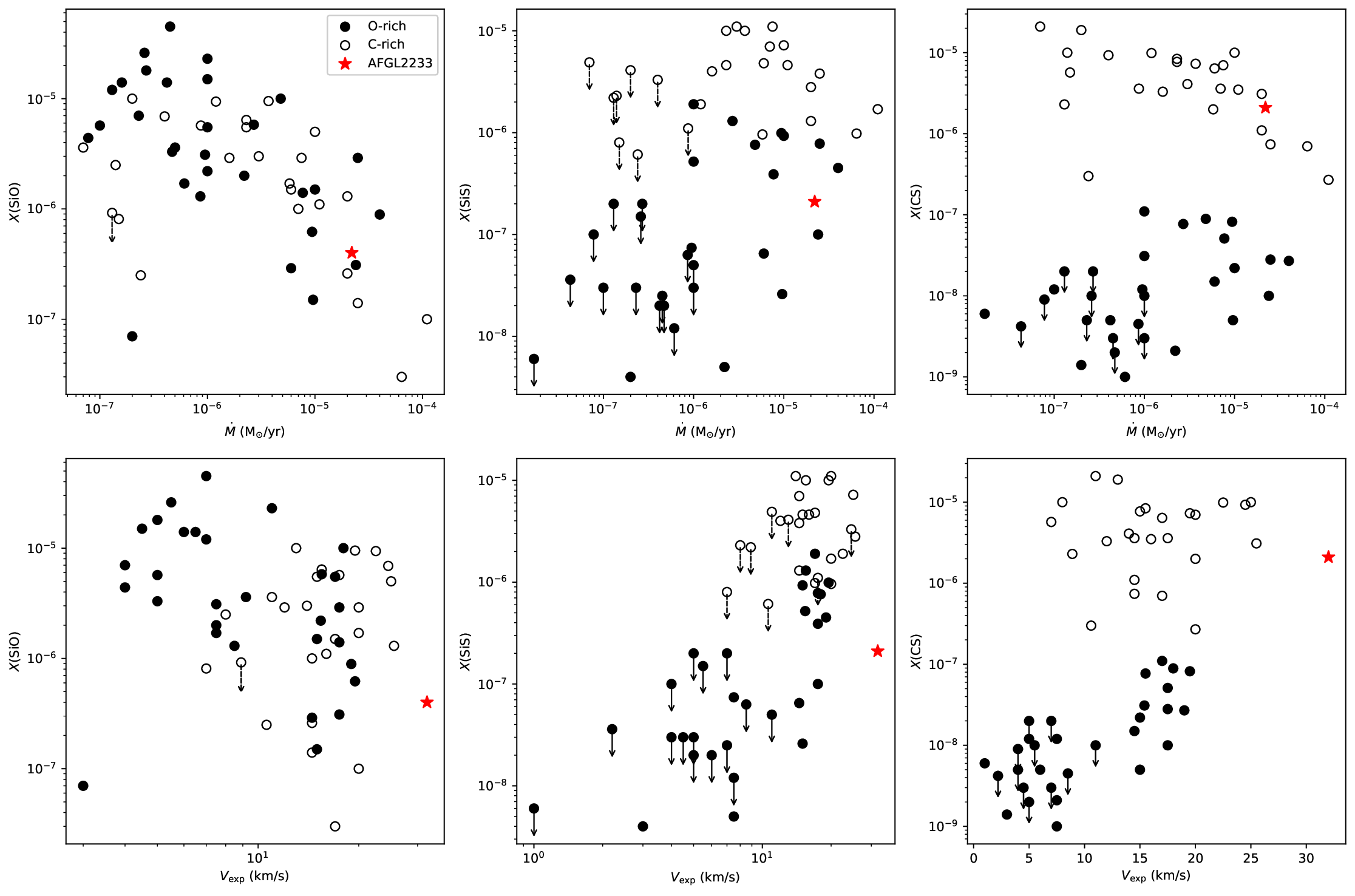}
      \caption{Comparison of the SiO, SiC and CS abundances obtained for C-rich AGB stars \citep{massalkhi-Crich} and O-rich AGBs
\citep{massalkhi-Orich} with those obtained for AFGL 2233, with respect to the mass-loss rates and the expansion velocities.
              }
         \label{sarah}
   \end{figure*}

{ We compared the abundances obtained for these species for AFGL 2233 with those 
derived for a large number of C-rich and O-rich AGB stars by \citet{massalkhi-Crich} and
\citet{massalkhi-Orich} respectively (see Fig.\,\ref{sarah}). The SiO and SiS abundances of AFGL 2233 do not clearly suggest an
O-rich or C-rich behaviour. Only for CS AFGL\,2233 falls clearly within the C-rich region. Even though it might seem
that CS abundace of AFGL\,2233 follows the $X_\mathrm{CS}(V_\mathrm{exp})$ trend of the O-rich AGBs, the values obtained for massive evolved
stars are not higher than several times 10$^{-7}$ \citep{chemyhg,survey10420,ziurys2025}, well bellow the
values found for C-rich stars (and AFGL\,2233).}

\subsubsection{C$_2$H \& C$_4$H}

These two species are only detected in sources presenting a C-rich chemistry in their circumstellar envelopes. 
However, the abundances here obtained for C$_2$H 
and C$_4$H are respectively higher and lower than in typically found in C-rich stars (such as CIT\,6 and IRC\,+10216).

\begin{figure}
   \centering
   \includegraphics[width=9cm]{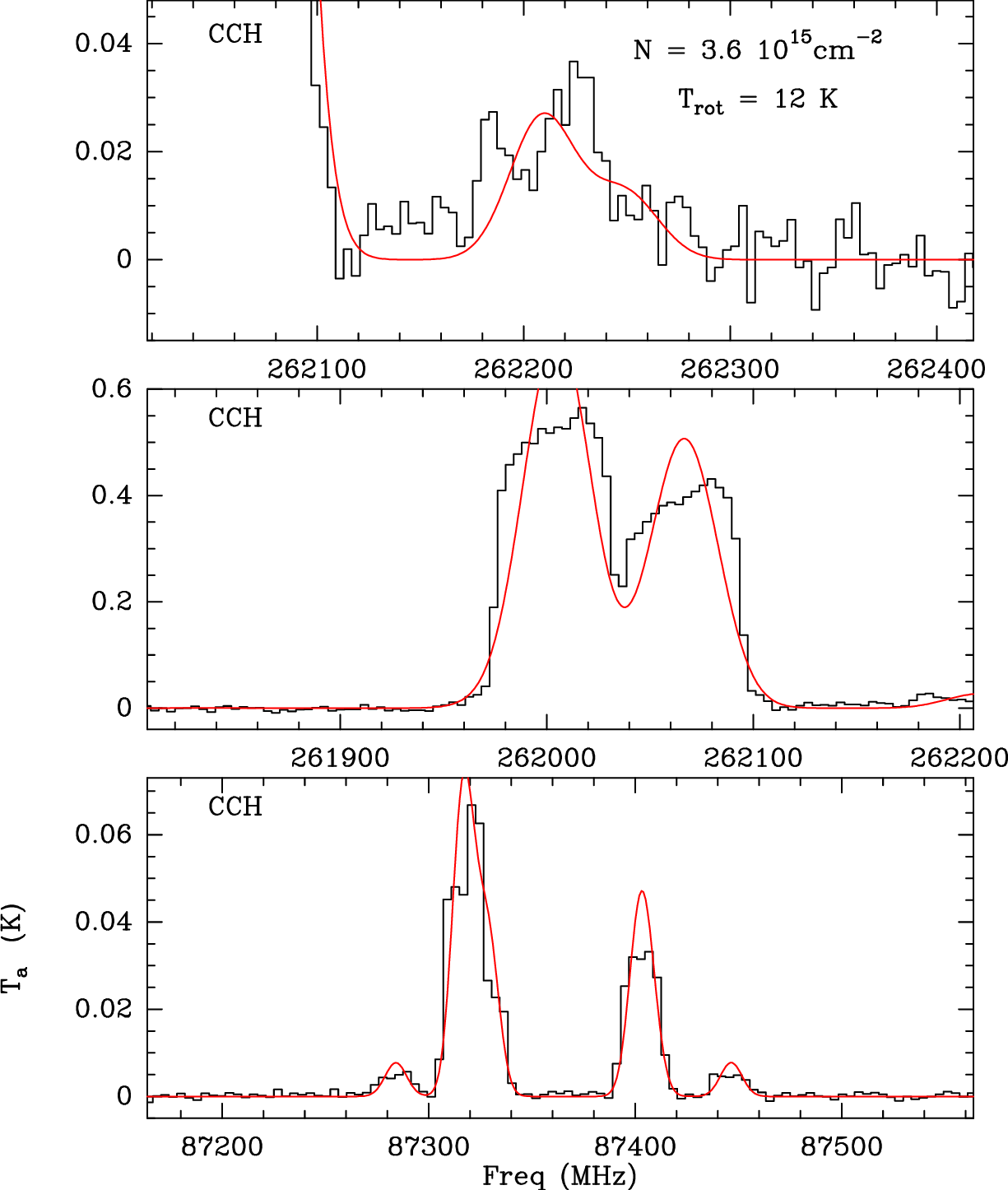}
      \caption{LTE fitting of the CCH line emission.
              }
         \label{madexLTE}
   \end{figure}

{ Despite the number of lines observed for C$_2$H (16), their $E_\mathrm{up}$ are 4.2\,K or $\sim$25\,K for all of them. Therefore, the values obtained via de rotational diagrams might be not accurate. We confirmed the values obtained modeling the observed lines using MADEX 
, in the LTE approach. The results so obtained are $N = 3.6\times 10^{15}$\,cm$^{-2}$ and $T_\mathrm{rot} = 12$\,K (see fitting in Fig.\,\ref{madexLTE}). These results are very similar to those presented in Table\,\ref{density}. The C$_2$H abundance is also very similar to the value derived by \citet{c2h2}.}

The formation of both species, C$_2$H and C$_4$H, is deeply connected. The formation paths are \citep[see][]{agundez17}:

\begin{eqnarray}
\rm{C_2H_2 + \emph{hv} \rightarrow C_2H + H}
\end{eqnarray}
\begin{eqnarray}
\rm{C_2H_2 + C_2H \rightarrow C_4H_2 + H}
\end{eqnarray}
\begin{eqnarray}
\rm{C_4H_2 + \emph{hv} \rightarrow C_4H + H}
\end{eqnarray}
\begin{eqnarray}
\rm{C_2H_2 + C_2 \rightarrow C_4H + H}
\end{eqnarray}

\citet{agundez17} found that, for IRC\,+10216, the regions where the emission from these species arise are similar. Thus, the processes taking place (i.e. photodissociation, eqs.1 and 3), should have similar impact for both molecules. However, as C$_4$H is relatively underabundant in our case, 
there should be a process -- or contrary a process is not taking place --, which as result avoids or limits the formation of C$_4$H  compared with C$_2$H.

\citet{c2h2} showed that C-rich AGBs with high expansion velocities, i.e. the HVC stars, present high C$_2$H abundances when compared with standard low-velocity C-rich AGB stars. These authors suggested that the higher expansion velocity { could prevent acetylene to stick to grains, compared to objects with low expansion velocities (as the depletion radius is proportional to $V_\mathrm{exp}^{3/2})$. This would result in more acetylene available to be photodisociated into C$_2$H via reaction (1).} However, this would also affect C$_4$H in the same manner (reactions (2) \& (3)), contrary to what is observed.

Other option would be that the formation of C$_2$H is enhanced. In particular, there is another path to form this latter species, related to the abundance of HC$_3$N.

\begin{eqnarray}
\rm{HC_3N + \emph{hv} \rightarrow C_3N + H}
\end{eqnarray}
\begin{eqnarray}
\rm{HC_3N + \emph{hv} \rightarrow C_2H + CN}
\end{eqnarray}

\begin{figure*}
   \centering
   \includegraphics[width=7cm]{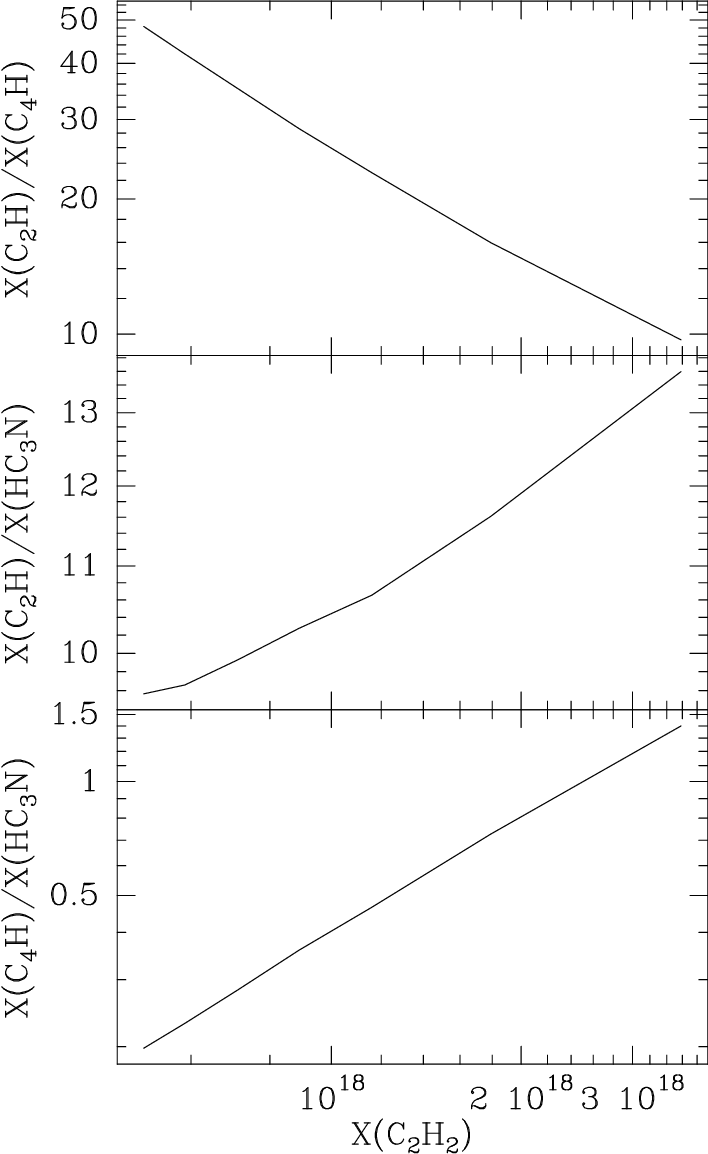}
   \includegraphics[width=7.2cm]{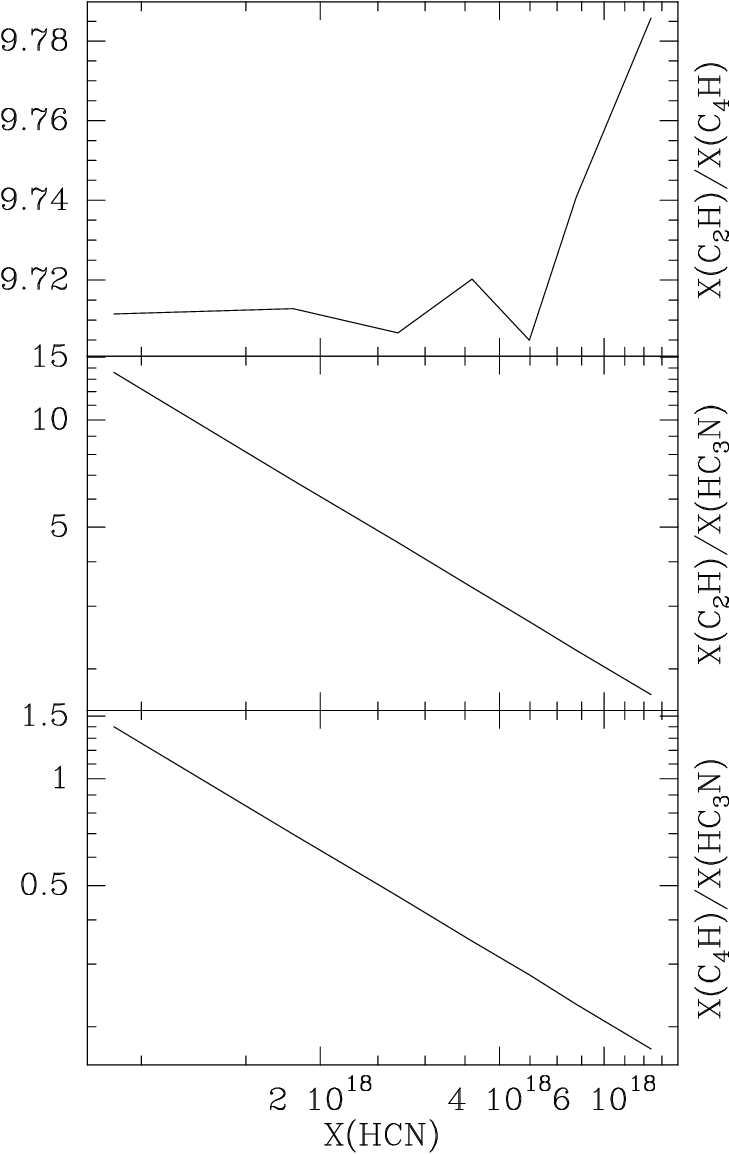}
      \caption{Variations of the abundance ratios X(C$_2$H)/X(C$_4$H) (top), X(C$_2$H)/X(HC$_3$N) (middle) and X(C$_4$H)/X(HC$_3$N) (bottom). Left: HCN abundance is kept constant while varying that of C$_2$H$_2$. Right: C$_2$H$_2$ abundance is kept constant while varying that of HCN.
              }
         \label{chemol}
   \end{figure*}

In order to try to understand the differences observed, we ran chemical models as those presented by \citet{agundez17}, varying both of the inputs for the reactions cited, i.e. acetylene and HCN. The characteristics of the circumstellar gas used for the chemical model is an average of the two inner shells fitted in Sect.\,\ref{model}. The results are presented in Fig.\,\ref{chemol}.

It can be seen that the impact of enhancing the HCN abundance has only a minor effect on the C$_2$H/C$_4$H abundance ratio, with a decrease of both species of $\sim$15\%, being this small difference the responsible for the (small) increase in the ratio observed in the figure. { On the contrary, note that varying the abundance of HCN has, however, a clear impact on the HC$_3$N abundance, which does not seem to severely affect the formation of C$_2$H via reaction (6).}

However, if we modify the amount of acetylene available, the C$_2$H/C$_4$H abundance ratio presents significant changes. In particular, the abundance ratio here derived is $\sim$58, while for CIT\,6 or IRC\,+10216 is close to 1. As observed in Fig.\,\ref{chemol}, the large ratios observed in AFGL\,2233 could be reached in cases of low acetylene abundances.

This increase in the ratio is due to that the reactions (1) to (4) are essential to the formation of C$_4$H; but while it also affects the formation of C$_2$H, this molecule could be formed via reaction (6), i.e. with no acetylene involved. Thus low abundance of acetylene would minimize reactions (1) to (4), severely affecting the abundance of C$_4$H, while C$_2$H could still be formed.

{ Despite this, ISO observations, as those presented by \citet{ISOSiC} showed that this object clearly shows a strong presence of acetylene, as predicted by \citep{c2h2}. 
}

\begin{figure*}
   \centering
   \includegraphics[width=15cm]{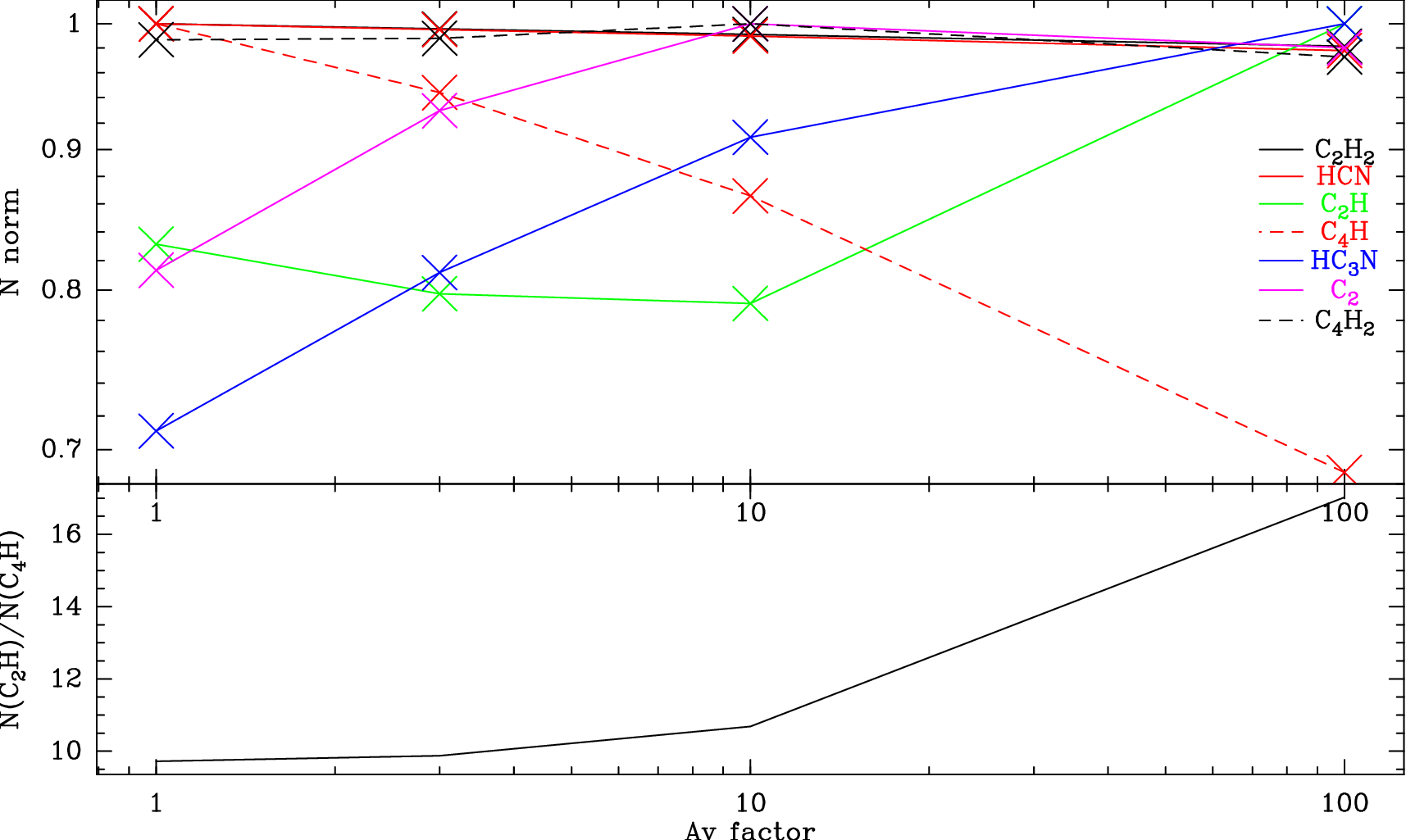}
         \caption{Chemical model result of the column densities (normalized for comparison purposes) 
         for different values of radiation field ($top$).
         		  Variation of the column density ratio C$_2$H/C$_4$H for different values of radiation field ($bottom$).
              }
         \label{Av}
   \end{figure*}

{
Interesingly, this ratio has been found to be high in PDRs compared with AGB stars \citep[see e.g.][]{Teyssier04}. Our chemical models
showed the same trend. We modeled the column densities resulting after increasing the standard interstellar radiation field \citep{Draine} by a factor  1, 3, 10 and 100 and found that the ratio C$_2$H/C$_4$H increases from $\sim 9$ to $\sim 17$ (Fig.\,\ref{Av}). This suggests that photochemistry might be the source of this high ratio. We note that the models used to check the variations of the X(C$_2$H)/X(C$_4$H) ratio assumed an average value of the gas properties obtained in Table\,\ref{result:1} for the regions where C$_2$H is observed, since we are interested in checking ratio trends qualitatively, rather than obtaining quantitative estimates of the UV field, for what we of lack essential information.
}

\subsubsection{NH$_3$}

Ammonia has been also detected toward this object with HIFI (see Table\,A.1). In particular, the transition $1_{0,0}-0_{0,1}$ of ortho-NH$_3$ was observed with HIFI. This is the only line observed of this species in AFGL\,2233. 

Since we count with density and temperature profiles, and collisional rates for this molecule, we modeled the line emission using the LVG approach with MADEX. 

As the regions and the mechanisms responsible for the NH$_3$ formation are not clear \citep[see e.g.][]{agundez20}, we assumed a similar distribution as found for IRC\,+10216 \citep{10216_NH3}, { scaled for the mass loss properties derived for this object, i.e. $R_\mathrm{out}\sim 5\times 10^{16}$cm}, and used the temperature and density profiles obtained above (see Table\,\ref{result:1}).

\begin{figure}
   \centering
   \includegraphics[width=\hsize]{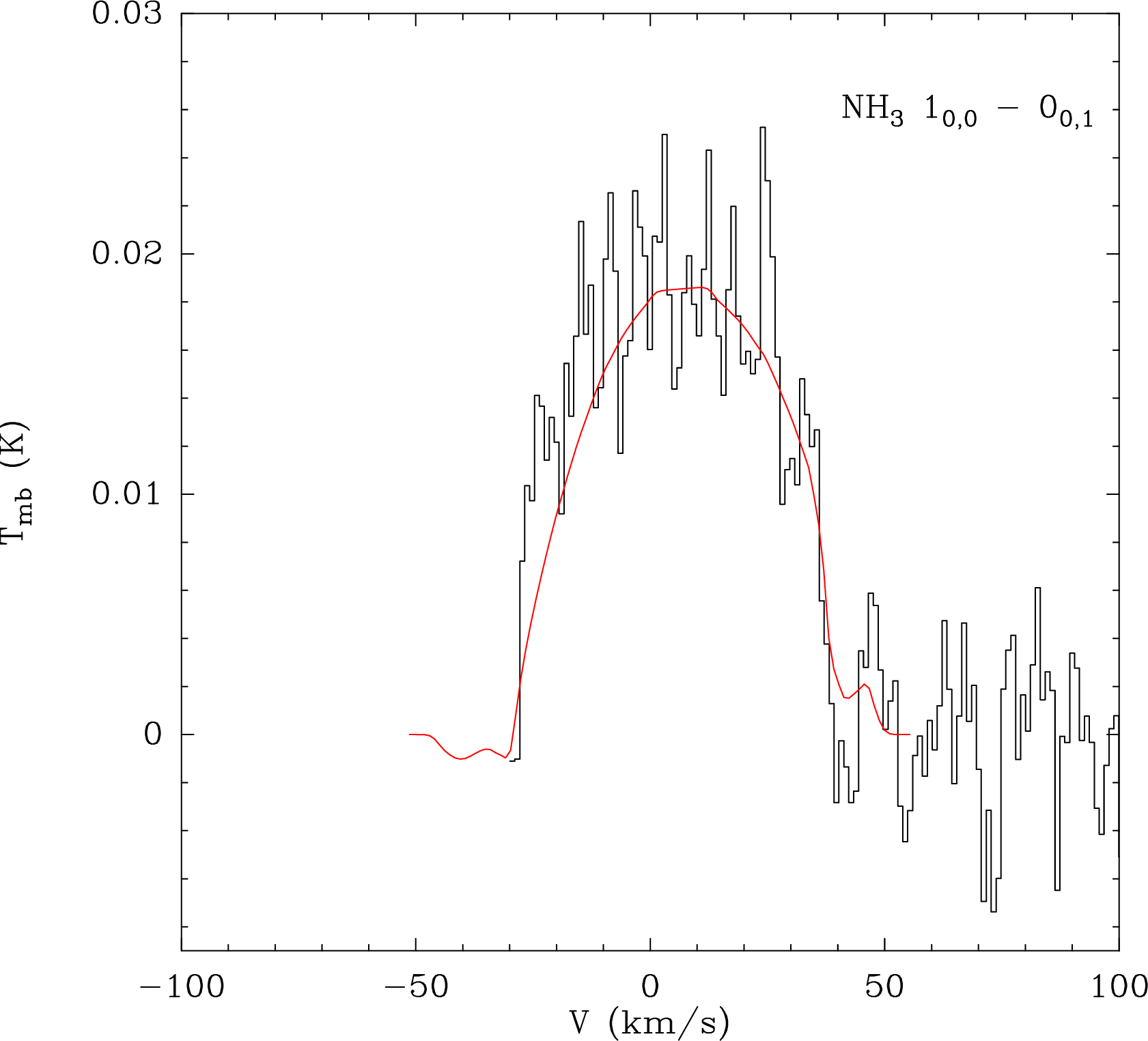}
      \caption{Fitting of the NH$_3\,1_{0,0}-0_{0,1}$ line.
              }
         \label{nh3}
   \end{figure}

The abundance value obtained, $X_\mathrm{NH_3}$ of $2.5\times 10^{-5}$, is high { compared with those derived for other objects.} The only source with a similar abundance is W\,Aql \citep[$1.7 \times 10^{-5}$,][]{debeck2020}. In case the emitting region is smaller the abundance needed to fit the profile must be higher. However, if we assume the extent of NH$_3$ is similar to that derived to CO, the abundance needed is similar, but in addition the synthetic profile obtained presents clear self-absorption. Therefore, the values of the abundance and outer radius reported above seem to be accurate. 
The result of the fitting can be seen in Fig.\,\ref{nh3}.

\citet{menten00nh3} studied the abundances of ammonia in different objects O-rich objects. They found that the abundance of ammonia was high ($2\times10^{-7}-3\times10^{-6}$). On the contrary, the value obtained for IRC\,+10216 \citep{10216_NH3} was significantly lower ($4\times10^{-8}$). The value obtained here is the higher reported for an evolved star. { However, it must be kept in mind that, as in the work by \citet{menten00nh3}, our model does not take into account the effect of the IR pumping, and thus could lead to an overestimate of the ammonia abundance. \cite{wong18} confirmed this fact, showing that the abundance estimate was few times lower. In any case, regardless of this possible modeling issue, since the correction factors obtained by \citet{wong18} are rarely an order of magnitude, the abundance of ammonia in this object, while not being accurate, would still be extremelly high.}

While the higher abundance observed in O-rich (high-mass) stars could be associated with an enhanced abundance of Nitrogen as expected for the Hot Bottom Burning (HBB) process taking place in these objects, \citet{wong18} found it to arise from clumpy structures. These authors suggested NH$_3$ emission traces swept-up material, suggesting that shocks might also contribute to the formation of ammonia.


\subsubsection{H$_2$O}

Two molecular lines of H$_2$O have been detected in the envelope of AFGL\,2233. 
If we compare the intensity of these transitions with those observed in IRC\,+10216 \citep{h2o_10216}, the prototypical C-rich AGB stars, and in O-rich objects \citep{h2o_orich}, we find that those of the HVC stars are more intense than expected. 

In particular, for IRC\,+10216, the intensity obtained for the 1$1_{1,0}-1_{0,1}$ transition is lower than 0.5\,K (in $T^*_\mathrm{a}$) and $\sim$0.6\,K for 1$1_{1,1}-0_{0,0}$ \citep{h2o_10216}. If we scale the fluxes found for AFGL\,2233 (table\,A.1) with the distance of IRC\,+10216 \citep[130pc,][]{agundez12} these fluxes would be 3.5 and 5.3\,K respectively, $\sim$ 7--9 times higher than those of IRC\,+10216. { Note that the mass loss rates are similar for both objects \cite{Velilla19}.}

If we compare the integrated intensities of these lines (see Table\,\ref{table:h2o}) with the distance-corrected integrated intensities presented by \citet{h2o_orich} we find that those reported here are higher than those of W\,Hya, R\,Dor, o\,Cet, R\,Cas, IK\,Tau and TX\,Cam, slightly lower than that of IRC\,+10011, AFGL\,5379 and OH\,26.5+0.6. Also, it is interesting to remark that the relation between the mass loss ($1.8 \times 10^{-5}$\my\ for the innermost shell where H$_2$O is expected to be located), and the intensity of the water lines follows the same trend as that observed in O-rich stars \citep[Fig\,2,][]{h2o_orich}.




In the case of C-rich evolved stars, it has been shown that H$_2$O is located in the inner regions of the envelopes, either if its formation is related with shocks \citep{h2o_shocks}, photochemistry due to clumpiness \citep{h2o_clumpy}, 
being the latter the most extended H$_2$O formation radius (40--400\,au $\sim 6 \times 10^{14}$ -- $6 \times 10^{15}$ cm). 

To obtain an estimate of the water abundance we run a LVG code \citep[colisional rates from][]{h2orates}, adopting the density and temperature profiles obtained for CO (table\,\ref{result:1}), and an ortho/para ratio of 3 \citep{h2o_10216}. We tried to model the emission adopting an outer radius for H$_2$O as that suggested by \citet{h2o_kuiper}, but we could not fit both line intensities simultaneously. The best fit was obtained restricting the emission to the innermost shell presented in Table\,\ref{result:1} ($R_\mathrm{out} = 5 \times 10^{16}$cm). Even though this extent is large for what has been found for the H$_2$O emission in evolved stars, the abundance obtained for H$_2$O was very high for a C-rich object and similar to what would be expected for an O-rich evolved star \citep[see e.g.][]{h2o_clumpy}: $7.5 \times 10^{-5}$ and $2.5 \times 10^{-5}$ for o-H$_2$O and p-H$_2$O respectively. It is worth noting that this value is directly related to the mass located within the shell assumed to be the origin of such emission; if the emitting region is more compact, this abundance will be even higher.

\begin{table}
\caption{Intensity of the water transitions observed in T$^*_\mathrm{a}$ (k), Jy, and corrected by distance, $I_d$.}             
\centering                          
\begin{tabular}{l c c c}        
\hline\hline                 
H$_2$O trans. & $I$(Jy) & $T^*_\mathrm{a}$ (mk) & $I_d$ (K\,\kms\,pc$^2$)\\
\hline 
$1_{1,0}-1_{0,1}$&	1.2&		36.8 &	2.0$\,\times\,10^{6}$\\
$1_{1,1}-0_{0,0}$&	4.9&		55.7 &	2.9$\,\times\,10^{6}$\\
\hline                                   
\label{table:h2o}
\end{tabular}
\end{table}

   \begin{figure}
   \centering
   \includegraphics[width=\hsize]{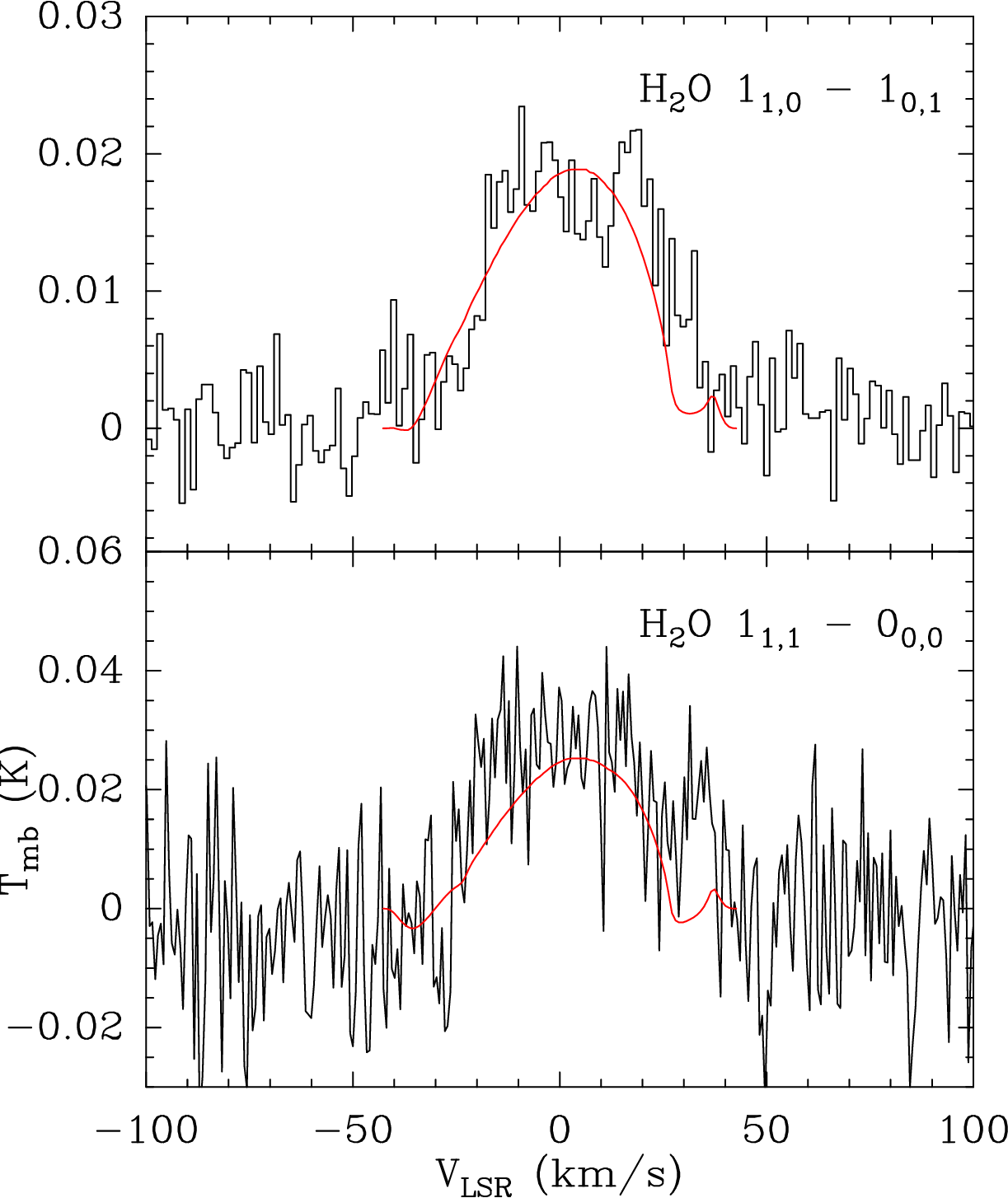}
      \caption{Fitting of the H$_2$O 1$1_{1,0}-1_{0,1}$ and 1$1_{1,1}-0_{0,0}$ lines.
              }
         \label{h2o}
   \end{figure}

\subsubsection{SiN}

SiN is a peculiar species. It has been found only in very few objects: W\,Aql \citep{debeck2020} -- a S-type star--, IRC\,+10216 \citep{SiN_10216} -- C-rich star--, and $\eta$\,Car \citep{Bordiu22} -- a massive evolved star. This would be the fourth detection of this species in the circumstellar gas. 

The origin of the formation of this species has been discussed and linked both to the triggering of chemical reactions by a binary \citep{SiN_WAql} or to the presence of 
a Nitrogen-enriched medium \citep{Bordiu22}.

Since the two transitions observed are too close in energy, the rotational diagram methodology is not a good approach to determine the abundance of SiN. Similarly to what we have done for ammonia and water, we used the temperature and density profiles obtained from CO observations, and model the data using an LTE approach (there are not SiN collisional rates available), assuming the most probable location of SiN.

It is worth noting that NH$_3$ has been found to be particularly abundant in this object. The abundance of ammonia might have a direct impact in the detection of SiN, since, as proposed by \citep{SiN_WAql}, the main path for the formation of SiN is:

\begin{eqnarray}
\rm{Si^+ + NH_3 \rightarrow SiNH_2^+ + H}
\end{eqnarray}
\begin{eqnarray}
\rm{SiNH_2^+ + e^- \rightarrow SiN + H_2}
\end{eqnarray}

Also, the main source of formation of Si$^+$ is the photodisociation of SiS. As a first approach we assume the same extent of SiO and SiS ($7 \times 10^{16}$\,cm, see table\,\ref{sizes}), which, as shown by \citet{Massalkhi24} is reasonable for objects with $\mloss / v_\mathrm{exp}>10^{-6}$ \ms\,yr$^{-1}$\,km$^{-1}$\,s (for AFGL\,2233 this value ranges from $7 \times 10^{-7}$ to $1.2 \times 10^{-6}$ \ms\,yr$^{-1}$\,km$^{-1}$\,s). { As we expect Si$^+$ to appear further than shell 1 (table\,\ref{result:1}) we ignored this shell in the fitting.}

{ Furthermore, we note that when assuming that the extent of SiN as arising from shells 2--4, the profiles presented a horned-like shape, suggesting that the extent assumed was larger than that observed -- the line profiles do not present that horned shape. Therefore, we constrained the extent to just shell 2.}

\begin{figure*}
   \centering
   \includegraphics[width=\hsize]{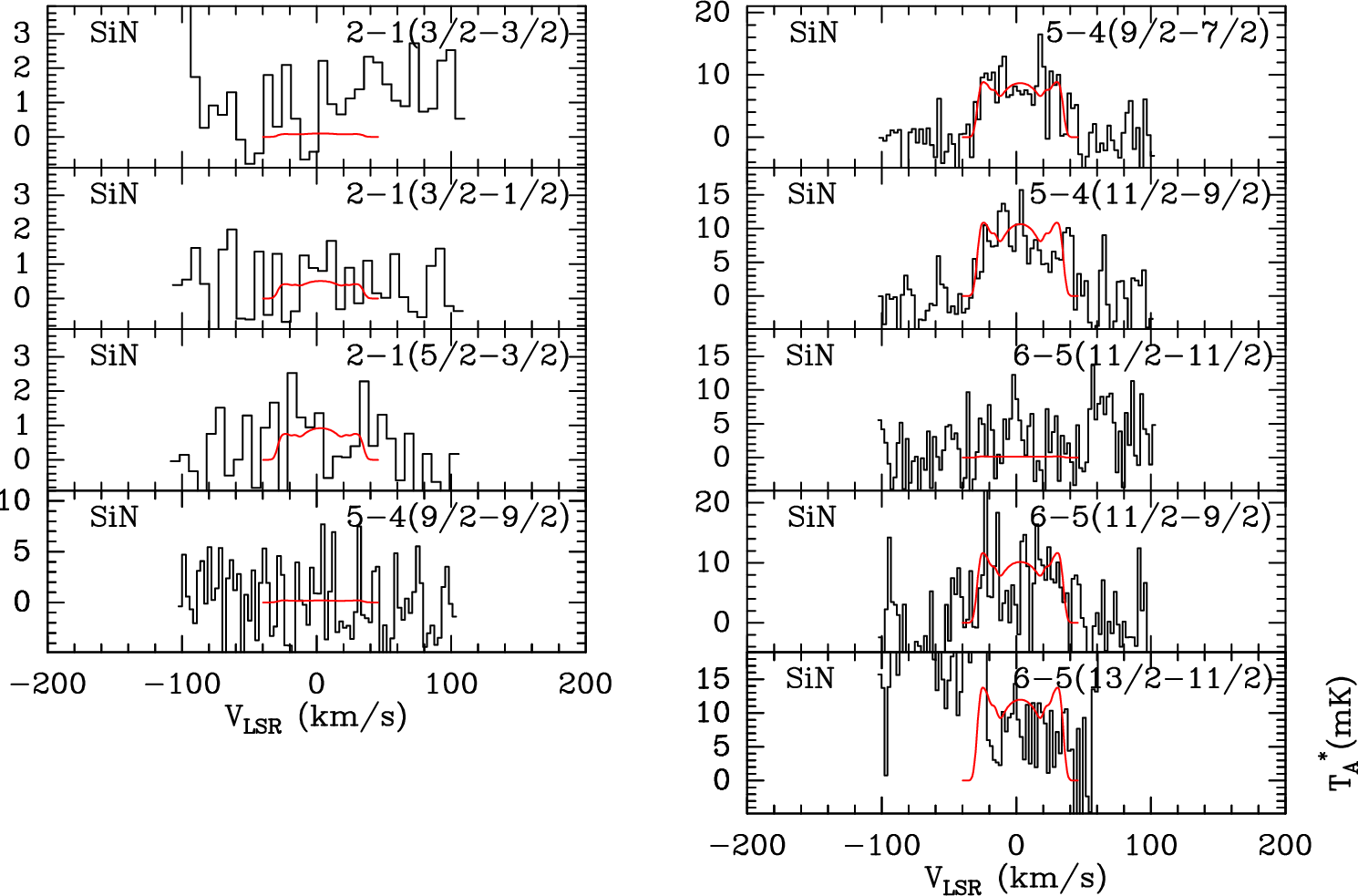} 
      \caption{Fitting of the SiN transitions covered by the line survey.
              }
         \label{sin}
\end{figure*}

The result of the fitting is presented in Fig.\ref{sin}. { To check the validity of our model fitting the two well-detected lines, we also obtained synthetic spectra for all the SiN transtions covered by our survey. As shown in the figure, the fitting also predicts the non-detections of the rest of the transitions -- tentative detection in the case of SiN 6--5(11/2--9/2) (at 261650.18\,MHz). }

The abundance derived is $X_\mathrm{SiN} = 7 \times 10^{-8}$. This abundance is { 2 times lower than} that observed in W\,Aql \citep{SiN_WAql}, and { 2 times larger} than that of $\eta$\,Car \citep{Bordiu22}. It is worth noting that SiN has not been detected in other massive objects such as IRC\,+10420 \citep{survey10420}, VY\,CMa or AFGL\,2343 (priv. comm.),{\  objects that, as observed in IRC\,+10420 \citep{10420no}, are expected to present a N-rich enrichment.}

\section{Discussion}

We have obtained the fractional abundances of different species toward the HVC star AFGL\,2233. At first sight, its chemistry and the abundances obtained are in general compatible with a standard C-rich star rather than with a massive star in its O-to-C chemical transition {  -- after the exhaustion of the HBB}. However certain results reveal the peculiar nature of this object.


While we found no strong signs of $^{14}$N enrichment compared with other objects, as the yellow hypergiant star IRC\,+10420, and the abundances of N-bearing species are compatible with a C-rich chemistry, certain facts might support the presence of a N-rich medium: the C$_2$H/C$_4$H ratio suggest a lack of acetylene in the circumstellar gas, compared with HC$_3$N, and the very high abundance of NH$_3$. Also SiN has been detected in this object, which has been related with the presence of N-enrichment in massive stars \citep{Bordiu22}.


{ The values obtained for $^{12}$C/$^{13}$C --those for which the abundances derived were trustworthy -- were 55--57.} These values are associated with CO, CS and HC$_3$N. Opacity effects have been discarded, and the quality of the fits was reliable. In the case of C$_3$N, the isotopic ratio relies on the assumption of similar rotational temperatures for both isotopologues, however these species are expected to arise from very similar regions, an extended and narrow shell \citep{agundez17}. Therefore the low value of $\sim 8$ probably accurate, at least at that region.

%
%

\citet{cisotopic} and \citet{Ramstedt14} show that low $^{12}$C/$^{13}$C ratios are typical of O-rich circumstellar envelopes, whereas higher ratios are generally associated with C-rich evolved stars. In addition, a high water abundance has been derived, comparable to that found in O-rich objects. These results are difficult to reconcile within a consistent evolutionary scenario.

We can divide these results into two categories according to the regions traced by the different species: 1) results obtained for gas extending from the innermost regions to the outer regions (CO, SiO, and in some extent HC$_3$N), and 2) results constrained to a thin shell structure (C$_2$H/C$_4$H, C$_3$N). Note that while the assumed ring-like structure assumed for HC$_3$N and C$_2$H is the same, interferometric maps of IRC\,+10216 \citep{agundez17} showed that the former species presents a smaller central hole, with a more smooth slope, while C$_2$H and C$_3$N present narrower and steeper slopes. In addition, as the same assumption is applied to all isotopologues, the abundance ratio does not depend on those assumptions. 


The results obtained for the species extending along the CSE could trace a mixture of photospheric initial abundances -- { for example, photospheric abundances evolving with time, creating a gradient in the CSE abundance ratios --}, being those of a C-rich media the ones dominating the results obtained. On the contrary, the results arising from the constrained region traced by a narrow outer ring does not suffer from this mixing. 

In any case, the "ring-constrained" species (C$_3$N, C$_2$H, C$_4$H,...) have not been detected in O-rich evolved stars. This region presents a C-rich chemistry (C/O $>$ 1) but with low $^{12}$C/$^{13}$C, which is characteristic of objects evolving from an O-rich regime to a C-rich regime after the exhaustion of the HBB \citep[e.g.][]{Karakas2017}. The further evolution of those objects would lead to C-rich envelopes with carbon isotopic ratios as those traced by CS or CO. 

We might wonder if the evolution is from a C-rich (extended) environment evolving towards a O-rich medium after the activation of the HBB, but models do not predict such scenario. However, these mixed characteristics have been previously observed in pre-planetary nebulae (pPNe) objects \citep{Garcia-Hernandez2016}. These authors observed OH and H$_2$O in C-rich pPNe with high mass losses. These results are interpreted as a result of a CO photodisociation, leading to a transitory O-rich chemistry as predicted by \citet{cerni04}.

This O-rich chemistry -- in particular the presence of H$_2$O --, the high abundance of NH$_3$ and the presence of SiN, which has been also related to a photoinduced chemistry due to the presence of a hot binary \citep{SiN_WAql}, might suggest that the anomalies found are related to the presence of a similar binary system rather to a nucleosynthesys-related C-to-O change. { Also, the presence of a hot binary-induced photochemistry could explain the high C$_2$H/C$_4$H abundance ratio derived. }

{ The effect of the IR-pumping has been found to be very important in the case of HCN, and, as suggested, it might also be important in the case of --at least-- NH$_3$. The abundace of this latter molecule has to be regarded as an upper limit.}

Further high-angular resolution observations of the different isotopologues might allow to study the gradient of $^{12}$C/$^{13}$C along the radius of the CSE, i.e. how this ratio has evolved during the recent evolution of the star. Also, probing the regions emitting SiN, and C$_2$H and C$_4$H could confirm the presence of the photoinduced chemistry and the possible presence of a companion. 

The impact of a companion, in addition to the chemical signatures, could also appear in the width of the profiles. It has been shown that the expansion velocity of the gas in the orbital plane could be significantly enhanced \citep{bermudez20}. It could be the case that the large line width observed in AFGL\,2233 is due to that the orbital plane is close to edge-on w.r.t. to us. However, the luminosity, even for the revisited distance is still too high for a C-rich intermediate-mass AGB star, { suggesting initial masses in the range $4-9\ms$ (see Sect.\,\ref{SED})}. 


This object might represent a mixture of both scenarios: a C-rich massive evolved stars resulting from exhausted HBB, combined with a binary companion triggering localized photoinduced chemistry. High angular-resolution interferometric molecular line maps could confirm this scenario, or suggest a new one.

{ It is worth noting that, as stated by \citet{Javadi11}, RSG stars cannot undergo a HBB process, thus, these objects would no undergo a O-to-C transition in their photospheres and subsequently in their ejecta. Only massive AGBs and super-AGBs would present this process. More massive stars, such as RSGs or yellow hypergiants -- which do not present TDU or HBB --  presenting N-enrichment are probable a result of intese mass loss events exposing inner layers rich in CNO processed elements or other processes such as e.g. convection or rotation-induced mixing \citep{Schootemeijer19}. Thus, since only AGB stars present HBB -- and therefore its exhaustion -- AFGL\,2233 should propably be a massive (super-)AGB rather than a RSG.}





\section{Conclusions}

{ We ﬁrst re-estimated the luminosity and temperature and
density proﬁles for AFGL 2233 according to the new Gaia DR3 distance. Even though
the new distance sets this source signiﬁcantly closer, its luminosity still suggests that AFGL 2233 is a massive AGB star, with
$M_\mathrm{init}$ = 4.5-9\ms. This would explain the origin of the wide
molecular-line proﬁles observed.}

We studied the chemistry of the HVC star AFGL\,2233. We aimed at finding hints of the massive nature of this object. However, we found a standard C-rich chemistry except for certain peculiar features, which are:

\begin{itemize}
\item { High C$_2$H/C$_4$H abundance ratio (probably related to photoinduced chemistry).}
\item Very high NH$_3$ abundance (the highest observed so far --  IR-pumping can lower this value).
\item High H$_2$O abundance (comparable with O-rich stars).
\item SiN detected (fourth time detected in an evolved star).
\end{itemize}

The isotopic ratios present larger changes (from 55 to 7) but currently it is hard to clearly determine if, as expected if this object is being enriched in $^{12}$C after the HBB has been exahusted, i.e. a higher $^{12}$C/$^{13}$C in the inner regions than in the outer shells. Dedicated high-angular resolution observations are needed to confirm if this is the case. 

In addition, the mentioned peculiarities seem to point toward the possible presence of a companion, triggering both the formation of SiN, and H$_2$O, { while also increasing the C$_2$H/C$_4$H abundance ratio}. Carefully tracing the regions where these species arise could help to confirm the presence of that companion and constrain the region where the suggested binary-induced photochemistry might take place.


\begin{acknowledgements}
      The research leading to these results has received funding 
 funding support from Spanish Ministerio de Ciencia, Innovación, y Universidades through grant PID2023-147545NB-I00.  
 JA is partially supported by I+D+i projects 1010 PID2019-105203GB-C21 and PID2023-146056NB-C21, funded by the Spanish MCIN/AEI/10.13039/501100011033 and EU/ERDF. LVP contribution is supported by project PID2020-117034RJ-I00.
 This publication is part of the grant RYC2023-045648-I funded by MICIU/AEI/10.13039/501100011033 and by ESF+.
This work has made use of data from the European Space Agency (ESA) mission
{\it Gaia} (\url{https://www.cosmos.esa.int/gaia}), processed by the {\it Gaia}
Data Processing and Analysis Consortium (DPAC,
\url{https://www.cosmos.esa.int/web/gaia/dpac/consortium}). Funding for the DPAC
has been provided by national institutions, in particular the institutions
participating in the {\it Gaia} Multilateral Agreement.
\end{acknowledgements}

%
%

\bibliographystyle{aa} 
\bibliography{mybib}

@ARTICLE{ccsio,
   author = {{Castro-Carrizo}, A. and {Lucas}, R. and {Bujarrabal}, V. and 
	{Colomer}, F. and {Alcolea}, J.},
    title = "{SiO emission from a huge, detached shell in IRC +10420}",
  journal = {\aap},
     year = 2001,
    month = mar,
   volume = 368,
    pages = {L34-L37},
      doi = {10.1051/0004-6361:20010211},
   adsurl = {http://adsabs.harvard.edu/abs/2001A\%26A...368L..34C}
}

@Article{chemyhg,
  author        = {{Quintana-Lacaci}, G. and {Bujarrabal}, V. and {Castro-Carrizo}, A. and {Alcolea}, J.},
  title         = {{The chemical composition of the circumstellar envelopes around yellow hypergiant stars}},
  journal       = {\aap},
  year          = {2007},
  volume        = {471},
  pages         = {551-560},
  month         = aug,
  adsurl        = {http://adsabs.harvard.edu/abs/2007A%26A...471..551Q},
  archiveprefix = {arXiv},
  doi           = {10.1051/0004-6361:20077244},
  eprint        = {0706.1639},
}

@Article{menten00nh3,
  author        = {{Menten}, K.~M. and {Wyrowski}, F. and {Alcolea}, J. and {De Beck}, E. and {Decin}, L. and {Marston}, A.~P. and {Bujarrabal}, V. and {Cernicharo}, J. and {Dominik}, C. and {Justtanont}, K. and {de Koter}, A. and {Melnick}, G. and {Neufeld}, D.~A. and {Olofsson}, H. and {Planesas}, P. and {Schmidt}, M. and {Sch{\"o}ier}, F.~L. and {Szczerba}, R. and {Teyssier}, D. and {Waters}, L.~B.~F.~M. and {Edwards}, K. and {Olberg}, M. and {Phillips}, T.~G. and {Morris}, P. and {Salez}, M. and {Caux}, E.},
  title         = {{Herschel/HIFI deepens the circumstellar NH$_{3}$ enigma}},
  journal       = {\aap},
  year          = {2010},
  volume        = {521},
  pages         = {L7},
  month         = oct,
  adsurl        = {http://adsabs.harvard.edu/abs/2010A%26A...521L...7M},
  archiveprefix = {arXiv},
  doi           = {10.1051/0004-6361/201015108},
  eid           = {L7},
  eprint        = {1007.1413},
  primaryclass  = {astro-ph.SR},
}

@InProceedings{MADEX,
  author    = {{Cernicharo}, J.},
  title     = {{Laboratory astrophysics and astrochemistry in the Herschel/ALMA era}},
  booktitle = {ECLA-2011: Proceedings of the European Conference on Laboratory Astrophysics},
  year      = {2012},
  series    = {European Astronomical Society Publications Series},
}

@Article{rtd,
  author   = {{Goldsmith}, P.~F. and {Langer}, W.~D.},
  title    = {{Population Diagram Analysis of Molecular Line Emission}},
  journal  = {\apj},
  year     = {1999},
  volume   = {517},
  pages    = {209-225},
  month    = may,
  adsurl   = {http://adsabs.harvard.edu/abs/1999ApJ...517..209G},
  doi      = {10.1086/307195},
}

@ARTICLE{10420no,
   author = {{Quintana-Lacaci}, G. and {Ag{\'u}ndez}, M. and {Cernicharo}, J. and 
	{Bujarrabal}, V. and {S{\'a}nchez Contreras}, C. and {Castro-Carrizo}, A. and 
	{Alcolea}, J.},
    title = "{Detection of circumstellar nitric oxide. Enhanced nitrogen abundance in IRC +10420}",
  journal = {\aap},
     year = 2013,
    month = dec,
   volume = 560,
      eid = {L2},
    pages = {L2},
      doi = {10.1051/0004-6361/201322728},
   adsurl = {http://adsabs.harvard.edu/abs/2013A%26A...560L...2Q}
}

@Article{cisotopic,
  author   = {{Milam}, S.~N. and {Woolf}, N.~J. and {Ziurys}, L.~M.},
  title    = {{Circumstellar $^{12}$C/$^{13}$C Isotope Ratios from Millimeter Observations of CN and CO: Mixing in Carbon- and Oxygen-Rich Stars}},
  journal  = {\apj},
  year     = {2009},
  volume   = {690},
  pages    = {837-849},
  month    = jan,
  adsurl   = {http://adsabs.harvard.edu/abs/2009ApJ...690..837M},
  doi      = {10.1088/0004-637X/690/1/837},
}

@Article{c2h2,
  author   = {{Fuente}, A. and {Cernicharo}, J. and {Omont}, A.},
  title    = {{Inferring acetylene abundances from C\_2H: the C\_2H\_2/HCN abundance ratio}},
  journal  = {\aap},
  year     = {1998},
  volume   = {330},
  pages    = {232-242},
  month    = feb,
  adsurl   = {http://adsabs.harvard.edu/abs/1998A%26A...330..232F},
}

@ARTICLE{Yuasa99,
   author = {{Yuasa}, M. and {Unno}, W. and {Magono}, S.},
    title = "{Distance Determination of Mass-Losing Stars}",
  journal = {\pasj},
     year = 1999,
    month = apr,
   volume = 51,
    pages = {197-209},
      doi = {10.1093/pasj/51.2.197},
   adsurl = {http://adsabs.harvard.edu/abs/1999PASJ...51..197Y}
}

@ARTICLE{Loup93,
   author = {{Loup}, C. and {Forveille}, T. and {Omont}, A. and {Paul}, J.~F.
	},
    title = "{CO and HCN observations of circumstellar envelopes. A catalogue - Mass loss rates and distributions}",
  journal = {\aaps},
     year = 1993,
    month = jun,
   volume = 99,
    pages = {291-377},
   adsurl = {http://adsabs.harvard.edu/abs/1993A%26AS...99..291L}
}

@Article{Likkel96,
  author   = {{Likkel}, L. and {Miao}, Y.},
  journal  = {\aj},
  title    = {{Circumstellar HCN Emission From Two Unusual Carbon Stars: IRC+10401 and AFGL 2233}},
  year     = {1996},
  month    = jul,
  pages    = {301},
  volume   = {112},
  adsurl   = {http://adsabs.harvard.edu/abs/1996AJ....112..301L},
  doi      = {10.1086/118015},
}

@PROCEEDINGS{AGBs2003,
    title = "{Asymptotic giant branch stars}",
booktitle = {Asymptotic giant branch stars, by Harm J. Habing and Hans Olofsson. Astronomy and astrophysics library, New York, Berlin: Springer, 2003},
     year = 2003,
   editor = {{Habing}, H.~J. and {Olofsson}, H.},
   adsurl = {http://adsabs.harvard.edu/abs/2003agbs.conf.....H}
}

@ARTICLE{Gaia2-1,
   author = {{Gaia Collaboration} and {Prusti}, T. and {de Bruijne}, J.~H.~J. and 
	{Brown}, A.~G.~A. and {Vallenari}, A. and {Babusiaux}, C. and 
	{Bailer-Jones}, C.~A.~L. and {Bastian}, U. and {Biermann}, M. and 
	{Evans}, D.~W. and et al.},
    title = "{The Gaia mission}",
  journal = {\aap},
archivePrefix = "arXiv",
   eprint = {1609.04153},
 primaryClass = "astro-ph.IM",
     year = 2016,
    month = nov,
   volume = 595,
      eid = {A1},
    pages = {A1},
      doi = {10.1051/0004-6361/201629272},
   adsurl = {http://adsabs.harvard.edu/abs/2016A%26A...595A...1G}
}

@ARTICLE{Meynet03,
   author = {{Meynet}, G. and {Maeder}, A.},
    title = "{Stellar evolution with rotation. X. Wolf-Rayet star populations at solar metallicity}",
  journal = {\aap},
   eprint = {astro-ph/0304069},
     year = 2003,
    month = jun,
   volume = 404,
    pages = {975-990},
      doi = {10.1051/0004-6361:20030512},
   adsurl = {http://adsabs.harvard.edu/abs/2003A%26A...404..975M}
}

@ARTICLE{tracks,
   author = {{Girardi}, L. and {Bressan}, A. and {Bertelli}, G. and {Chiosi}, C.
	},
    title = "{Evolutionary tracks and isochrones for low- and intermediate-mass stars: From 0.15 to 7 M$_{sun}$, and from Z=0.0004 to 0.03}",
  journal = {\aaps},
   eprint = {astro-ph/9910164},
     year = 2000,
    month = feb,
   volume = 141,
    pages = {371-383},
      doi = {10.1051/aas:2000126},
   adsurl = {http://adsabs.harvard.edu/abs/2000A%26AS..141..371G}
}

@Article{karakas2017,
  author        = {{Karakas}, Amanda I. and {Lugaro}, Maria and {Carlos}, Mar{\'\i}lia and {Cseh}, Borb{\'a}la and {Kamath}, Devika and {Garc{\'\i}a-Hern{\'a}ndez}, D.~A.},
  title         = {{Heavy-element yields and abundances of asymptotic giant branch models with a Small Magellanic Cloud metallicity}},
  journal       = {\mnras},
  year          = {2018},
  volume        = {477},
  number        = {1},
  pages         = {421-437},
  month         = {Jun},
  adsurl        = {https://ui.adsabs.harvard.edu/abs/2018MNRAS.477..421K},
  archiveprefix = {arXiv},
  doi           = {10.1093/mnras/sty625},
  eprint        = {1803.02028},
  primaryclass  = {astro-ph.SR},
}

@Article{survey10420,
  author        = {{Quintana-Lacaci}, G. and {Ag{\'u}ndez}, M. and {Cernicharo}, J. and {Bujarrabal}, V. and {S{\'a}nchez Contreras}, C. and {Castro-Carrizo}, A. and {Alcolea}, J.},
  title         = {{A {\ensuremath{\lambda}} 3 mm and 1 mm line survey toward the yellow hypergiant IRC +10420. N-rich chemistry and IR flux variations}},
  journal       = {\aap},
  year          = {2016},
  volume        = {592},
  pages         = {A51},
  month         = {Jul},
  adsurl        = {https://ui.adsabs.harvard.edu/abs/2016A&A...592A..51Q},
  archiveprefix = {arXiv},
  doi           = {10.1051/0004-6361/201527688},
  eid           = {A51},
  eprint        = {1605.09183},
  primaryclass  = {astro-ph.SR},
}

@Article{agundez17,
  author        = {{Ag{\'u}ndez}, M. and {Cernicharo}, J. and {Quintana-Lacaci}, G. and {Castro-Carrizo}, A. and {Velilla Prieto}, L. and {Marcelino}, N. and {Gu{\'e}lin}, M. and {Joblin}, C. and {Mart{\'\i}n-Gago}, J.~A. and {Gottlieb}, C.~A. and {Patel}, N.~A. and {McCarthy}, M.~C.},
  title         = {{Growth of carbon chains in IRC +10216 mapped with ALMA}},
  journal       = {\aap},
  year          = {2017},
  volume        = {601},
  pages         = {A4},
  month         = {May},
  adsurl        = {https://ui.adsabs.harvard.edu/abs/2017A&A...601A...4A},
  archiveprefix = {arXiv},
  doi           = {10.1051/0004-6361/201630274},
  eid           = {A4},
  eprint        = {1702.04429},
  primaryclass  = {astro-ph.SR},
}

@Article{Velilla19,
  author        = {{Velilla-Prieto}, L. and {Cernicharo}, J. and {Ag{\'u}ndez}, M. and {Fonfr{\'\i}a}, J.~P. and {Quintana-Lacaci}, G. and {Marcelino}, N. and {Castro-Carrizo}, A.},
  title         = {{IRC + 10{\textdegree}216 mass loss properties through the study of {\ensuremath{\lambda}}3 mm emission. Large spatial scale distribution of SiO, SiS, and CS}},
  journal       = {\aap},
  year          = {2019},
  volume        = {629},
  pages         = {A146},
  month         = {Sep},
  adsurl        = {https://ui.adsabs.harvard.edu/abs/2019A&A...629A.146V},
  archiveprefix = {arXiv},
  doi           = {10.1051/0004-6361/201834717},
  eid           = {A146},
  eprint        = {1908.05652},
  primaryclass  = {astro-ph.SR},
}

@Article{Pardo01,
  author   = {{Pardo}, J.~R. and {Cernicharo}, J. and {Serabyn}, E.},
  title    = {Atmospheric transmission at microwaves (ATM): an improved model for millimeter/submillimeter applications},
  journal  = {IEEE Trans. on Antennas and Propagation},
  year     = {2001},
  volume   = {49/12},
  pages    = {1683-1694},
  month    = jan,
}

@Article{Cerni85,
  author  = {{Cernicharo}, J.},
  journal = {IRAM internal report No. 52},
  year    = {1985},
}

@Article{AlcoleaVYCMa,
  author        = {{Alcolea}, J. and {Bujarrabal}, V. and {Planesas}, P. and {Teyssier}, D. and {Cernicharo}, J. and {De Beck}, E. and {Decin}, L. and {Dominik}, C. and {Justtanont}, K. and {de Koter}, A. and {Marston}, A.~P. and {Melnick}, G. and {Menten}, K.~M. and {Neufeld}, D.~A. and {Olofsson}, H. and {Schmidt}, M. and {Sch{\"o}ier}, F.~L. and {Szczerba}, R. and {Waters}, L.~B.~F.~M.},
  title         = {{HIFISTARS Herschel/HIFI observations of VY Canis Majoris. Molecular-line inventory of the envelope around the largest known star}},
  journal       = {\aap},
  year          = {2013},
  volume        = {559},
  pages         = {A93},
  month         = nov,
  adsurl        = {https://ui.adsabs.harvard.edu/abs/2013A&A...559A..93A},
  archiveprefix = {arXiv},
  doi           = {10.1051/0004-6361/201321683},
  eid           = {A93},
  eprint        = {1310.2400},
  primaryclass  = {astro-ph.SR},
}

@Article{gaia3,
  author        = {{Gaia Collaboration} and {Vallenari}, A. and {Brown}, A.~G.~A. and {Prusti}, T. and {de Bruijne}, J.~H.~J. and {Arenou}, F. and {Babusiaux}, C. and {Biermann}, M. and {Creevey}, O.~L. and {Ducourant}, C. and {Evans}, D.~W. and {Eyer}, L. and {Guerra}, R. and {Hutton}, A. and {Jordi}, C. and {Klioner}, S.~A. and {Lammers}, U.~L. and {Lindegren}, L. and {Luri}, X. and {Mignard}, F. and {Panem}, C. and {Pourbaix}, D. and {Randich}, S. and {Sartoretti}, P. and {Soubiran}, C. and {Tanga}, P. and {Walton}, N.~A. and {Bailer-Jones}, C.~A.~L. and {Bastian}, U. and {Drimmel}, R. and {Jansen}, F. and {Katz}, D. and {Lattanzi}, M.~G. and {van Leeuwen}, F. and {Bakker}, J. and {Cacciari}, C. and {Casta{\~n}eda}, J. and {De Angeli}, F. and {Fabricius}, C. and {Fouesneau}, M. and {Fr{\'e}mat}, Y. and {Galluccio}, L. and {Guerrier}, A. and {Heiter}, U. and {Masana}, E. and {Messineo}, R. and {Mowlavi}, N. and {Nicolas}, C. and {Nienartowicz}, K. and {Pailler}, F. and {Panuzzo}, P. and {Riclet}, F. and {Roux}, W. and {Seabroke}, G.~M. and {Sordo}, R. and {Th{\'e}venin}, F. and {Gracia-Abril}, G. and {Portell}, J. and {Teyssier}, D. and {Altmann}, M. and {Andrae}, R. and {Audard}, M. and {Bellas-Velidis}, I. and {Benson}, K. and {Berthier}, J. and {Blomme}, R. and {Burgess}, P.~W. and {Busonero}, D. and {Busso}, G. and {C{\'a}novas}, H. and {Carry}, B. and {Cellino}, A. and {Cheek}, N. and {Clementini}, G. and {Damerdji}, Y. and {Davidson}, M. and {de Teodoro}, P. and {Nu{\~n}ez Campos}, M. and {Delchambre}, L. and {Dell'Oro}, A. and {Esquej}, P. and {Fern{\'a}ndez-Hern{\'a}ndez}, J. and {Fraile}, E. and {Garabato}, D. and {Garc{\'\i}a-Lario}, P. and {Gosset}, E. and {Haigron}, R. and {Halbwachs}, J. -L. and {Hambly}, N.~C. and {Harrison}, D.~L. and {Hern{\'a}ndez}, J. and {Hestroffer}, D. and {Hodgkin}, S.~T. and {Holl}, B. and {Jan{\ss}en}, K. and {Jevardat de Fombelle}, G. and {Jordan}, S. and {Krone-Martins}, A. and {Lanzafame}, A.~C. and {L{\"o}ffler}, W. and {Marchal}, O. and {Marrese}, P.~M. and {Moitinho}, A. and {Muinonen}, K. and {Osborne}, P. and {Pancino}, E. and {Pauwels}, T. and {Recio-Blanco}, A. and {Reyl{\'e}}, C. and {Riello}, M. and {Rimoldini}, L. and {Roegiers}, T. and {Rybizki}, J. and {Sarro}, L.~M. and {Siopis}, C. and {Smith}, M. and {Sozzetti}, A. and {Utrilla}, E. and {van Leeuwen}, M. and {Abbas}, U. and {{\'A}brah{\'a}m}, P. and {Abreu Aramburu}, A. and {Aerts}, C. and {Aguado}, J.~J. and {Ajaj}, M. and {Aldea-Montero}, F. and {Altavilla}, G. and {{\'A}lvarez}, M.~A. and {Alves}, J. and {Anders}, F. and {Anderson}, R.~I. and {Anglada Varela}, E. and {Antoja}, T. and {Baines}, D. and {Baker}, S.~G. and {Balaguer-N{\'u}{\~n}ez}, L. and {Balbinot}, E. and {Balog}, Z. and {Barache}, C. and {Barbato}, D. and {Barros}, M. and {Barstow}, M.~A. and {Bartolom{\'e}}, S. and {Bassilana}, J. -L. and {Bauchet}, N. and {Becciani}, U. and {Bellazzini}, M. and {Berihuete}, A. and {Bernet}, M. and {Bertone}, S. and {Bianchi}, L. and {Binnenfeld}, A. and {Blanco-Cuaresma}, S. and {Blazere}, A. and {Boch}, T. and {Bombrun}, A. and {Bossini}, D. and {Bouquillon}, S. and {Bragaglia}, A. and {Bramante}, L. and {Breedt}, E. and {Bressan}, A. and {Brouillet}, N. and {Brugaletta}, E. and {Bucciarelli}, B. and {Burlacu}, A. and {Butkevich}, A.~G. and {Buzzi}, R. and {Caffau}, E. and {Cancelliere}, R. and {Cantat-Gaudin}, T. and {Carballo}, R. and {Carlucci}, T. and {Carnerero}, M.~I. and {Carrasco}, J.~M. and {Casamiquela}, L. and {Castellani}, M. and {Castro-Ginard}, A. and {Chaoul}, L. and {Charlot}, P. and {Chemin}, L. and {Chiaramida}, V. and {Chiavassa}, A. and {Chornay}, N. and {Comoretto}, G. and {Contursi}, G. and {Cooper}, W.~J. and {Cornez}, T. and {Cowell}, S. and {Crifo}, F. and {Cropper}, M. and {Crosta}, M. and {Crowley}, C. and {Dafonte}, C. and {Dapergolas}, A. and {David}, M. and {David}, P. and {de Laverny}, P. and {De Luise}, F. and {De March}, R. and {De Ridder}, J. and {de Souza}, R. and {de Torres}, A. and {del Peloso}, E.~F. and {del Pozo}, E. and {Delbo}, M. and {Delgado}, A. and {Delisle}, J. -B. and {Demouchy}, C. and {Dharmawardena}, T.~E. and {Di Matteo}, P. and {Diakite}, S. and {Diener}, C. and {Distefano}, E. and {Dolding}, C. and {Edvardsson}, B. and {Enke}, H. and {Fabre}, C. and {Fabrizio}, M. and {Faigler}, S. and {Fedorets}, G. and {Fernique}, P. and {Fienga}, A. and {Figueras}, F. and {Fournier}, Y. and {Fouron}, C. and {Fragkoudi}, F. and {Gai}, M. and {Garcia-Gutierrez}, A. and {Garcia-Reinaldos}, M. and {Garc{\'\i}a-Torres}, M. and {Garofalo}, A. and {Gavel}, A. and {Gavras}, P. and {Gerlach}, E. and {Geyer}, R. and {Giacobbe}, P. and {Gilmore}, G. and {Girona}, S. and {Giuffrida}, G. and {Gomel}, R. and {Gomez}, A. and {Gonz{\'a}lez-N{\'u}{\~n}ez}, J. and {Gonz{\'a}lez-Santamar{\'\i}a}, I. and {Gonz{\'a}lez-Vidal}, J.~J. and {Granvik}, M. and {Guillout}, P. and {Guiraud}, J. and {Guti{\'e}rrez-S{\'a}nchez}, R. and {Guy}, L.~P. and {Hatzidimitriou}, D. and {Hauser}, M. and {Haywood}, M. and {Helmer}, A. and {Helmi}, A. and {Sarmiento}, M.~H. and {Hidalgo}, S.~L. and {Hilger}, T. and {H{\l}adczuk}, N. and {Hobbs}, D. and {Holland}, G. and {Huckle}, H.~E. and {Jardine}, K. and {Jasniewicz}, G. and {Jean-Antoine Piccolo}, A. and {Jim{\'e}nez-Arranz}, {\'O}. and {Jorissen}, A. and {Juaristi Campillo}, J. and {Julbe}, F. and {Karbevska}, L. and {Kervella}, P. and {Khanna}, S. and {Kontizas}, M. and {Kordopatis}, G. and {Korn}, A.~J. and {K{\'o}sp{\'a}l}, {\'A}. and {Kostrzewa-Rutkowska}, Z. and {Kruszy{\'n}ska}, K. and {Kun}, M. and {Laizeau}, P. and {Lambert}, S. and {Lanza}, A.~F. and {Lasne}, Y. and {Le Campion}, J. -F. and {Lebreton}, Y. and {Lebzelter}, T. and {Leccia}, S. and {Leclerc}, N. and {Lecoeur-Taibi}, I. and {Liao}, S. and {Licata}, E.~L. and {Lindstr{\o}m}, H.~E.~P. and {Lister}, T.~A. and {Livanou}, E. and {Lobel}, A. and {Lorca}, A. and {Loup}, C. and {Madrero Pardo}, P. and {Magdaleno Romeo}, A. and {Managau}, S. and {Mann}, R.~G. and {Manteiga}, M. and {Marchant}, J.~M. and {Marconi}, M. and {Marcos}, J. and {Marcos Santos}, M.~M.~S. and {Mar{\'\i}n Pina}, D. and {Marinoni}, S. and {Marocco}, F. and {Marshall}, D.~J. and {Martin Polo}, L. and {Mart{\'\i}n-Fleitas}, J.~M. and {Marton}, G. and {Mary}, N. and {Masip}, A. and {Massari}, D. and {Mastrobuono-Battisti}, A. and {Mazeh}, T. and {McMillan}, P.~J. and {Messina}, S. and {Michalik}, D. and {Millar}, N.~R. and {Mints}, A. and {Molina}, D. and {Molinaro}, R. and {Moln{\'a}r}, L. and {Monari}, G. and {Mongui{\'o}}, M. and {Montegriffo}, P. and {Montero}, A. and {Mor}, R. and {Mora}, A. and {Morbidelli}, R. and {Morel}, T. and {Morris}, D. and {Muraveva}, T. and {Murphy}, C.~P. and {Musella}, I. and {Nagy}, Z. and {Noval}, L. and {Oca{\~n}a}, F. and {Ogden}, A. and {Ordenovic}, C. and {Osinde}, J.~O. and {Pagani}, C. and {Pagano}, I. and {Palaversa}, L. and {Palicio}, P.~A. and {Pallas-Quintela}, L. and {Panahi}, A. and {Payne-Wardenaar}, S. and {Pe{\~n}alosa Esteller}, X. and {Penttil{\"a}}, A. and {Pichon}, B. and {Piersimoni}, A.~M. and {Pineau}, F. -X. and {Plachy}, E. and {Plum}, G. and {Poggio}, E. and {Pr{\v{s}}a}, A. and {Pulone}, L. and {Racero}, E. and {Ragaini}, S. and {Rainer}, M. and {Raiteri}, C.~M. and {Rambaux}, N. and {Ramos}, P. and {Ramos-Lerate}, M. and {Re Fiorentin}, P. and {Regibo}, S. and {Richards}, P.~J. and {Rios Diaz}, C. and {Ripepi}, V. and {Riva}, A. and {Rix}, H. -W. and {Rixon}, G. and {Robichon}, N. and {Robin}, A.~C. and {Robin}, C. and {Roelens}, M. and {Rogues}, H.~R.~O. and {Rohrbasser}, L. and {Romero-G{\'o}mez}, M. and {Rowell}, N. and {Royer}, F. and {Ruz Mieres}, D. and {Rybicki}, K.~A. and {Sadowski}, G. and {S{\'a}ez N{\'u}{\~n}ez}, A. and {Sagrist{\`a} Sell{\'e}s}, A. and {Sahlmann}, J. and {Salguero}, E. and {Samaras}, N. and {Sanchez Gimenez}, V. and {Sanna}, N. and {Santove{\~n}a}, R. and {Sarasso}, M. and {Schultheis}, M. and {Sciacca}, E. and {Segol}, M. and {Segovia}, J.~C. and {S{\'e}gransan}, D. and {Semeux}, D. and {Shahaf}, S. and {Siddiqui}, H.~I. and {Siebert}, A. and {Siltala}, L. and {Silvelo}, A. and {Slezak}, E. and {Slezak}, I. and {Smart}, R.~L. and {Snaith}, O.~N. and {Solano}, E. and {Solitro}, F. and {Souami}, D. and {Souchay}, J. and {Spagna}, A. and {Spina}, L. and {Spoto}, F. and {Steele}, I.~A. and {Steidelm{\"u}ller}, H. and {Stephenson}, C.~A. and {S{\"u}veges}, M. and {Surdej}, J. and {Szabados}, L. and {Szegedi-Elek}, E. and {Taris}, F. and {Taylor}, M.~B. and {Teixeira}, R. and {Tolomei}, L. and {Tonello}, N. and {Torra}, F. and {Torra}, J. and {Torralba Elipe}, G. and {Trabucchi}, M. and {Tsounis}, A.~T. and {Turon}, C. and {Ulla}, A. and {Unger}, N. and {Vaillant}, M.~V. and {van Dillen}, E. and {van Reeven}, W. and {Vanel}, O. and {Vecchiato}, A. and {Viala}, Y. and {Vicente}, D. and {Voutsinas}, S. and {Weiler}, M. and {Wevers}, T. and {Wyrzykowski}, {\L}. and {Yoldas}, A. and {Yvard}, P. and {Zhao}, H. and {Zorec}, J. and {Zucker}, S. and {Zwitter}, T.},
  title         = {{Gaia Data Release 3. Summary of the content and survey properties}},
  journal       = {\aap},
  year          = {2023},
  volume        = {674},
  pages         = {A1},
  month         = jun,
  adsurl        = {https://ui.adsabs.harvard.edu/abs/2023A&A...674A...1G},
  archiveprefix = {arXiv},
  doi           = {10.1051/0004-6361/202243940},
  eid           = {A1},
  eprint        = {2208.00211},
  primaryclass  = {astro-ph.GA},
}

@Article{Bujarrabal94a,
  author   = {{Bujarrabal}, V. and {Fuente}, A. and {Omont}, A.},
  title    = {{Molecular observations of O- and C-rich circumstellar envelopes}},
  journal  = {\aap},
  year     = {1994},
  volume   = {285},
  pages    = {247-271},
  month    = may,
  adsurl   = {https://ui.adsabs.harvard.edu/abs/1994A&A...285..247B},
}

@Article{zhang2009,
  author        = {{Zhang}, Yong and {Kwok}, Sun and {Dinh-V-Trung}},
  title         = {{A Molecular Line Survey of the Highly Evolved Carbon Star CIT 6}},
  journal       = {\apj},
  year          = {2009},
  volume        = {691},
  number        = {2},
  pages         = {1660-1677},
  month         = feb,
  adsurl        = {https://ui.adsabs.harvard.edu/abs/2009ApJ...691.1660Z},
  archiveprefix = {arXiv},
  doi           = {10.1088/0004-637X/691/2/1660},
  eprint        = {0808.3226},
  primaryclass  = {astro-ph},
}

@Article{agundez12,
  author        = {{Ag{\'u}ndez}, M. and {Fonfr{\'\i}a}, J.~P. and {Cernicharo}, J. and {Kahane}, C. and {Daniel}, F. and {Gu{\'e}lin}, M.},
  title         = {{Molecular abundances in the inner layers of IRC +10216}},
  journal       = {\aap},
  year          = {2012},
  volume        = {543},
  pages         = {A48},
  month         = jul,
  adsurl        = {https://ui.adsabs.harvard.edu/abs/2012A&A...543A..48A},
  archiveprefix = {arXiv},
  doi           = {10.1051/0004-6361/201218963},
  eid           = {A48},
  eprint        = {1204.4720},
  primaryclass  = {astro-ph.GA},
}

@InProceedings{2014apn6.confE..88S,
  author    = {{S{\'a}nchez Contreras}, C. and {Velilla}, L. and {Alcolea}, J. and {Quintana-Lacaci}, G. and {Cernicharo}, J. and {Agundez}, M. and {Teyssier}, D. and {Bujarrabal}, V. and {Castro-Carrizo}, A. and {Daniel}, F. and {Fonfria}, J.~P. and {Garcia-Lario}, P. and {Goicoechea}, J.~R. and {Herpin}, F. and {Barlow}, M. and {Cherchneff}, I. and {Comito}, C. and {Cordiner}, M. and {Decin}, L. and {Halfen}, D. \raisebox{-0.5ex}\textasciitildeT. and {Justtanont}, K. and {Latter}, W. and {Malloci}, G. and {Matsuura}, M. and {Menten}, K. and {Mulas}, G. and {Muller}, H.~S.~P. and {Pardo}, J.~R. and {Pearson}, J. and {Swinyward}, B. and {Tenenbaum}, E. and {Wesson}, R. and {Wyrowski}, F. and {Ziurys}, L.},
  title     = {{Mm-wave and far-IR Molecular line survey of OH 231.8+4.2: Hard-boiled rotten eggs}},
  booktitle = {Asymmetrical Planetary Nebulae VI Conference},
  year      = {2014},
  editor    = {{Morisset}, C. and {Delgado-Inglada}, G. and {Torres-Peimbert}, S.},
  pages     = {88},
  month     = apr,
  adsurl    = {https://ui.adsabs.harvard.edu/abs/2014apn6.confE..88S},
  eid       = {88},
}

@Article{debeck2020,
  author        = {{De Beck}, E. and {Olofsson}, H.},
  journal       = {\aap},
  title         = {{The surprisingly carbon-rich environment of the S-type star W Aql}},
  year          = {2020},
  month         = oct,
  pages         = {A20},
  volume        = {642},
  adsurl        = {https://ui.adsabs.harvard.edu/abs/2020A&A...642A..20D},
  archiveprefix = {arXiv},
  doi           = {10.1051/0004-6361/202038335},
  eid           = {A20},
  eprint        = {2007.01756},
  primaryclass  = {astro-ph.SR},
}

@Article{Guelin78,
  author   = {{Guelin}, M. and {Green}, S. and {Thaddeus}, P.},
  title    = {{Detection of the C$_{4}$H radical toward IRC +10216.}},
  journal  = {\apjl},
  year     = {1978},
  volume   = {224},
  pages    = {L27-L30},
  month    = aug,
  adsurl   = {https://ui.adsabs.harvard.edu/abs/1978ApJ...224L..27G},
  doi      = {10.1086/182751},
}

@Article{chemol,
  author        = {{Ag{\'u}ndez}, M. and {Mart{\'\i}nez}, J.~I. and {de Andres}, P.~L. and {Cernicharo}, J. and {Mart{\'\i}n-Gago}, J.~A.},
  title         = {{Chemical equilibrium in AGB atmospheres: successes, failures, and prospects for small molecules, clusters, and condensates}},
  journal       = {\aap},
  year          = {2020},
  volume        = {637},
  pages         = {A59},
  month         = may,
  adsurl        = {https://ui.adsabs.harvard.edu/abs/2020A&A...637A..59A},
  archiveprefix = {arXiv},
  doi           = {10.1051/0004-6361/202037496},
  eid           = {A59},
  eprint        = {2004.00519},
  primaryclass  = {astro-ph.SR},
}

@Article{10216_NH3,
  author        = {{Schmidt}, M.~R. and {He}, J.~H. and {Szczerba}, R. and {Bujarrabal}, V. and {Alcolea}, J. and {Cernicharo}, J. and {Decin}, L. and {Justtanont}, K. and {Teyssier}, D. and {Menten}, K.~M. and {Neufeld}, D.~A. and {Olofsson}, H. and {Planesas}, P. and {Marston}, A.~P. and {Sobolev}, A.~M. and {de Koter}, A. and {Sch{\"o}ier}, F.~L.},
  title         = {{Herschel/HIFI observations of the circumstellar ammonia lines in IRC+10216}},
  journal       = {\aap},
  year          = {2016},
  volume        = {592},
  pages         = {A131},
  month         = aug,
  adsurl        = {https://ui.adsabs.harvard.edu/abs/2016A&A...592A.131S},
  archiveprefix = {arXiv},
  doi           = {10.1051/0004-6361/201527290},
  eid           = {A131},
  eprint        = {1606.01878},
  primaryclass  = {astro-ph.SR},
}

@Article{agundez20,
  author        = {{Ag{\'u}ndez}, M. and {Mart{\'\i}nez}, J.~I. and {de Andres}, P.~L. and {Cernicharo}, J. and {Mart{\'\i}n-Gago}, J.~A.},
  title         = {{Chemical equilibrium in AGB atmospheres: successes, failures, and prospects for small molecules, clusters, and condensates}},
  journal       = {\aap},
  year          = {2020},
  volume        = {637},
  pages         = {A59},
  month         = may,
  adsurl        = {https://ui.adsabs.harvard.edu/abs/2020A&A...637A..59A},
  archiveprefix = {arXiv},
  doi           = {10.1051/0004-6361/202037496},
  eid           = {A59},
  eprint        = {2004.00519},
  primaryclass  = {astro-ph.SR},
}

@Article{wong18,
  author        = {{Wong}, K.~T. and {Menten}, K.~M. and {Kami{\'n}ski}, T. and {Wyrowski}, F. and {Lacy}, J.~H. and {Greathouse}, T.~K.},
  title         = {{Circumstellar ammonia in oxygen-rich evolved stars}},
  journal       = {\aap},
  year          = {2018},
  volume        = {612},
  pages         = {A48},
  month         = apr,
  adsurl        = {https://ui.adsabs.harvard.edu/abs/2018A&A...612A..48W},
  archiveprefix = {arXiv},
  doi           = {10.1051/0004-6361/201731873},
  eid           = {A48},
  eprint        = {1710.01027},
  primaryclass  = {astro-ph.SR},
}

@Article{h2o_orich,
  author        = {{Justtanont}, K. and {Khouri}, T. and {Maercker}, M. and {Alcolea}, J. and {Decin}, L. and {Olofsson}, H. and {Sch{\"o}ier}, F.~L. and {Bujarrabal}, V. and {Marston}, A.~P. and {Teyssier}, D. and {Cernicharo}, J. and {Dominik}, C. and {de Koter}, A. and {Melnick}, G. and {Menten}, K.~M. and {Neufeld}, D. and {Planesas}, P. and {Schmidt}, M. and {Szczerba}, R. and {Waters}, R.},
  journal       = {\aap},
  title         = {{Herschel/HIFI observations of O-rich AGB stars: molecular inventory}},
  year          = {2012},
  month         = jan,
  pages         = {A144},
  volume        = {537},
  adsurl        = {https://ui.adsabs.harvard.edu/abs/2012A&A...537A.144J},
  archiveprefix = {arXiv},
  doi           = {10.1051/0004-6361/201117524},
  eid           = {A144},
  eprint        = {1111.5156},
  primaryclass  = {astro-ph.SR},
}

@Article{h2o_10216,
  author        = {{Neufeld}, David A. and {Gonz{\'a}lez-Alfonso}, Eduardo and {Melnick}, Gary J. and {Szczerba}, Ryszard and {Schmidt}, Miroslaw and {Decin}, Leen and {de Koter}, Alex and {Sch{\"o}ier}, Fredrik and {Cernicharo}, Jos{\'e}},
  journal       = {\apjl},
  title         = {{Herschel/HIFI Observations of IRC+10216: Water Vapor in the Inner Envelope of a Carbon-rich Asymptotic Giant Branch Star}},
  year          = {2011},
  month         = feb,
  number        = {2},
  pages         = {L28},
  volume        = {727},
  adsurl        = {https://ui.adsabs.harvard.edu/abs/2011ApJ...727L..28N},
  archiveprefix = {arXiv},
  doi           = {10.1088/2041-8205/727/2/L28},
  eid           = {L28},
  eprint        = {1012.1854},
  primaryclass  = {astro-ph.SR},
}

@Article{h2o_kuiper,
  author        = {{Saavik Ford}, K.~E. and {Neufeld}, David A.},
  journal       = {\apjl},
  title         = {{Water Vapor in Carbon-rich Asymptotic Giant Branch Stars from the Vaporization of Icy Orbiting Bodies}},
  year          = {2001},
  month         = aug,
  number        = {2},
  pages         = {L113-L116},
  volume        = {557},
  adsurl        = {https://ui.adsabs.harvard.edu/abs/2001ApJ...557L.113S},
  archiveprefix = {arXiv},
  doi           = {10.1086/323268},
  eprint        = {astro-ph/0107223},
  primaryclass  = {astro-ph},
}

@Article{h2o_clumpy,
  author        = {{Ag{\'u}ndez}, Marcelino and {Cernicharo}, Jos{\'e} and {Gu{\'e}lin}, Michel},
  journal       = {\apjl},
  title         = {{Photochemistry in the Inner Layers of Clumpy Circumstellar Envelopes: Formation of Water in C-rich Objects and of C-bearing Molecules in O-rich Objects}},
  year          = {2010},
  month         = dec,
  number        = {2},
  pages         = {L133-L136},
  volume        = {724},
  adsurl        = {https://ui.adsabs.harvard.edu/abs/2010ApJ...724L.133A},
  archiveprefix = {arXiv},
  doi           = {10.1088/2041-8205/724/2/L133},
  eprint        = {1010.2093},
  primaryclass  = {astro-ph.GA},
}

@Article{h2o_shocks,
  author        = {{Cherchneff}, I.},
  journal       = {\aap},
  title         = {{Water in IRC+10216: a genuine formation process by shock-induced chemistry in the inner wind}},
  year          = {2011},
  month         = feb,
  pages         = {L11},
  volume        = {526},
  adsurl        = {https://ui.adsabs.harvard.edu/abs/2011A&A...526L..11C},
  archiveprefix = {arXiv},
  doi           = {10.1051/0004-6361/201016035},
  eid           = {L11},
  eprint        = {1012.5076},
  primaryclass  = {astro-ph.SR},
}

@Article{SiN_WAql,
  author        = {{Danilovich}, T. and {Malfait}, J. and {Van de Sande}, M. and {Montarg{\`e}s}, M. and {Kervella}, P. and {De Ceuster}, F. and {Coenegrachts}, A. and {Millar}, T.~J. and {Richards}, A.~M.~S. and {Decin}, L. and {Gottlieb}, C.~A. and {Pinte}, C. and {De Beck}, E. and {Price}, D.~J. and {Wong}, K.~T. and {Bolte}, J. and {Menten}, K.~M. and {Baudry}, A. and {de Koter}, A. and {Etoka}, S. and {Gobrecht}, D. and {Gray}, M. and {Herpin}, F. and {Jeste}, M. and {Lagadec}, E. and {Maes}, S. and {McDonald}, I. and {Marinho}, L. and {M{\"u}ller}, H.~S.~P. and {Pimpanuwat}, B. and {Plane}, J.~M.~C. and {Sahai}, R. and {Wallstr{\"o}m}, S.~H.~J. and {Yates}, J. and {Zijlstra}, A.},
  journal       = {Nature Astronomy},
  title         = {{Chemical tracers of a highly eccentric AGB-main-sequence star binary}},
  year          = {2024},
  month         = mar,
  pages         = {308-327},
  volume        = {8},
  adsurl        = {https://ui.adsabs.harvard.edu/abs/2024NatAs...8..308D},
  archiveprefix = {arXiv},
  doi           = {10.1038/s41550-023-02154-y},
  eprint        = {2407.16979},
  primaryclass  = {astro-ph.SR},
}

@Article{SiN_10216,
  author   = {{Turner}, B.~E.},
  journal  = {\apjl},
  title    = {{Detection of SiN in IRC +10216}},
  year     = {1992},
  month    = mar,
  pages    = {L35},
  volume   = {388},
  adsurl   = {https://ui.adsabs.harvard.edu/abs/1992ApJ...388L..35T},
  doi      = {10.1086/186324},
}

@Article{Bordiu22,
  author        = {{Bordiu}, C. and {Rizzo}, J.~R. and {Bufano}, F. and {Quintana-Lacaci}, G. and {Buemi}, C. and {Leto}, P. and {Cavallaro}, F. and {Cerrigone}, L. and {Ingallinera}, A. and {Loru}, S. and {Riggi}, S. and {Trigilio}, C. and {Umana}, G. and {Sciacca}, E.},
  journal       = {\apjl},
  title         = {{First Detection of Silicon-bearing Molecules in {\ensuremath{\eta}} Car}},
  year          = {2022},
  month         = nov,
  number        = {2},
  pages         = {L30},
  volume        = {939},
  adsurl        = {https://ui.adsabs.harvard.edu/abs/2022ApJ...939L..30B},
  archiveprefix = {arXiv},
  doi           = {10.3847/2041-8213/ac9b10},
  eid           = {L30},
  eprint        = {2210.09774},
  primaryclass  = {astro-ph.GA},
}

@Article{Massalkhi24,
  author        = {{Massalkhi}, S. and {Ag{\'u}ndez}, M. and {Fonfr{\'\i}a}, J.~P. and {Pardo}, J.~R. and {Velilla-Prieto}, L. and {Cernicharo}, J.},
  journal       = {\aap},
  title         = {{Multiline study of the radial extent of SiO, CS, and SiS in asymptotic giant branch envelopes}},
  year          = {2024},
  month         = aug,
  pages         = {A16},
  volume        = {688},
  adsurl        = {https://ui.adsabs.harvard.edu/abs/2024A&A...688A..16M},
  archiveprefix = {arXiv},
  doi           = {10.1051/0004-6361/202450188},
  eid           = {A16},
  eprint        = {2405.19922},
  primaryclass  = {astro-ph.SR},
}

@InProceedings{Garcia-Hernandez2016,
  author    = {{Garc{\'\i}a-Hern{\'a}ndez}, D.~A. and {Garc{\'\i}a-Lario}, P. and {Cernicharo}, J. and {Engels}, D. and {Perea-Calder{\'o}n}, J.~V.},
  booktitle = {Journal of Physics Conference Series},
  title     = {{Transitory O-rich chemistry in heavily obscured C-rich post-AGB stars}},
  year      = {2016},
  month     = jul,
  pages     = {052003},
  publisher = {IOP},
  series    = {Journal of Physics Conference Series},
  volume    = {728},
  adsurl    = {https://ui.adsabs.harvard.edu/abs/2016JPhCS.728e2003G},
  doi       = {10.1088/1742-6596/728/5/052003},
  eid       = {052003},
}

@Article{cerni04,
  author   = {{Cernicharo}, Jos{\'e}},
  journal  = {\apjl},
  title    = {{The Polymerization of Acetylene, Hydrogen Cyanide, and Carbon Chains in the Neutral Layers of Carbon-rich Proto-planetary Nebulae}},
  year     = {2004},
  month    = jun,
  number   = {1},
  pages    = {L41-L44},
  volume   = {608},
  adsurl   = {https://ui.adsabs.harvard.edu/abs/2004ApJ...608L..41C},
  doi      = {10.1086/422170},
}

@Article{bermudez20,
  author        = {{Berm{\'u}dez-Bustamante}, Luis C. and {Garc{\'\i}a-Segura}, G. and {Steffen}, W. and {Sabin}, L.},
  journal       = {\mnras},
  title         = {{AGB winds in interacting binary stars}},
  year          = {2020},
  month         = apr,
  number        = {2},
  pages         = {2606-2617},
  volume        = {493},
  adsurl        = {https://ui.adsabs.harvard.edu/abs/2020MNRAS.493.2606B},
  archiveprefix = {arXiv},
  doi           = {10.1093/mnras/staa403},
  eprint        = {2002.02570},
  primaryclass  = {astro-ph.SR},
}

@Article{ISOSiC,
  author   = {{Yang}, Xiaohong and {Chen}, Peisheng and {He}, Jinhua},
  journal  = {\aap},
  title    = {{Molecular and dust features of 29 SiC carbon AGB stars}},
  year     = {2004},
  month    = feb,
  pages    = {1049-1063},
  volume   = {414},
  adsurl   = {https://ui.adsabs.harvard.edu/abs/2004A&A...414.1049Y},
  doi      = {10.1051/0004-6361:20031673},
}

@Article{Ramstedt14,
  author        = {{Ramstedt}, S. and {Olofsson}, H.},
  journal       = {\aap},
  title         = {{The $^{12}$CO/$^{13}$CO ratio in AGB stars of different chemical type. Connection to the $^{12}$C/$^{13}$C ratio and the evolution along the AGB}},
  year          = {2014},
  month         = jun,
  pages         = {A145},
  volume        = {566},
  adsurl        = {https://ui.adsabs.harvard.edu/abs/2014A&A...566A.145R},
  archiveprefix = {arXiv},
  doi           = {10.1051/0004-6361/201423721},
  eid           = {A145},
  eprint        = {1405.6404},
  primaryclass  = {astro-ph.SR},
}

@Article{Teyssier04,
  author        = {{Teyssier}, D. and {Foss{\'e}}, D. and {Gerin}, M. and {Pety}, J. and {Abergel}, A. and {Roueff}, E.},
  journal       = {\aap},
  title         = {{Carbon budget and carbon chemistry in Photon Dominated Regions}},
  year          = {2004},
  month         = apr,
  pages         = {135-149},
  volume        = {417},
  adsurl        = {https://ui.adsabs.harvard.edu/abs/2004A&A...417..135T},
  archiveprefix = {arXiv},
  doi           = {10.1051/0004-6361:20034534},
  eprint        = {astro-ph/0401309},
  primaryclass  = {astro-ph},
}

@Article{HIFIobs,
  author   = {{Roelfsema}, P.~R. and {Helmich}, F.~P. and {Teyssier}, D. and {Ossenkopf}, V. and {Morris}, P. and {Olberg}, M. and {Shipman}, R. and {Risacher}, C. and {Akyilmaz}, M. and {Assendorp}, R. and {Avruch}, I.~M. and {Beintema}, D. and {Biver}, N. and {Boogert}, A. and {Borys}, C. and {Braine}, J. and {Caris}, M. and {Caux}, E. and {Cernicharo}, J. and {Coeur-Joly}, O. and {Comito}, C. and {de Lange}, G. and {Delforge}, B. and {Dieleman}, P. and {Dubbeldam}, L. and {de Graauw}, Th. and {Edwards}, K. and {Fich}, M. and {Flederus}, F. and {Gal}, C. and {di Giorgio}, A. and {Herpin}, F. and {Higgins}, D.~R. and {Hoac}, A. and {Huisman}, R. and {Jarchow}, C. and {Jellema}, W. and {de Jonge}, A. and {Kester}, D. and {Klein}, T. and {Kooi}, J. and {Kramer}, C. and {Laauwen}, W. and {Larsson}, B. and {Leinz}, C. and {Lord}, S. and {Lorenzani}, A. and {Luinge}, W. and {Marston}, A. and {Mart{\'\i}n-Pintado}, J. and {McCoey}, C. and {Melchior}, M. and {Michalska}, M. and {Moreno}, R. and {M{\"u}ller}, H. and {Nowosielski}, W. and {Okada}, Y. and {Orlea{\'n}ski}, P. and {Phillips}, T.~G. and {Pearson}, J. and {Rabois}, D. and {Ravera}, L. and {Rector}, J. and {Rengel}, M. and {Sagawa}, H. and {Salomons}, W. and {S{\'a}nchez-Su{\'a}rez}, E. and {Schieder}, R. and {Schl{\"o}der}, F. and {Schm{\"u}lling}, F. and {Soldati}, M. and {Stutzki}, J. and {Thomas}, B. and {Tielens}, A.~G.~G.~M. and {Vastel}, C. and {Wildeman}, K. and {Xie}, Q. and {Xilouris}, M. and {Wafelbakker}, C. and {Whyborn}, N. and {Zaal}, P. and {Bell}, T. and {Bjerkeli}, P. and {De Beck}, E. and {Cavali{\'e}}, T. and {Crockett}, N.~R. and {Hily-Blant}, P. and {Kama}, M. and {Kaminski}, T. and {Lefl{\'o}ch}, B. and {Lombaert}, R. and {de Luca}, M. and {Makai}, Z. and {Marseille}, M. and {Nagy}, Z. and {Pacheco}, S. and {van der Wiel}, M.~H.~D. and {Wang}, S. and {Y{\i}ld{\i}z}, U.},
  journal  = {\aap},
  title    = {{In-orbit performance of Herschel-HIFI}},
  year     = {2012},
  month    = jan,
  pages    = {A17},
  volume   = {537},
  adsurl   = {https://ui.adsabs.harvard.edu/abs/2012A&A...537A..17R},
  doi      = {10.1051/0004-6361/201015120},
  eid      = {A17},
}

@Article{HIFI,
  author   = {{de Graauw}, Th. and {Helmich}, F.~P. and {Phillips}, T.~G. and {Stutzki}, J. and {Caux}, E. and {Whyborn}, N.~D. and {Dieleman}, P. and {Roelfsema}, P.~R. and {Aarts}, H. and {Assendorp}, R. and {Bachiller}, R. and {Baechtold}, W. and {Barcia}, A. and {Beintema}, D.~A. and {Belitsky}, V. and {Benz}, A.~O. and {Bieber}, R. and {Boogert}, A. and {Borys}, C. and {Bumble}, B. and {Ca{\"\i}s}, P. and {Caris}, M. and {Cerulli-Irelli}, P. and {Chattopadhyay}, G. and {Cherednichenko}, S. and {Ciechanowicz}, M. and {Coeur-Joly}, O. and {Comito}, C. and {Cros}, A. and {de Jonge}, A. and {de Lange}, G. and {Delforges}, B. and {Delorme}, Y. and {den Boggende}, T. and {Desbat}, J. -M. and {Diez-Gonz{\'a}lez}, C. and {di Giorgio}, A.~M. and {Dubbeldam}, L. and {Edwards}, K. and {Eggens}, M. and {Erickson}, N. and {Evers}, J. and {Fich}, M. and {Finn}, T. and {Franke}, B. and {Gaier}, T. and {Gal}, C. and {Gao}, J.~R. and {Gallego}, J. -D. and {Gauffre}, S. and {Gill}, J.~J. and {Glenz}, S. and {Golstein}, H. and {Goulooze}, H. and {Gunsing}, T. and {G{\"u}sten}, R. and {Hartogh}, P. and {Hatch}, W.~A. and {Higgins}, R. and {Honingh}, E.~C. and {Huisman}, R. and {Jackson}, B.~D. and {Jacobs}, H. and {Jacobs}, K. and {Jarchow}, C. and {Javadi}, H. and {Jellema}, W. and {Justen}, M. and {Karpov}, A. and {Kasemann}, C. and {Kawamura}, J. and {Keizer}, G. and {Kester}, D. and {Klapwijk}, T.~M. and {Klein}, Th. and {Kollberg}, E. and {Kooi}, J. and {Kooiman}, P. -P. and {Kopf}, B. and {Krause}, M. and {Krieg}, J. -M. and {Kramer}, C. and {Kruizenga}, B. and {Kuhn}, T. and {Laauwen}, W. and {Lai}, R. and {Larsson}, B. and {Leduc}, H.~G. and {Leinz}, C. and {Lin}, R.~H. and {Liseau}, R. and {Liu}, G.~S. and {Loose}, A. and {L{\'o}pez-Fernandez}, I. and {Lord}, S. and {Luinge}, W. and {Marston}, A. and {Mart{\'\i}n-Pintado}, J. and {Maestrini}, A. and {Maiwald}, F.~W. and {McCoey}, C. and {Mehdi}, I. and {Megej}, A. and {Melchior}, M. and {Meinsma}, L. and {Merkel}, H. and {Michalska}, M. and {Monstein}, C. and {Moratschke}, D. and {Morris}, P. and {Muller}, H. and {Murphy}, J.~A. and {Naber}, A. and {Natale}, E. and {Nowosielski}, W. and {Nuzzolo}, F. and {Olberg}, M. and {Olbrich}, M. and {Orfei}, R. and {Orleanski}, P. and {Ossenkopf}, V. and {Peacock}, T. and {Pearson}, J.~C. and {Peron}, I. and {Phillip-May}, S. and {Piazzo}, L. and {Planesas}, P. and {Rataj}, M. and {Ravera}, L. and {Risacher}, C. and {Salez}, M. and {Samoska}, L.~A. and {Saraceno}, P. and {Schieder}, R. and {Schlecht}, E. and {Schl{\"o}der}, F. and {Schm{\"u}lling}, F. and {Schultz}, M. and {Schuster}, K. and {Siebertz}, O. and {Smit}, H. and {Szczerba}, R. and {Shipman}, R. and {Steinmetz}, E. and {Stern}, J.~A. and {Stokroos}, M. and {Teipen}, R. and {Teyssier}, D. and {Tils}, T. and {Trappe}, N. and {van Baaren}, C. and {van Leeuwen}, B. -J. and {van de Stadt}, H. and {Visser}, H. and {Wildeman}, K.~J. and {Wafelbakker}, C.~K. and {Ward}, J.~S. and {Wesselius}, P. and {Wild}, W. and {Wulff}, S. and {Wunsch}, H. -J. and {Tielens}, X. and {Zaal}, P. and {Zirath}, H. and {Zmuidzinas}, J. and {Zwart}, F.},
  journal  = {\aap},
  title    = {{The Herschel-Heterodyne Instrument for the Far-Infrared (HIFI)}},
  year     = {2010},
  month    = jul,
  pages    = {L6},
  volume   = {518},
  adsurl   = {https://ui.adsabs.harvard.edu/abs/2010A&A...518L...6D},
  doi      = {10.1051/0004-6361/201014698},
  eid      = {L6},
}

@Article{HVCs,
  author    = {Guillermo Quintana-Lacaci and Jos{\'{e}} Cernicharo and Marcelino Ag{\'{u}}ndez and Jos{\'{e}} Pablo Fonfr{\'{\i}}a and Luis Velilla-Prieto and Carmen S{\'{a}}nchez Contreras and Valent{\'{\i}}n Bujarrabal and Arancha Castro-Carrizo and Javier Alcolea},
  title     = {Hints of the Existence of C-rich Massive Evolved Stars},
  journal   = {The Astrophysical Journal},
  year      = {2019},
  volume    = {876},
  number    = {2},
  pages     = {116},
  month     = {may},
  doi       = {10.3847/1538-4357/ab133e},
  publisher = {American Astronomical Society},
  url       = {https://doi.org/10.3847%2F1538-4357%2Fab133e},
}

@ARTICLE{ziurys2025,
       author = {{Ziurys}, Lucy M. and {Richards}, Anita M.~S.},
        title = "{Molecules and Chemistry in Red Supergiants}",
      journal = {Galaxies},
         year = 2025,
        month = jul,
       volume = {13},
       number = {4},
          eid = {82},
        pages = {82},
          doi = {10.3390/galaxies13040082},
archivePrefix = {arXiv},
       eprint = {2507.15968},
 primaryClass = {astro-ph.SR},
       adsurl = {https://ui.adsabs.harvard.edu/abs/2025Galax..13...82Z}
}

@ARTICLE{massalkhi-Orich,
       author = {{Massalkhi}, S. and {Ag{\'u}ndez}, M. and {Cernicharo}, J. and {Velilla-Prieto}, L.},
        title = "{The abundance of S- and Si-bearing molecules in O-rich circumstellar envelopes of AGB stars}",
      journal = {\aap},
         year = 2020,
        month = sep,
       volume = {641},
          eid = {A57},
        pages = {A57},
          doi = {10.1051/0004-6361/202037900},
archivePrefix = {arXiv},
       eprint = {2007.00572},
 primaryClass = {astro-ph.SR},
       adsurl = {https://ui.adsabs.harvard.edu/abs/2020A&A...641A..57M}
}

@ARTICLE{massalkhi-Crich,
       author = {{Massalkhi}, S. and {Ag{\'u}ndez}, M. and {Cernicharo}, J.},
        title = "{Study of CS, SiO, and SiS abundances in carbon star envelopes: assessing their role as gas-phase precursors of dust}",
      journal = {\aap},
         year = 2019,
        month = aug,
       volume = {628},
          eid = {A62},
        pages = {A62},
          doi = {10.1051/0004-6361/201935069},
archivePrefix = {arXiv},
       eprint = {1906.09461},
 primaryClass = {astro-ph.SR},
       adsurl = {https://ui.adsabs.harvard.edu/abs/2019A&A...628A..62M}
}

@Article{HSO,
  author        = {{Pilbratt}, G.~L. and {Riedinger}, J.~R. and {Passvogel}, T. and {Crone}, G. and {Doyle}, D. and {Gageur}, U. and {Heras}, A.~M. and {Jewell}, C. and {Metcalfe}, L. and {Ott}, S. and {Schmidt}, M.},
  journal       = {\aap},
  title         = {{Herschel Space Observatory. An ESA facility for far-infrared and submillimetre astronomy}},
  year          = {2010},
  month         = jul,
  pages         = {L1},
  volume        = {518},
  adsurl        = {https://ui.adsabs.harvard.edu/abs/2010A&A...518L...1P},
  archiveprefix = {arXiv},
  doi           = {10.1051/0004-6361/201014759},
  eid           = {L1},
  eprint        = {1005.5331},
  primaryclass  = {astro-ph.IM},
}

@Article{GSP-Phot,
  author        = {{Andrae}, R. and {Fouesneau}, M. and {Sordo}, R. and {Bailer-Jones}, C.~A.~L. and {Dharmawardena}, T.~E. and {Rybizki}, J. and {De Angeli}, F. and {Lindstr{\o}m}, H.~E.~P. and {Marshall}, D.~J. and {Drimmel}, R. and {Korn}, A.~J. and {Soubiran}, C. and {Brouillet}, N. and {Casamiquela}, L. and {Rix}, H.-W. and {Abreu Aramburu}, A. and {{\'A}lvarez}, M.~A. and {Bakker}, J. and {Bellas-Velidis}, I. and {Bijaoui}, A. and {Brugaletta}, E. and {Burlacu}, A. and {Carballo}, R. and {Chaoul}, L. and {Chiavassa}, A. and {Contursi}, G. and {Cooper}, W.~J. and {Creevey}, O.~L. and {Dafonte}, C. and {Dapergolas}, A. and {de Laverny}, P. and {Delchambre}, L. and {Demouchy}, C. and {Edvardsson}, B. and {Fr{\'e}mat}, Y. and {Garabato}, D. and {Garc{\'\i}a-Lario}, P. and {Garc{\'\i}a-Torres}, M. and {Gavel}, A. and {Gomez}, A. and {Gonz{\'a}lez-Santamar{\'\i}a}, I. and {Hatzidimitriou}, D. and {Heiter}, U. and {Jean-Antoine Piccolo}, A. and {Kontizas}, M. and {Kordopatis}, G. and {Lanzafame}, A.~C. and {Lebreton}, Y. and {Licata}, E.~L. and {Livanou}, E. and {Lobel}, A. and {Lorca}, A. and {Magdaleno Romeo}, A. and {Manteiga}, M. and {Marocco}, F. and {Mary}, N. and {Nicolas}, C. and {Ordenovic}, C. and {Pailler}, F. and {Palicio}, P.~A. and {Pallas-Quintela}, L. and {Panem}, C. and {Pichon}, B. and {Poggio}, E. and {Recio-Blanco}, A. and {Riclet}, F. and {Robin}, C. and {Santove{\~n}a}, R. and {Sarro}, L.~M. and {Schultheis}, M.~S. and {Segol}, M. and {Silvelo}, A. and {Slezak}, I. and {Smart}, R.~L. and {S{\"u}veges}, M. and {Th{\'e}venin}, F. and {Torralba Elipe}, G. and {Ulla}, A. and {Utrilla}, E. and {Vallenari}, A. and {van Dillen}, E. and {Zhao}, H. and {Zorec}, J.},
  journal       = {\aap},
  title         = {{Gaia Data Release 3. Analysis of the Gaia BP/RP spectra using the General Stellar Parameterizer from Photometry}},
  year          = {2023},
  month         = jun,
  pages         = {A27},
  volume        = {674},
  adsurl        = {https://ui.adsabs.harvard.edu/abs/2023A&A...674A..27A},
  archiveprefix = {arXiv},
  doi           = {10.1051/0004-6361/202243462},
  eid           = {A27},
  eprint        = {2206.06138},
  primaryclass  = {astro-ph.SR},
}

@Article{h2orates,
  author   = {{Daniel}, F. and {Dubernet}, M.-L. and {Grosjean}, A.},
  journal  = {\aap},
  title    = {{Rotational excitation of 45 levels of ortho/para-H$_{2}$O by excited ortho/para-H$_{2}$ from 5 K to 1500 K: state-to-state, effective, and thermalized rate coefficients}},
  year     = {2011},
  month    = dec,
  pages    = {A76},
  volume   = {536},
  adsurl   = {https://ui.adsabs.harvard.edu/abs/2011A&A...536A..76D},
  doi      = {10.1051/0004-6361/201118049},
  eid      = {A76},
}

@Article{Schoier13,
  author        = {{Sch{\"o}ier}, F.~L. and {Ramstedt}, S. and {Olofsson}, H. and {Lindqvist}, M. and {Bieging}, J.~H. and {Marvel}, K.~B.},
  journal       = {\aap},
  title         = {{The abundance of HCN in circumstellar envelopes of AGB stars of different chemical type}},
  year          = {2013},
  month         = feb,
  pages         = {A78},
  volume        = {550},
  adsurl        = {https://ui.adsabs.harvard.edu/abs/2013A&A...550A..78S},
  archiveprefix = {arXiv},
  doi           = {10.1051/0004-6361/201220400},
  eid           = {A78},
  eprint        = {1301.2129},
  primaryclass  = {astro-ph.SR},
}

@Article{HCN_cols,
  author   = {{Hern{\'a}ndez Vera}, M. and {Lique}, F. and {Dumouchel}, F. and {Hily-Blant}, P. and {Faure}, A.},
  journal  = {\mnras},
  title    = {{The rotational excitation of the HCN and HNC molecules by H$_{2}$ revisited}},
  year     = {2017},
  month    = jun,
  number   = {1},
  pages    = {1084-1091},
  volume   = {468},
  adsurl   = {https://ui.adsabs.harvard.edu/abs/2017MNRAS.468.1084H},
  doi      = {10.1093/mnras/stx422},
}

@Article{Draine,
  author   = {{Draine}, B.~T.},
  journal  = {\apjs},
  title    = {{Photoelectric heating of interstellar gas.}},
  year     = {1978},
  month    = apr,
  pages    = {595-619},
  volume   = {36},
  adsurl   = {https://ui.adsabs.harvard.edu/abs/1978ApJS...36..595D},
  doi      = {10.1086/190513},
}

@Article{Javadi11,
  author        = {{Javadi}, Atefeh and {van Loon}, Jacco Th. and {Mirtorabi}, Mohammad Taghi},
  journal       = {\mnras},
  title         = {{The UK Infrared Telescope M33 monitoring project - II. The star formation history in the central square kiloparsec}},
  year          = {2011},
  month         = jul,
  number        = {4},
  pages         = {3394-3409},
  volume        = {414},
  adsurl        = {https://ui.adsabs.harvard.edu/abs/2011MNRAS.414.3394J},
  archiveprefix = {arXiv},
  doi           = {10.1111/j.1365-2966.2011.18638.x},
  eprint        = {1103.0755},
  primaryclass  = {astro-ph.SR},
}

@Article{Schootemeijer19,
  author        = {{Schootemeijer}, A. and {Langer}, N. and {Grin}, N.~J. and {Wang}, C.},
  journal       = {\aap},
  title         = {{Constraining mixing in massive stars in the Small Magellanic Cloud}},
  year          = {2019},
  month         = may,
  pages         = {A132},
  volume        = {625},
  adsurl        = {https://ui.adsabs.harvard.edu/abs/2019A&A...625A.132S},
  archiveprefix = {arXiv},
  doi           = {10.1051/0004-6361/201935046},
  eid           = {A132},
  eprint        = {1903.10423},
  primaryclass  = {astro-ph.SR},
}

\begin{appendix}
\section{Line properties}

Below we list the main properties of the lines observed with IRAM 30m and HIFI.

\vspace{1cm}
\onecolumn
\begin{longtable}{lccccc}
\caption{Line transitions detected toward AFGL\,2233.}\\
\hline\hline
$\nu_{rest}$ (MHz)      &Molecule       &Transition     &\tas (mK) &Area (K km/s) &Notes\\    
\hline
\endfirsthead
\caption{continued.}\\
\hline\hline
\hline
\endhead
\hline
\endfoot
      71889.595    &           HC$_{5}$N &                        27--26         &     10.00 $\pm$  2.8 &  0.65 &       T$_\mathrm{A}^{*}$ \\
      72475.060    &        HC$^{13}$CCN &                          8--7         &      5.00 $\pm$  2.3 &  0.38 &       T$_\mathrm{A}^{*}$ \\
      72618.112    &                 SiS &                          4--3         &      5.00 $\pm$  2.4 &  0.41 &       T$_\mathrm{A}^{*}$ \\
      72783.818    &           HC$_{3}$N &                          8--7         &     80.00 $\pm$  2.6 &  4.73 &       T$_\mathrm{A}^{*}$ \\
      74551.987    &           HC$_{5}$N &                        28--27         &      7.00 $\pm$  2.0 &  0.56 &       T$_\mathrm{A}^{*}$ \\
      76117.477    &            C$_{4}$H &                          8--7         &      3.00 $\pm$  2.0 &  0.23 &       T$_\mathrm{A}^{*}$ \\
      76158.375    &            C$_{4}$H &                          8--7         &      4.00 $\pm$  2.0 &  0.25 &       T$_\mathrm{A}^{*}$ \\
      76338.016    &         $^{13}$CCCN &                          8--7         &      7.00 $\pm$  1.9 &  0.48 &       T$_\mathrm{A}^{*}$ \\
      77214.359    &           HC$_{5}$N &                        29--28         &      9.00 $\pm$  2.2 &  0.61 &       T$_\mathrm{A}^{*}$ \\
      79160.367    &              C$_3$N &                          8--7         &      5.00 $\pm$  1.6 &  0.76 &       T$_\mathrm{A}^{*}$ \\
      79876.710    &           HC$_{5}$N &                        30--29         &      8.00 $\pm$  1.4 &  0.66 &       T$_\mathrm{A}^{*}$ \\
      80082.377    &             HC$_7$N &                        71--70         &      3.00 $\pm$  1.4 &  0.19 &       T$_\mathrm{A}^{*}$ \\
      81505.202    &                 CCS &                  6$_7$--5$_6$         &      2.00 $\pm$  1.0 &  0.20 &       T$_\mathrm{A}^{*}$ \\
      81534.111    &        HC$^{13}$CCN &                          9--8         &      4.00 $\pm$  1.1 &  0.26 &       T$_\mathrm{A}^{*}$ \\
      81881.462    &           HC$_{3}$N &                          9--8         &    100.00 $\pm$  1.1 &  6.78 &       T$_\mathrm{A}^{*}$ \\
      82539.039    &           HC$_{5}$N &                        31--30         &      8.00 $\pm$  1.1 &  0.64 &       T$_\mathrm{A}^{*}$ \\
      84746.165    &          $^{30}$SiO &                          2--1         &      5.00 $\pm$  0.9 &  0.34 &       T$_\mathrm{A}^{*}$ \\
      85201.346    &           HC$_{5}$N &                        32--31         &      8.00 $\pm$  1.0 &  0.59 &       T$_\mathrm{A}^{*}$ \\
      85338.900    &          C$^{13}$CH &          2$_{1,2}$--1$_{0,1}$         &      5.00 $\pm$  1.0 &  0.36 &       T$_\mathrm{A}^{*}$ \\
      85634.039    &            C$_{4}$H &                          9--8         &      5.00 $\pm$  0.9 &  0.38 &       T$_\mathrm{A}^{*}$ \\
      85674.672    &            C$_{4}$H &                          9--8         &      5.00 $\pm$  0.9 &  0.41 &       T$_\mathrm{A}^{*}$ \\
      85759.194    &          $^{29}$SiO &                          3--2         &      7.00 $\pm$  1.0 &  0.49 &       T$_\mathrm{A}^{*}$ \\
      86339.921    &          H$^{13}$CN &                          1--0         &     57.00 $\pm$  1.2 &  4.00 &       T$_\mathrm{A}^{*}$ \\
      86846.986    &                 SiO &                          2--1         &     98.00 $\pm$  1.0 &  6.38 &       T$_\mathrm{A}^{*}$ \\
      87284.105    &                 CCH &                1--0(3/2--1/2)         &      5.00 $\pm$  1.0 &  0.36 &       T$_\mathrm{A}^{*}$ \\
      87316.898    &                 CCH &                1--0(3/2--1/2)         &     46.00 $\pm$  1.0 &  3.87 &       T$_\mathrm{A}^{*}$ \\
      87328.585    &                 CCH &                1--0(3/2--1/2)         &     23.00 $\pm$  1.0 &  1.64 &       T$_\mathrm{A}^{*}$ \\
      87401.989    &                 CCH &                1--0(1/2--1/2)         &     24.00 $\pm$  1.0 &  1.98 &       T$_\mathrm{A}^{*}$ \\
      87407.165    &                 CCH &                1--0(1/2--1/2)         &      8.00 $\pm$  1.0 &  0.67 &       T$_\mathrm{A}^{*}$ \\
      87446.470    &                 CCH &                1--0(1/2--1/2)         &      4.00 $\pm$  1.0 &  0.36 &       T$_\mathrm{A}^{*}$ \\
      87863.630    &           HC$_{5}$N &                        33--32         &      7.00 $\pm$  1.0 &  0.59 &       T$_\mathrm{A}^{*}$ \\
      88166.794    &        H$^{13}$CCCN &                         10--9         &      1.00 $\pm$  1.1 &  0.21 &       T$_\mathrm{A}^{*}$ \\
      88631.602    &                 HCN &                          1--0         &    863.00 $\pm$ 17.5 & 54.94 &       T$_\mathrm{A}^{*}$ \\
      89054.953    &              C$_3$N &                          9--8         &      7.00 $\pm$  1.1 &  1.09 &       T$_\mathrm{A}^{*}$ \\
      90525.889    &           HC$_{5}$N &                        34--33         &      6.00 $\pm$  0.9 &  0.53 &       T$_\mathrm{A}^{*}$ \\
      90593.045    &        HCC$^{13}$CN &                         10--9         &      3.00 $\pm$  1.0 &  0.25 &       T$_\mathrm{A}^{*}$ \\
      90663.563    &                 HNC &                          1--0         &     37.00 $\pm$  1.1 &  3.04 &       T$_\mathrm{A}^{*}$ \\
      90771.566    &                 SiS &                          5--4         &      9.00 $\pm$  0.9 &  0.75 &       T$_\mathrm{A}^{*}$ \\
      90978.989    &           HC$_{3}$N &                         10--9         &     95.00 $\pm$  1.1 &  7.14 &       T$_\mathrm{A}^{*}$ \\
      91979.994    &          CH$_{3}$CN &                  5$_2$--4$_2$         &      3.00 $\pm$  1.2 &  0.25 &       T$_\mathrm{A}^{*}$ \\
      92494.271    &           $^{13}$CS &                          2--1         &      4.00 $\pm$  1.1 &  0.37 &       T$_\mathrm{A}^{*}$ \\
      93063.623    &           SiC$_{2}$ &          4$_{0,4}$--3$_{0,3}$         &     20.00 $\pm$  1.1 &  1.58 &       T$_\mathrm{A}^{*}$ \\
      93188.125    &           HC$_{5}$N &               $\emph{35--34}$         &      6.00 $\pm$  1.0 &  0.45 &       T$_\mathrm{A}^{*}$ \\
      93870.091    &                 CCS &                  7$_8$--6$_7$         &      2.00 $\pm$  1.0 &  0.20 &       T$_\mathrm{A}^{*}$ \\
      94245.355    &           SiC$_{2}$ &          4$_{2,3}$--3$_{2,2}$         &     13.00 $\pm$  1.2 &  1.14 &       T$_\mathrm{A}^{*}$ \\
      95150.414    &            C$_{4}$H &             10--9(21/2--19/2)         &      6.00 $\pm$  1.0 &  0.50 &       T$_\mathrm{A}^{*}$ \\
      95190.820    &            C$_{4}$H &             10--9(19/2--17/2)         &      6.00 $\pm$  1.0 &  0.48 &       T$_\mathrm{A}^{*}$ \\
      95579.416    &           SiC$_{2}$ &          4$_{2,2}$--3$_{2,1}$         &     14.00 $\pm$  1.1 &  1.14 &       T$_\mathrm{A}^{*}$ \\
      95850.335    &           HC$_{5}$N &                        36--35         &      6.00 $\pm$  1.0 &  0.43 &       T$_\mathrm{A}^{*}$ \\
      96412.952    &           C$^{34}$S &                          2--1         &     16.00 $\pm$  0.9 &  1.28 &       T$_\mathrm{A}^{*}$ \\
      96983.001    &        H$^{13}$CCCN &                        11--10         &      2.00 $\pm$  0.8 &  0.17 &       T$_\mathrm{A}^{*}$ \\
      97172.064    &           C$^{33}$S &                          2--1         &      3.00 $\pm$  0.8 &  0.28 &       T$_\mathrm{A}^{*}$ \\
      97980.953    &                  CS &                          2--1         &    264.00 $\pm$  1.2 & 20.51 &       T$_\mathrm{A}^{*}$ \\
      98512.519    &           HC$_{5}$N &                        37--36         &      5.00 $\pm$  0.7 &  0.48 &       T$_\mathrm{A}^{*}$ \\
      98949.383    &              C$_3$N &                         10--9         &     11.00 $\pm$  0.8 &  2.00 &       T$_\mathrm{A}^{*}$ \\
      99661.467    &        HCC$^{13}$CN &                        11--10         &      3.00 $\pm$  0.8 &  0.31 &       T$_\mathrm{A}^{*}$ \\
      99866.503    &                 CCS &                  8$_7$--7$_6$         &      2.00 $\pm$  0.8 &  0.16 &       T$_\mathrm{A}^{*}$ \\
     100076.385    &           HC$_{3}$N &                        11--10         &    114.00 $\pm$  0.9 &  9.69 &       T$_\mathrm{A}^{*}$ \\
     101174.676    &           HC$_{5}$N &                        38--37         &      4.00 $\pm$  0.8 &  0.44 &       T$_\mathrm{A}^{*}$ \\
     103836.806    &           HC$_{5}$N &                        39--38         &      4.00 $\pm$  1.4 &  0.38 &       T$_\mathrm{A}^{*}$ \\
     104666.586    &            C$_{4}$H &            11--10(23/2--21/2)         &      7.00 $\pm$  1.5 &  0.66 &       T$_\mathrm{A}^{*}$ \\
     104706.867    &            C$_{4}$H &            11--10(21/2--19/2)         &      7.00 $\pm$  1.5 &  0.66 &       T$_\mathrm{A}^{*}$ \\
     106347.731    &                 CCS &                  8$_9$--7$_8$         &      3.00 $\pm$  1.4 &  0.22 &       T$_\mathrm{A}^{*}$ \\
     106498.908    &           HC$_{5}$N &                        40--39         &      3.00 $\pm$  1.5 &  0.36 &       T$_\mathrm{A}^{*}$ \\
     108647.523    &           $^{13}$CN &    1$_{*,1,*}$--0$_{1/2,*,*}$         &      2.00 $\pm$  1.8 &  0.30 &       T$_\mathrm{A}^{*}$ \\
     108721.003    &        HCC$^{13}$CN &                        12--11         &      3.00 $\pm$  1.2 &  0.27 &       T$_\mathrm{A}^{*}$ \\
     108784.203    &           $^{13}$CN &1$_{1,3/2,2,*}$--0$_{1/2,1,*}$         &      2.70 $\pm$  1.8 &  0.31 &       T$_\mathrm{A}^{*}$ \\
     108843.633    &              C$_3$N &                        11--10         &     11.00 $\pm$  1.8 &  1.64 &       T$_\mathrm{A}^{*}$ \\
     108924.303    &                 SiS &                          6--5         &     16.00 $\pm$  1.6 &  1.55 &       T$_\mathrm{A}^{*}$ \\
     109173.637    &           HC$_{3}$N &                        12--11         &     85.00 $\pm$  1.9 &  7.86 &       T$_\mathrm{A}^{*}$ \\
     109218.148    &           $^{13}$CN &  1$_{1/2,1,*}$--0$_{1/2,0,1}$         &      1.50 $\pm$  1.8 &  0.09 &       T$_\mathrm{A}^{*}$ \\
     110201.354    &           $^{13}$CO &                          1--0         &      8.00 $\pm$  2.2 &  1.40 &       T$_\mathrm{A}^{*}$ \\
     110383.499    &          CH$_{3}$CN &                  6$_0$--5$_0$         &      4.00 $\pm$  1.8 &  0.44 &       T$_\mathrm{A}^{*}$ \\
     111823.024    &           HC$_{5}$N &                        42--41         &      4.00 $\pm$  1.6 &  0.40 &       T$_\mathrm{A}^{*}$ \\
     113168.781    &                  CN &          1$_{1/2}$--0$_{1/2}$         &     50.00 $\pm$  2.1 &  2.97 &       T$_\mathrm{A}^{*}$ \\
     113496.430    &                  CN &          1$_{3/2}$--0$_{1/2}$         &    179.00 $\pm$  2.1 &  4.71 &       T$_\mathrm{A}^{*}$ \\
     113659.508    &                   U &                            --         &     10.00 $\pm$  2.3 &  0.73 &       T$_\mathrm{A}^{*}$ \\
     114182.531    &            C$_{4}$H &            12--11(25/2--23/2)         &      9.00 $\pm$  2.9 &  0.90 &       T$_\mathrm{A}^{*}$ \\
     114221.023    &            C$_{4}$H &            12--11(23/2--21/2)         &     11.00 $\pm$  2.9 &  0.99 &       T$_\mathrm{A}^{*}$ \\
     114318.532    &          C$_8$H$^-$ &                         98-97         &      8.00 $\pm$  2.7 &  0.17 &       T$_\mathrm{A}^{*}$ \\
     115271.202    &                  CO &                          1--0         &    549.00 $\pm$ 26.7 & 67.18 &       T$_\mathrm{A}^{*}$ \\
     115382.377    &           SiC$_{2}$ &           5$_{0,5}$-4$_{0,4}$         &     32.00 $\pm$  3.7 &  3.46 &       T$_\mathrm{A}^{*}$ \\
     199812.375    &            C$_{4}$H &            21--20(43/2--41/2)         &     14.00 $\pm$  4.5 &  0.90 &       T$_\mathrm{A}^{*}$ \\
     199826.156    &         C$_4$H $v7$ & 21$_{-1,41/2}$--20$_{1,39/2}$         &      8.00 $\pm$  4.5 &  0.64 &       T$_\mathrm{A}^{*}$ \\
     199850.781    &            C$_{4}$H &            21--20(41/2--39/2)         &     18.00 $\pm$  4.5 &  1.35 &       T$_\mathrm{A}^{*}$ \\
     200135.388    &           HC$_{3}$N &                        22--21         &     42.00 $\pm$  4.3 &  2.47 &       T$_\mathrm{A}^{*}$ \\
     200162.407    &           SiC$_{2}$ &          9$_{0,9}$--8$_{0,8}$         &     58.00 $\pm$  4.3 &  3.46 &       T$_\mathrm{A}^{*}$ \\
     202355.507    &          CH$_{3}$CN &                11$_0$--10$_0$         &      7.00 $\pm$  4.7 &  0.37 &       T$_\mathrm{A}^{*}$ \\
     207772.172    &              C$_3$N &                        21--20         &      9.00 $\pm$  3.0 &  0.69 &       T$_\mathrm{A}^{*}$ \\
     207851.962    &   $^{29}$Si$^{34}$S &                        12--11         &      9.00 $\pm$  3.0 &  0.60 &       T$_\mathrm{A}^{*}$ \\
     207867.625    &                   U &                            --         &      9.00 $\pm$  3.0 &  0.69 &       T$_\mathrm{A}^{*}$ \\
     209230.199    &           HC$_{3}$N &                        23--22         &     23.00 $\pm$  6.2 &  1.52 &       T$_\mathrm{A}^{*}$ \\
     209324.906    &            C$_{4}$H &            22--21(45/2--43/2)         &     11.00 $\pm$  5.5 &  0.93 &       T$_\mathrm{A}^{*}$ \\
     209363.297    &            C$_{4}$H &            22--21(43/2--41/2)         &     15.00 $\pm$  5.5 &  1.13 &       T$_\mathrm{A}^{*}$ \\
     209891.991    &           SiC$_{2}$ &          9$_{2,8}$--8$_{2,7}$         &     39.00 $\pm$  5.3 &  2.42 &       T$_\mathrm{A}^{*}$ \\
     211023.281    &           SiC$_{2}$ &          9$_{8,2}$--8$_{8,1}$         &     10.00 $\pm$  4.8 &  0.50 &       T$_\mathrm{A}^{*}$ \\
     211853.467    &          $^{30}$SiO &                          5--4         &     21.00 $\pm$  5.4 &  1.34 &       T$_\mathrm{A}^{*}$ \\
     212031.873    &           SiC$_{2}$ &          9$_{6,4}$--8$_{6,3}$         &     24.00 $\pm$  5.5 &  1.42 &       T$_\mathrm{A}^{*}$ \\
     213208.022    &           SiC$_{2}$ &          9$_{4,6}$--8$_{4,5}$         &     28.00 $\pm$  4.9 &  1.88 &       T$_\mathrm{A}^{*}$ \\
     213292.328    &           SiC$_{2}$ &          9$_{4,5}$--8$_{4,4}$         &     24.00 $\pm$  4.9 &  1.49 &       T$_\mathrm{A}^{*}$ \\
     214385.752    &          $^{29}$SiO &                          5--4         &     28.00 $\pm$  6.7 &  1.58 &       T$_\mathrm{A}^{*}$ \\
     217104.920    &                 SiO &                          5--4         &    403.00 $\pm$  3.4 & 21.01 &       T$_\mathrm{A}^{*}$ \\
     217297.125    &           $^{13}$CN &                2--1(3/2--1/2)         &      2.00 $\pm$  2.1 &  0.21 &       T$_\mathrm{A}^{*}$ \\
     217456.594    &           $^{13}$CN &                2--1(5/2--3/2)         &      4.00 $\pm$  2.1 &  0.31 &       T$_\mathrm{A}^{*}$ \\
     217663.203    &              C$_3$N &                        22--21         &      4.00 $\pm$  2.1 &  0.32 &       T$_\mathrm{A}^{*}$ \\
     217817.656    &                 SiS &                        12--11         &     76.00 $\pm$  4.7 &  5.04 &       T$_\mathrm{A}^{*}$ \\
     218007.543    &                 SiN &                5--4(9/2--7/2)         &      7.00 $\pm$  3.6 &  0.52 &       T$_\mathrm{A}^{*}$ \\
     218324.709    &           HC$_{3}$N &                        24--23         &     24.00 $\pm$  3.5 &  1.69 &       T$_\mathrm{A}^{*}$ \\
     218512.915    &                 SiN &               5--4(11/2--9/2)         &     10.00 $\pm$  3.6 &  0.67 &       T$_\mathrm{A}^{*}$ \\
     218837.000    &            C$_{4}$H &            23--22(47/2--45/2)         &     13.00 $\pm$  3.8 &  0.77 &       T$_\mathrm{A}^{*}$ \\
     218875.344    &            C$_{4}$H &            23--22(45/2--43/2)         &     13.00 $\pm$  3.8 &  1.01 &       T$_\mathrm{A}^{*}$ \\
     220398.684    &           $^{13}$CO &                          2--1         &     20.00 $\pm$  4.1 &  2.35 &       T$_\mathrm{A}^{*}$ \\
     220747.259    &          CH$_{3}$CN &                12$_0$--11$_0$         &     10.00 $\pm$  3.6 &  1.05 &       T$_\mathrm{A}^{*}$ \\
     220773.676    &           SiC$_{2}$ &        10$_{0,10}$--9$_{0,9}$         &     34.00 $\pm$  3.6 &  2.48 &       T$_\mathrm{A}^{*}$ \\
     222009.399    &           SiC$_{2}$ &          9$_{2,7}$--8$_{2,6}$         &     43.00 $\pm$  3.9 &  2.81 &       T$_\mathrm{A}^{*}$ \\
     224692.531    &                   U &                            --         &     10.00 $\pm$  3.4 &  0.74 &       T$_\mathrm{A}^{*}$ \\
     225554.609    &            CH$_2$NH &          1$_{1,1}$--0$_{0,0}$         &      5.00 $\pm$  2.1 &  0.28 &       T$_\mathrm{A}^{*}$ \\
     226333.094    &                  CN &                2--1(3/2--3/2)         &     24.00 $\pm$  3.4 &  2.80 &       T$_\mathrm{A}^{*}$ \\
     226658.922    &                  CN &                2--1(3/2--1/2)         &    103.00 $\pm$  3.0 &  5.70 &       T$_\mathrm{A}^{*}$ \\
     226876.453    &                  CN &                2--1(5/2--3/2)         &    240.00 $\pm$  2.9 & 10.40 &       T$_\mathrm{A}^{*}$ \\
     227418.906    &           HC$_{3}$N &                        25--24         &     22.00 $\pm$  3.6 &  1.61 &       T$_\mathrm{A}^{*}$ \\
     228348.594    &            C$_{4}$H &            24--23(49/2--47/2)         &     10.00 $\pm$  2.9 &  0.77 &       T$_\mathrm{A}^{*}$ \\
     228386.953    &            C$_{4}$H &            24--23(47/2--45/2)         &     12.00 $\pm$  2.9 &  0.97 &       T$_\mathrm{A}^{*}$ \\
     228557.844    &         C$_4$H $v7$ &                        24--23         &      4.00 $\pm$  1.8 &  0.55 &       T$_\mathrm{A}^{*}$ \\
     230463.712    &                   U &                            --         &    404.00 $\pm$  8.9 &  0.54 &       T$_\mathrm{A}^{*}$ \\
     230538.000    &                  CO &                          2--1         &   1540.00 $\pm$ 47.5 &113.97 &       T$_\mathrm{A}^{*}$ \\
     231220.685    &           $^{13}$CS &                          5--4         &     16.00 $\pm$ 10.9 &  0.99 &       T$_\mathrm{A}^{*}$ \\
     232534.063    &           SiC$_{2}$ &         10$_{2,9}$--9$_{2,8}$         &     28.00 $\pm$  8.9 &  2.08 &       T$_\mathrm{A}^{*}$ \\
     235961.366    &                 SiS &                        13--12         &     68.00 $\pm$  9.9 &  4.70 &       T$_\mathrm{A}^{*}$ \\
     236512.776    &           HC$_{3}$N &                        26--25         &     19.00 $\pm$  8.4 &  1.22 &       T$_\mathrm{A}^{*}$ \\
     236962.031    &                   U &                            --         &     16.00 $\pm$  7.7 &  0.94 &       T$_\mathrm{A}^{*}$ \\
     237150.009    &           SiC$_{2}$ &         10$_{4,7}$--9$_{4,6}$         &     29.00 $\pm$  8.7 &  2.20 &       T$_\mathrm{A}^{*}$ \\
     237331.303    &           SiC$_{2}$ &         10$_{4,6}$--9$_{4,5}$         &     18.00 $\pm$  5.7 &  1.40 &       T$_\mathrm{A}^{*}$ \\
     237859.688    &            C$_{4}$H &            25--24(51/2--49/2)         &      3.00 $\pm$  4.0 &  0.43 &       T$_\mathrm{A}^{*}$ \\
     237898.031    &            C$_{4}$H &            25--24(49/2--47/2)         &      4.00 $\pm$  4.0 &  0.38 &       T$_\mathrm{A}^{*}$ \\
     239073.047    &                   U &                            --         &     49.00 $\pm$  3.4 &  1.76 &       T$_\mathrm{A}^{*}$ \\
     239098.406    &                   U &                            --         &     57.00 $\pm$  3.4 &  1.73 &       T$_\mathrm{A}^{*}$ \\
     241016.089    &           C$^{34}$S &                          5--4         &     60.00 $\pm$  3.8 &  4.24 &       T$_\mathrm{A}^{*}$ \\
     241367.704    &           SiC$_{2}$ &      11$_{0,11}$--10$_{0,10}$         &     33.00 $\pm$  3.5 &  2.43 &       T$_\mathrm{A}^{*}$ \\
     242913.610    &           C$^{33}$S &                          5--4         &     13.00 $\pm$  3.7 &  0.96 &       T$_\mathrm{A}^{*}$ \\
     244935.555    &                  CS &                          5--4         &    879.00 $\pm$  4.6 & 51.10 &       T$_\mathrm{A}^{*}$ \\
     245606.307    &           HC$_{3}$N &                        27--26         &     14.00 $\pm$  3.0 &  1.15 &       T$_\mathrm{A}^{*}$ \\
     247529.132    &           SiC$_{2}$ &         10$_{2,8}$--9$_{2,7}$         &     46.00 $\pm$  6.2 &  3.20 &       T$_\mathrm{A}^{*}$ \\
     254103.212    &                 SiS &                        14--13         &     53.00 $\pm$  4.8 &  3.95 &       T$_\mathrm{A}^{*}$ \\
     254216.652    &          $^{30}$SiO &                          6--5         &     17.00 $\pm$  4.7 &  1.28 &       T$_\mathrm{A}^{*}$ \\
     254699.486    &           HC$_{3}$N &                        28--27         &     19.00 $\pm$  5.0 &  1.59 &       T$_\mathrm{A}^{*}$ \\
     254981.478    &           SiC$_{2}$ &       11$_{2,10}$--10$_{2,9}$         &     23.00 $\pm$  4.6 &  1.67 &       T$_\mathrm{A}^{*}$ \\
     257255.213    &          $^{29}$SiO &                          6--5         &     24.00 $\pm$  5.1 &  1.83 &       T$_\mathrm{A}^{*}$ \\
     258065.049    &           SiC$_{2}$ &11$_{8,(4-3)}$--10$_{8,(3-2)}$         &      8.00 $\pm$  5.6 &  0.50 &       T$_\mathrm{A}^{*}$ \\
     259011.798    &          H$^{13}$CN &                          3--2         &    179.00 $\pm$  5.1 & 11.80 &       T$_\mathrm{A}^{*}$ \\
     259433.413    &           SiC$_{2}$ &        11$_{6,5}$--10$_{6,4}$         &     16.00 $\pm$  4.9 &  1.22 &       T$_\mathrm{A}^{*}$ \\
     260518.018    &                 SiO &                          6--5         &    277.00 $\pm$  6.2 & 17.65 &       T$_\mathrm{A}^{*}$ \\
     261150.699    &           SiC$_{2}$ &        11$_{4,8}$--10$_{4,7}$         &     12.00 $\pm$  6.0 &  0.98 &       T$_\mathrm{A}^{*}$ \\
     261509.322    &           SiC$_{2}$ &        11$_{4,7}$--10$_{4,6}$         &     15.00 $\pm$  5.5 &  1.12 &       T$_\mathrm{A}^{*}$ \\
     261978.120    &                 CCH &      3$_{7/2,3}$--2$_{5/2,3}$         &     21.00 $\pm$  5.7 &  1.19 &       T$_\mathrm{A}^{*}$ \\
     262004.260    &                 CCH &      3$_{7/2,4}$--2$_{5/2,3}$         &    244.00 $\pm$  5.7 & 23.22 &       T$_\mathrm{A}^{*}$ \\
     262006.482    &                 CCH &      3$_{7/2,3}$--2$_{5/2,2}$         &    293.00 $\pm$  5.7 & 18.69 &       T$_\mathrm{A}^{*}$ \\
     262064.986    &                 CCH &      3$_{5/2,3}$--2$_{3/2,2}$         &    237.00 $\pm$  5.7 & 21.51 &       T$_\mathrm{A}^{*}$ \\
     262067.469    &                 CCH &      3$_{5/2,2}$--2$_{3/2,1}$         &    150.00 $\pm$  5.7 &  9.71 &       T$_\mathrm{A}^{*}$ \\
     262078.934    &                 CCH &      3$_{5/2,2}$--2$_{3/2,1}$         &     41.00 $\pm$  5.7 &  2.38 &       T$_\mathrm{A}^{*}$ \\
     262208.614    &                 CCH &      3$_{5/2,3}$--2$_{5/2,3}$         &      9.00 $\pm$  9.6 &  1.19 &       T$_\mathrm{A}^{*}$ \\
     262222.585    &                 CCH &      3$_{5/2,2}$--2$_{5/2,3}$         &      5.00 $\pm$  9.6 &  0.34 &       T$_\mathrm{A}^{*}$ \\
     262236.958    &                 CCH &      3$_{5/2,3}$--2$_{5/2,2}$         &      0.00 $\pm$  9.6 &  0.39 &       T$_\mathrm{A}^{*}$ \\
     262250.929    &                 CCH &      3$_{5/2,2}$--2$_{5/2,2}$         &      0.00 $\pm$  9.6 &  0.32 &       T$_\mathrm{A}^{*}$ \\
     265886.433    &                 HCN &                          3--2         &   1759.00 $\pm$ 19.2 &101.93 &       T$_\mathrm{A}^{*}$ \\
     268479.867    &                   U &                            --         &     26.00 $\pm$  3.5 &  1.04 &       T$_\mathrm{A}^{*}$ \\
     271981.107    &                 HNC &                          3--2         &     53.00 $\pm$ 13.2 &  4.58 &       T$_\mathrm{A}^{*}$ \\
     272243.053    &                 SiS &                        15--14         &     57.00 $\pm$ 11.4 &  4.03 &       T$_\mathrm{A}^{*}$ \\
     272787.823    &           SiC$_{2}$ &        11$_{2,9}$--10$_{2,8}$         &     27.00 $\pm$ 15.0 &  2.48 &       T$_\mathrm{A}^{*}$ \\
     556936.002    &              H$_2$O &          1$_{1,0}$--1$_{0,1}$         &     28.00 $\pm$  3.0 &  1.26 &          T$_\mathrm{mb}$ \\
     572498.160    &              NH$_3$ &         1$_{0,1}$--0$_{0,-1}$         &     22.00 $\pm$  2.4 &  1.33 &          T$_\mathrm{mb}$ \\
     690552.086    &          H$^{13}$CN &                          8--7         &      9.00 $\pm$  5.5 &  0.60 &          T$_\mathrm{mb}$ \\
     691473.076    &                  CO &                          6--5         &    156.00 $\pm$  5.8 &  8.51 &          T$_\mathrm{mb}$ \\
     694294.129    &                 SiO &                        16--15         &     12.00 $\pm$  5.6 &  0.98 &          T$_\mathrm{mb}$ \\
    1113342.964    &              H$_2$O &          1$_{1,1}$--0$_{0,0}$         &     35.00 $\pm$ 12.1 &  1.87 &          T$_\mathrm{mb}$ \\
    1151449.090    &                 HCN &                        13--12         &    112.00 $\pm$ 29.7 &  4.19 &          T$_\mathrm{mb}$ \\
    1151985.444    &                  CO &			   10--9	 &    170.00 $\pm$ 24.3 &  8.61 &          T$_\mathrm{mb}$ \\
    1841345.514    &                  CO &                        16--15         &    155.24 $\pm$ 46.6 &  7.72 &          T$_\mathrm{mb}$ \\
\hline
\end{longtable}

\section{Rotational Diagrams}

   \begin{figure}
   \centering
   \includegraphics[width=8cm]{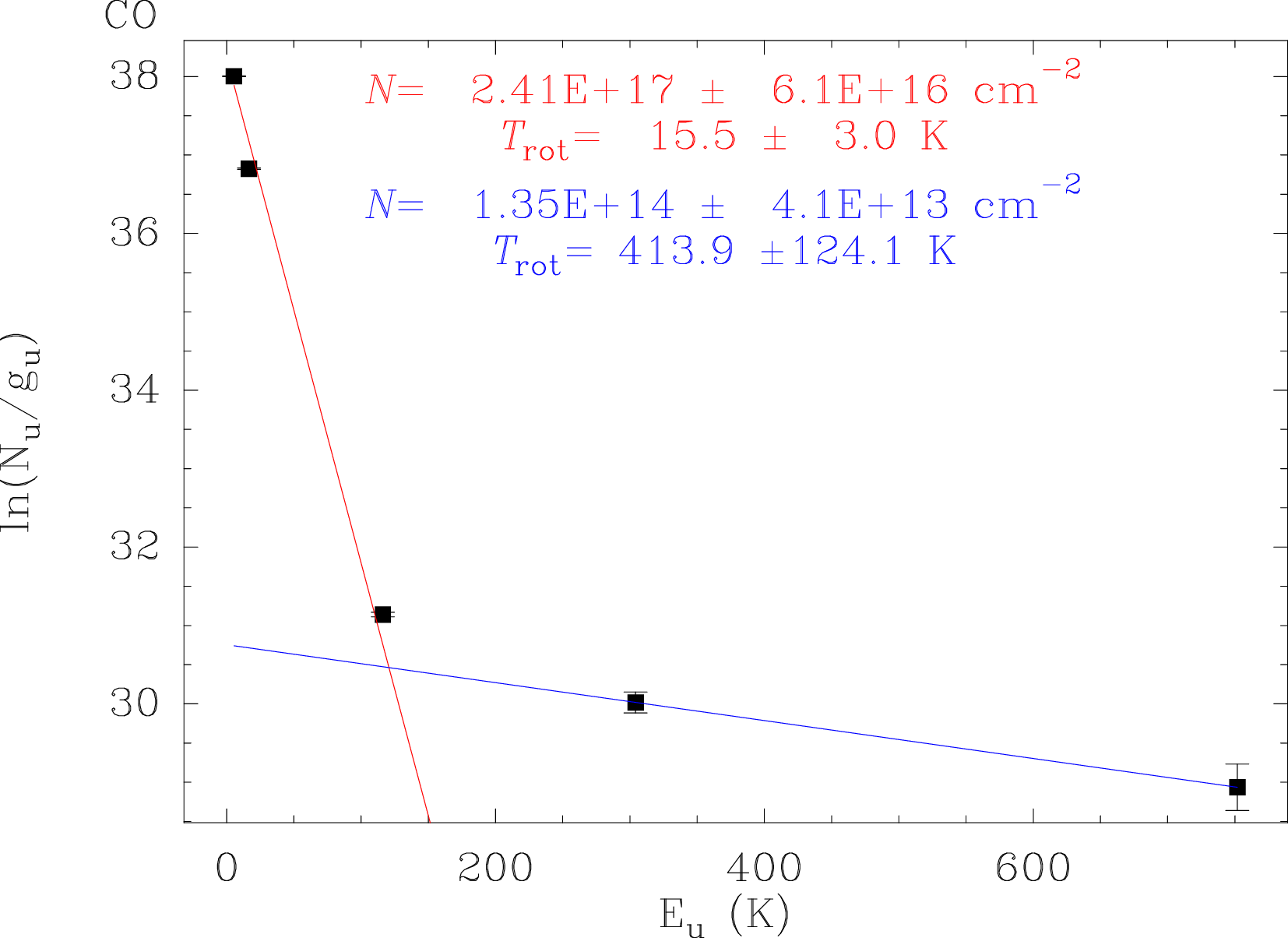}
   \includegraphics[width=8cm]{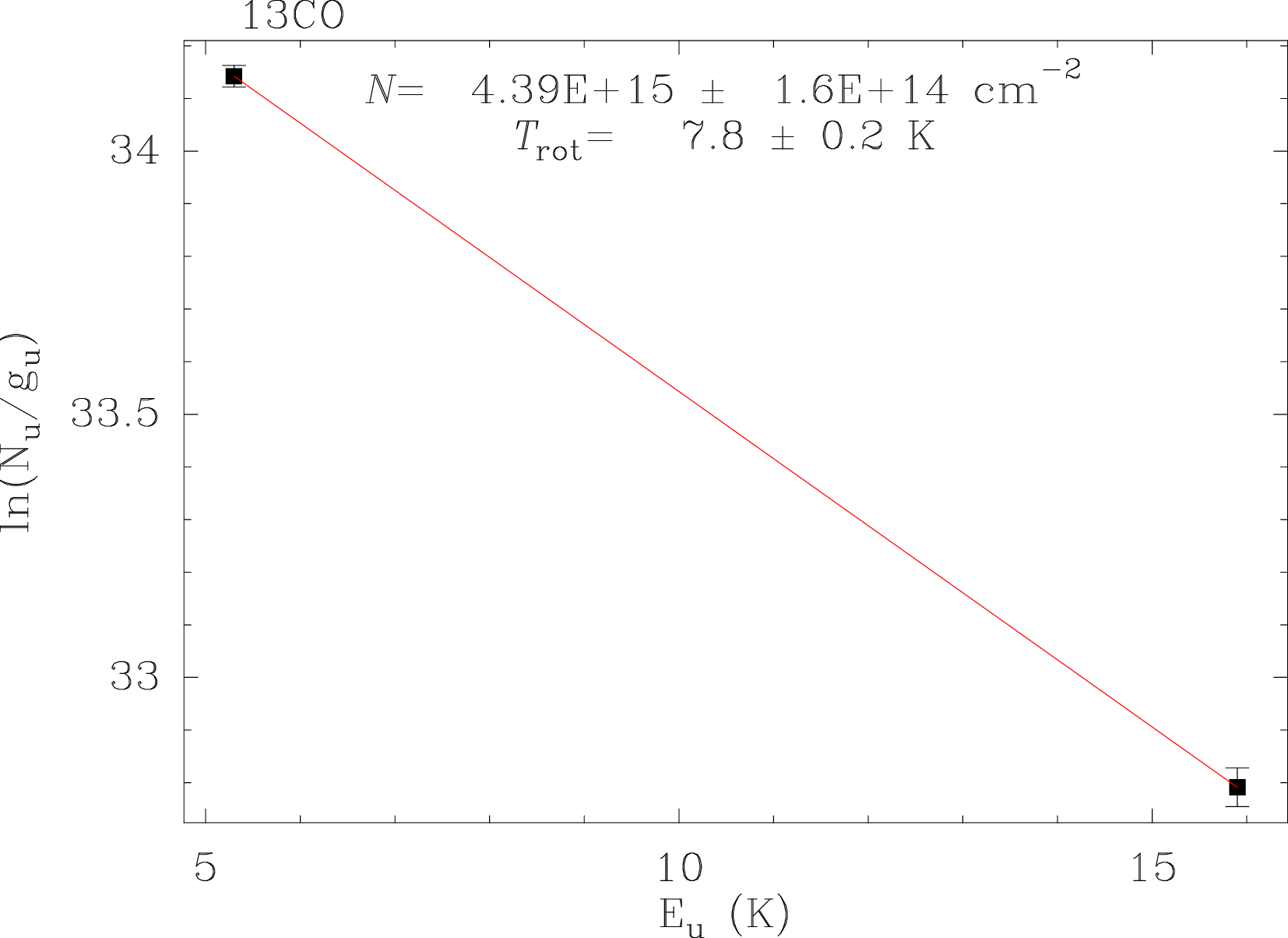}
   \includegraphics[width=8cm]{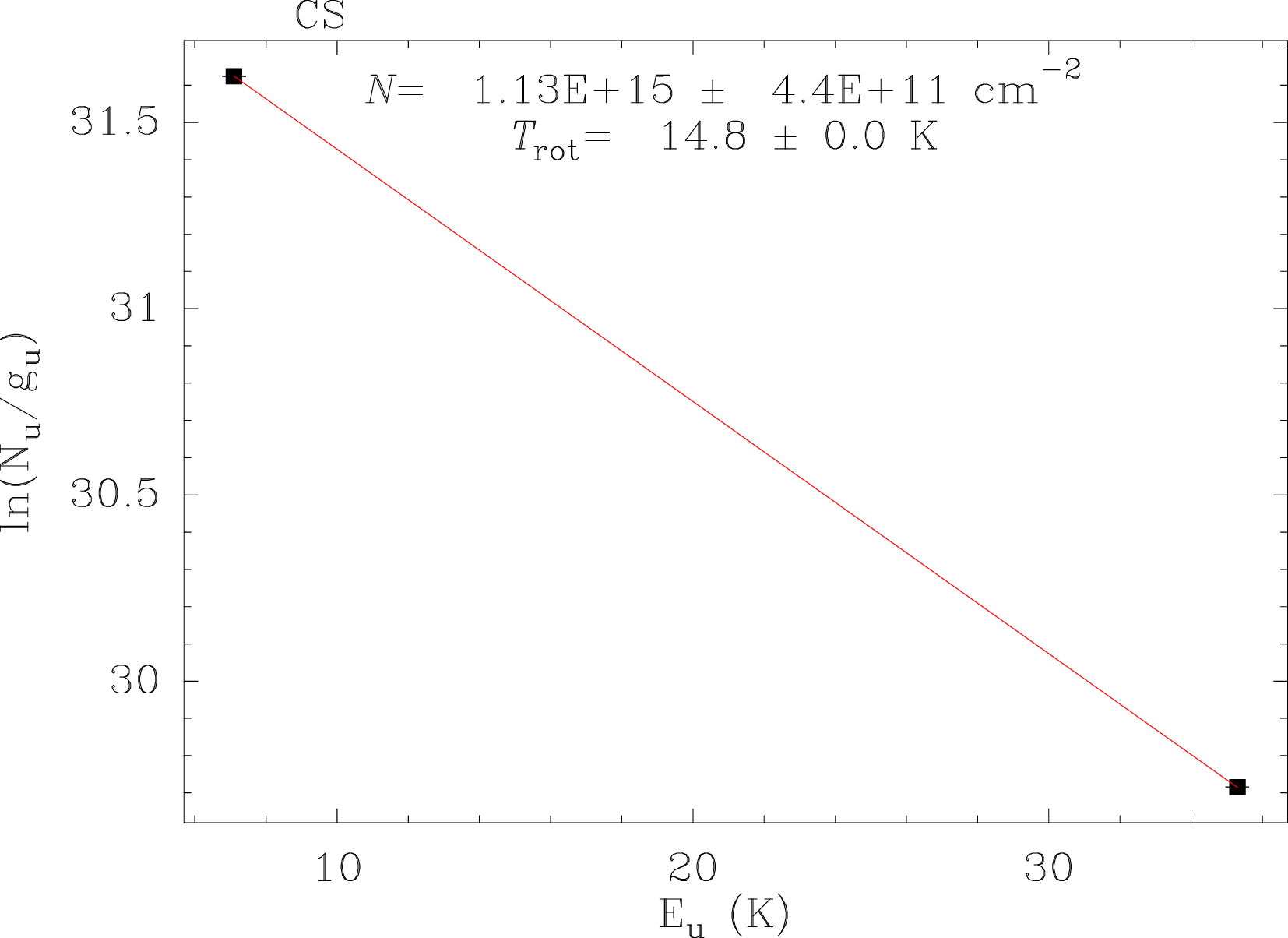}
   \includegraphics[width=8cm]{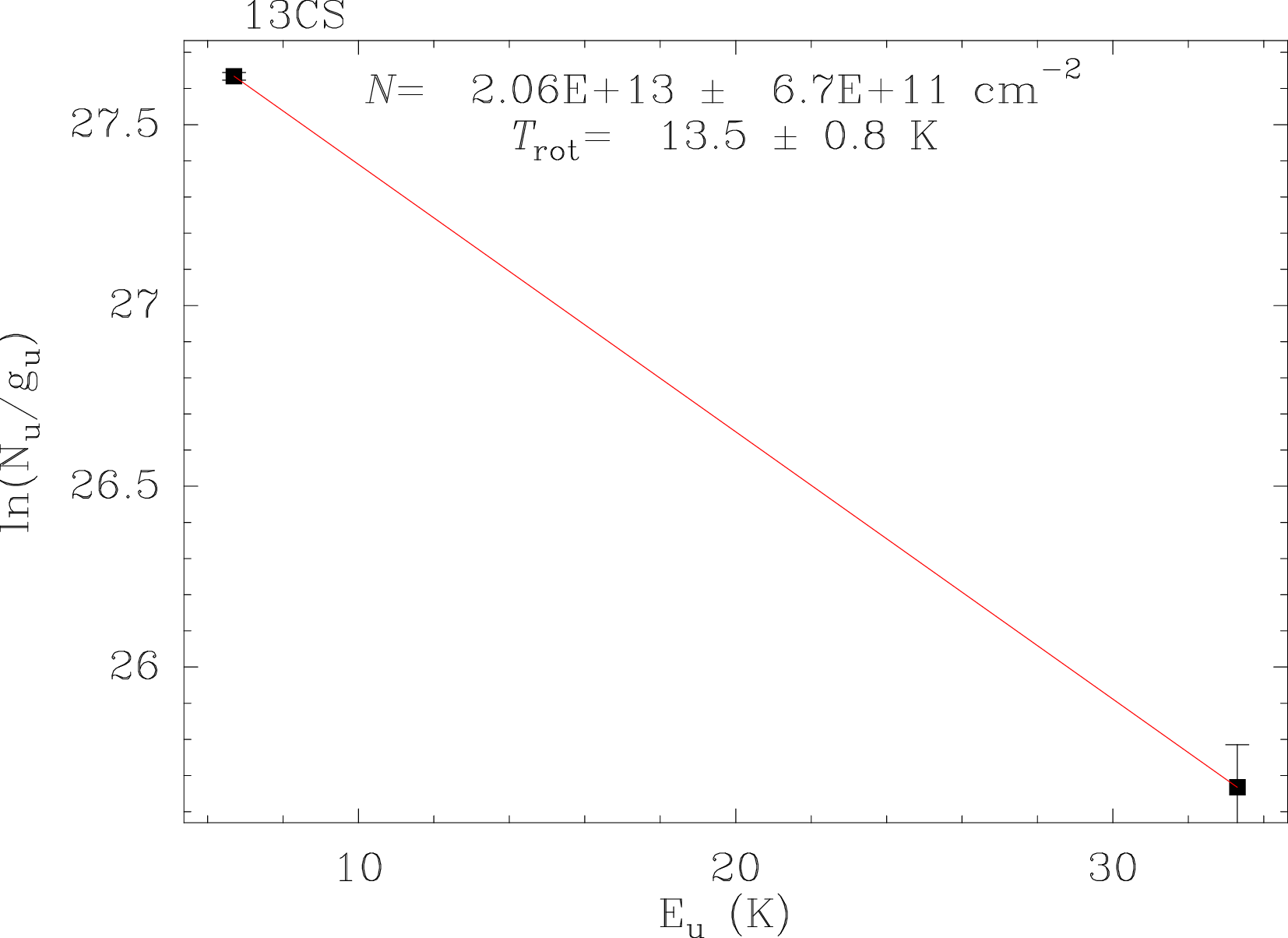}
   \includegraphics[width=8cm]{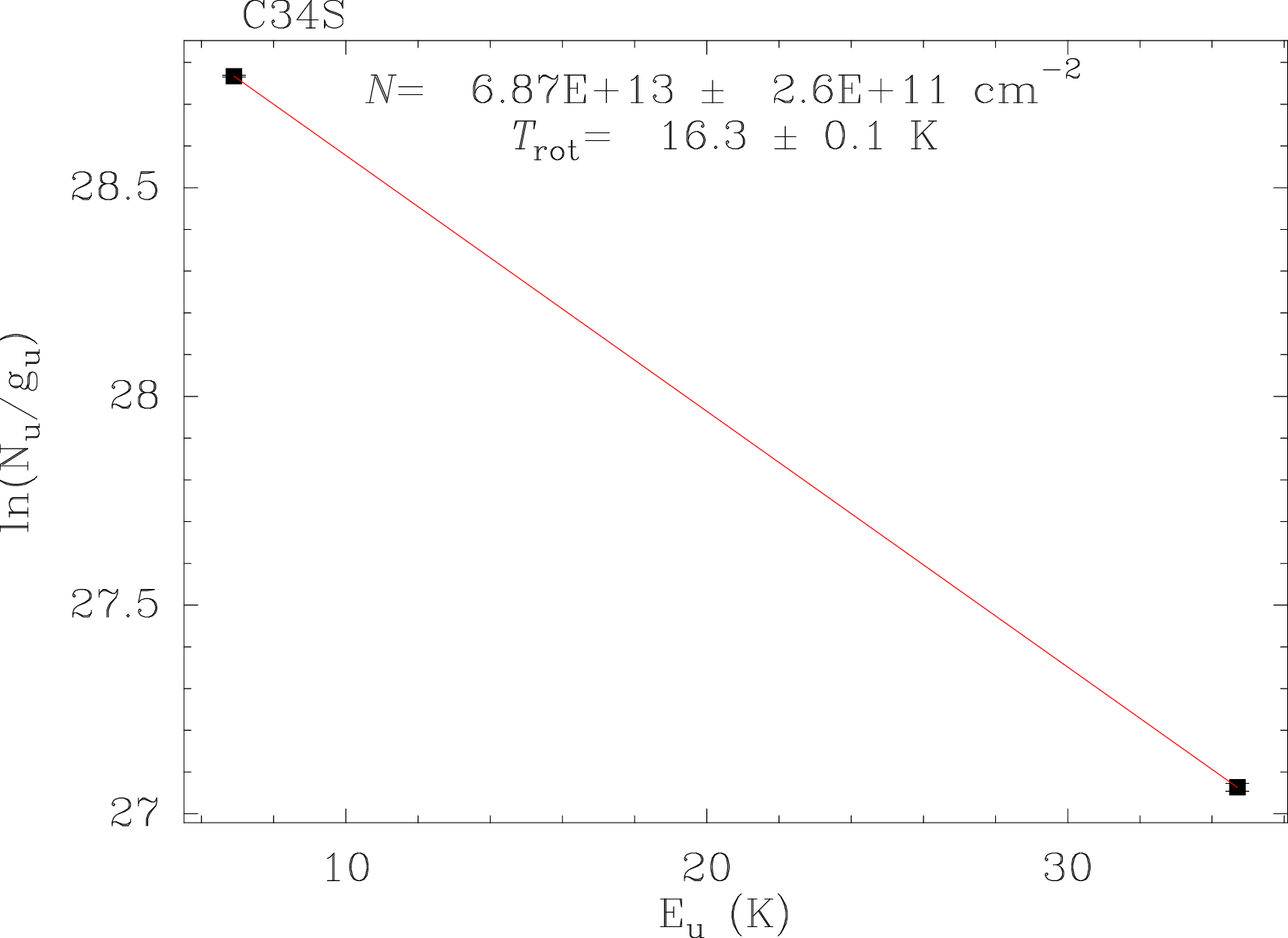}
   \includegraphics[width=8cm]{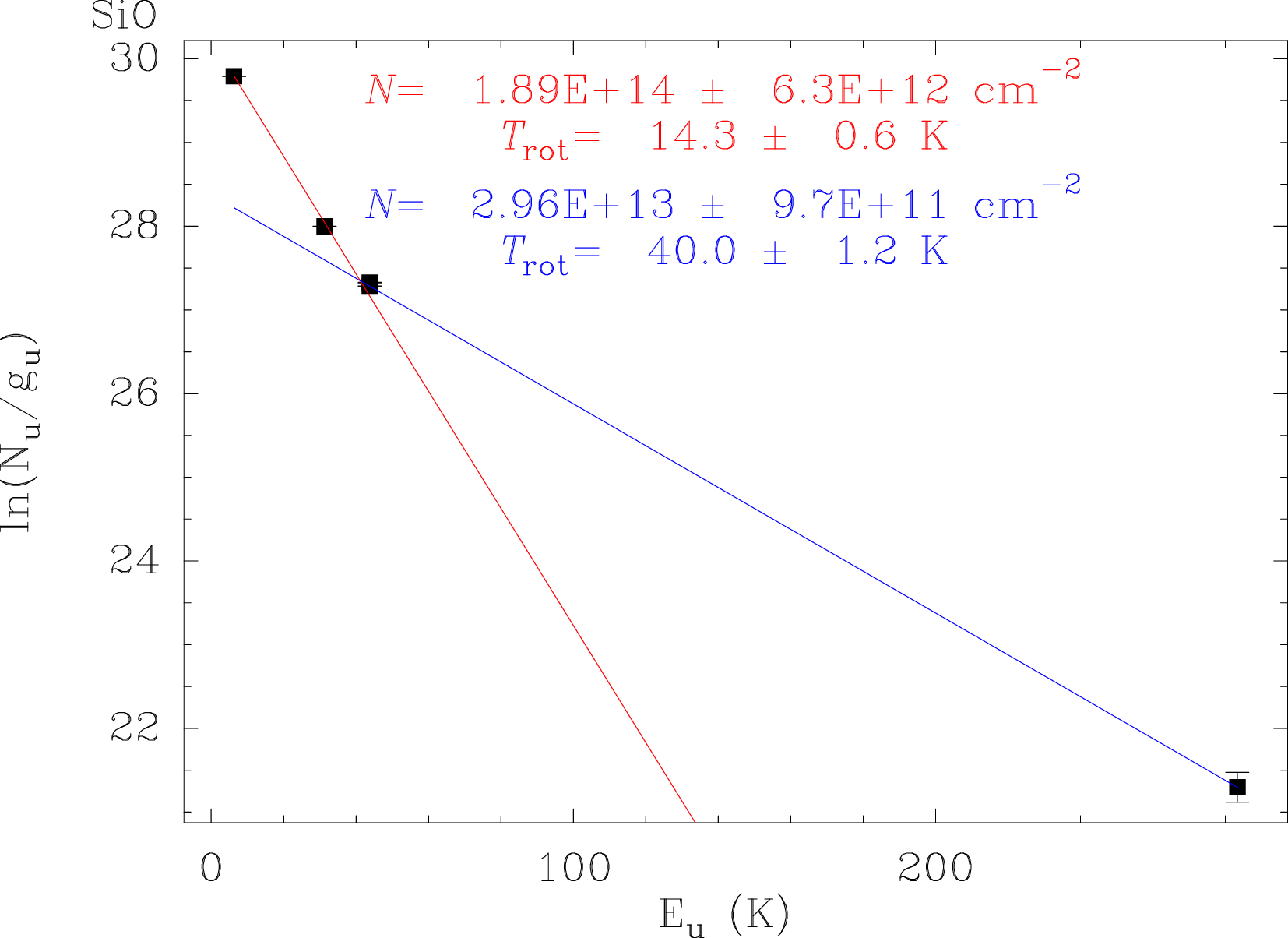}
   \includegraphics[width=8cm]{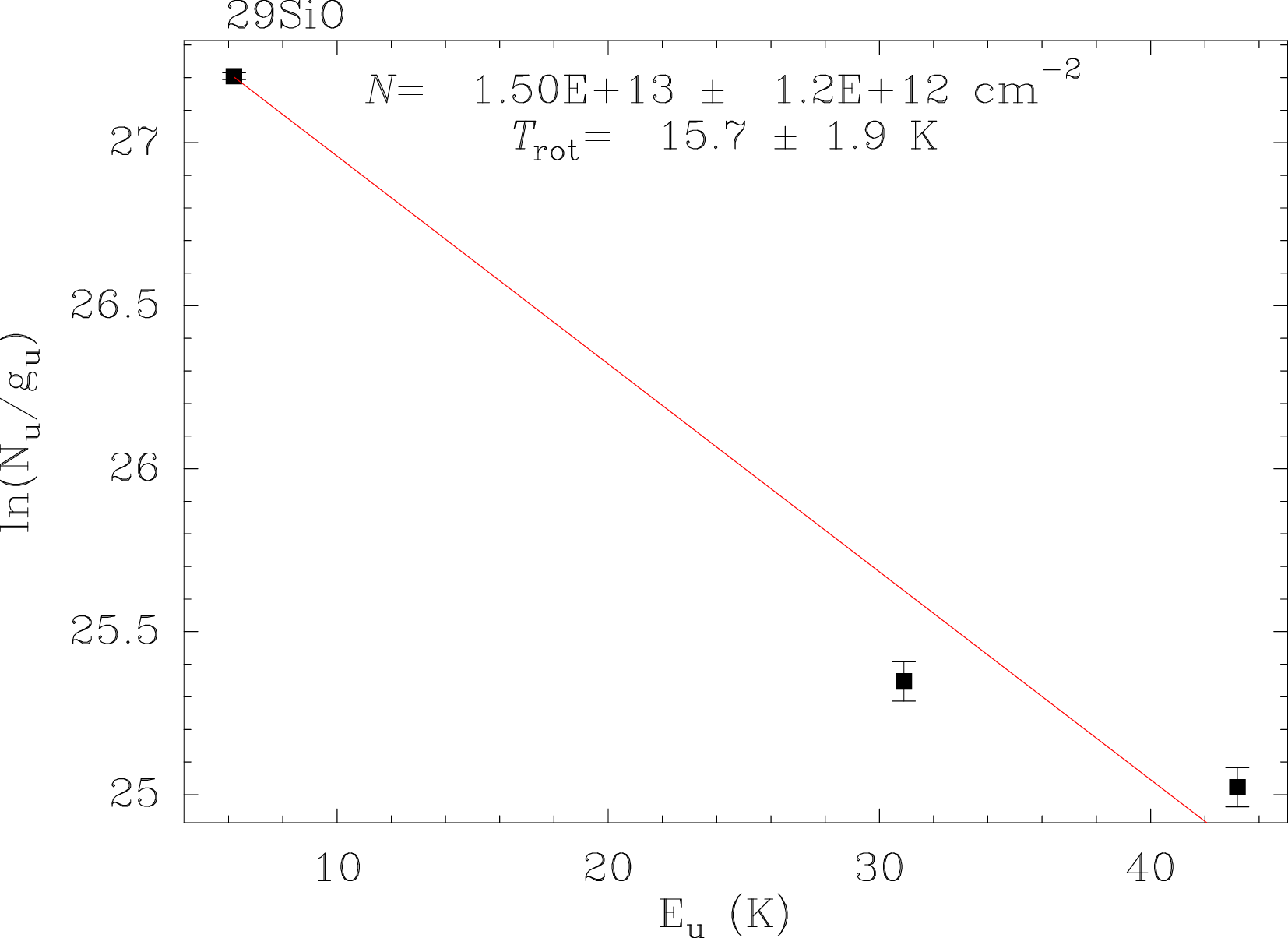}
   \includegraphics[width=8cm]{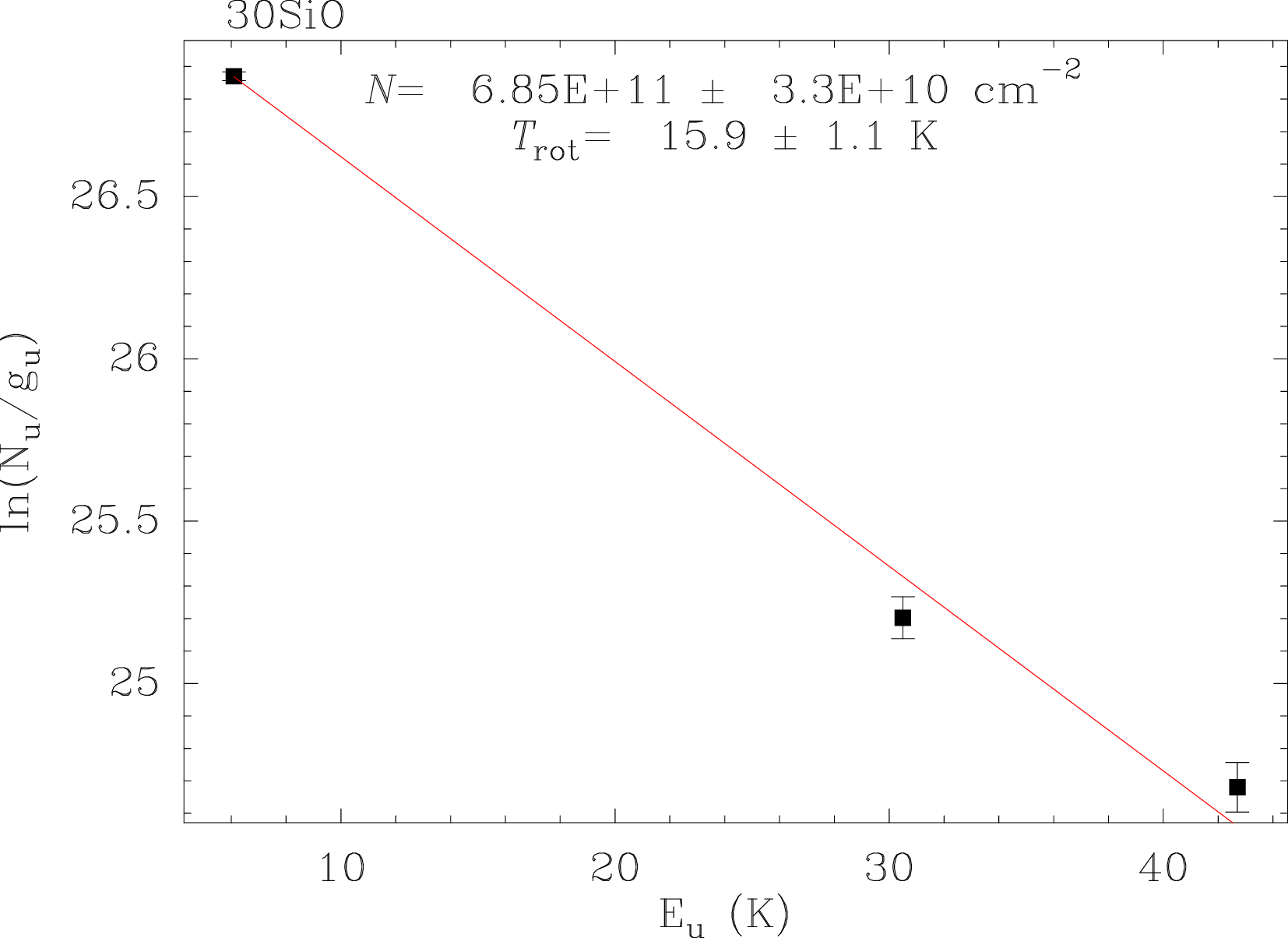}
  
      \caption{Results of the rotational diagram fitting I.}
         \label{rot1}
   \end{figure}

   \begin{figure}
   \centering

   \includegraphics[width=8cm]{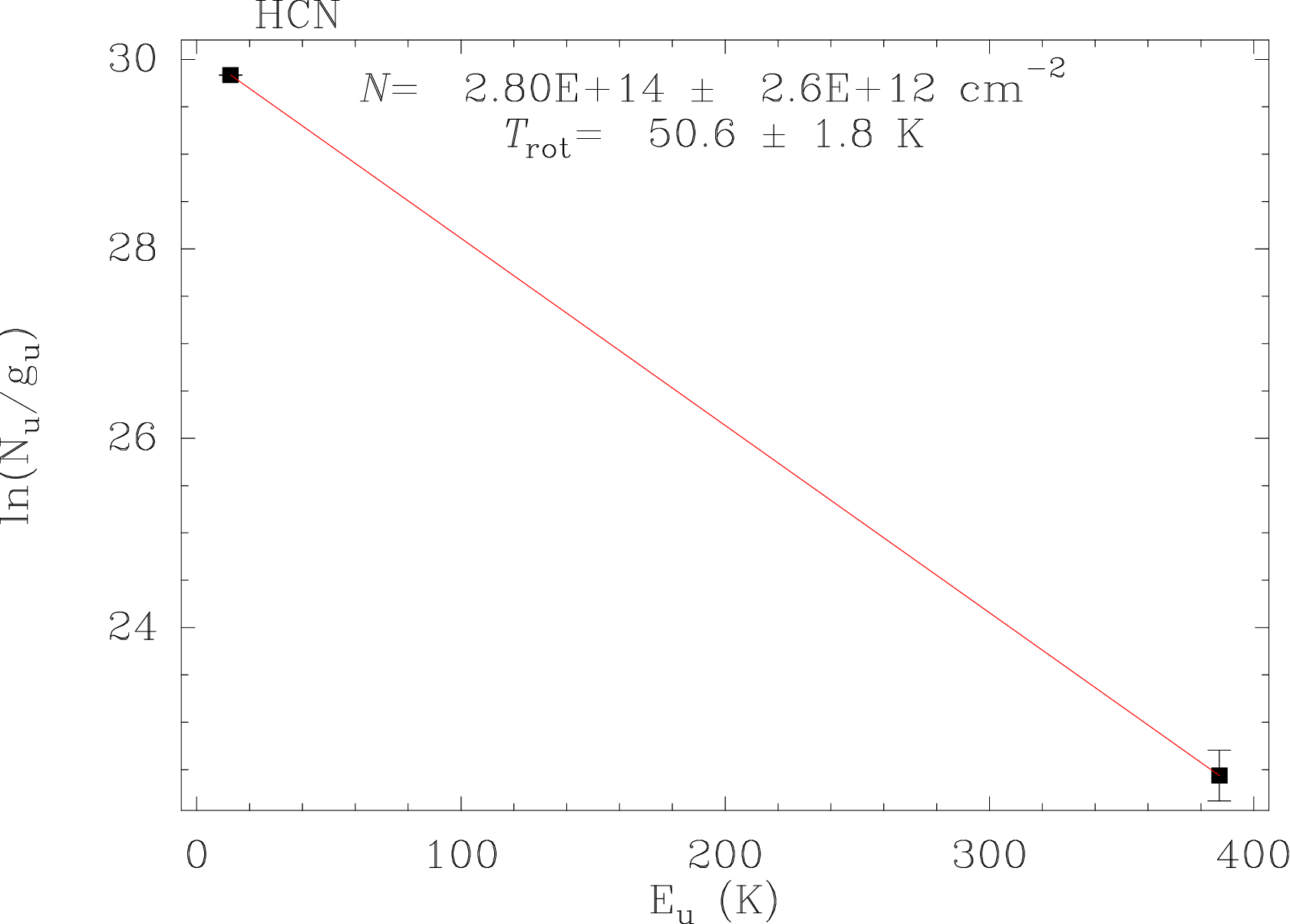}	
   \includegraphics[width=8cm]{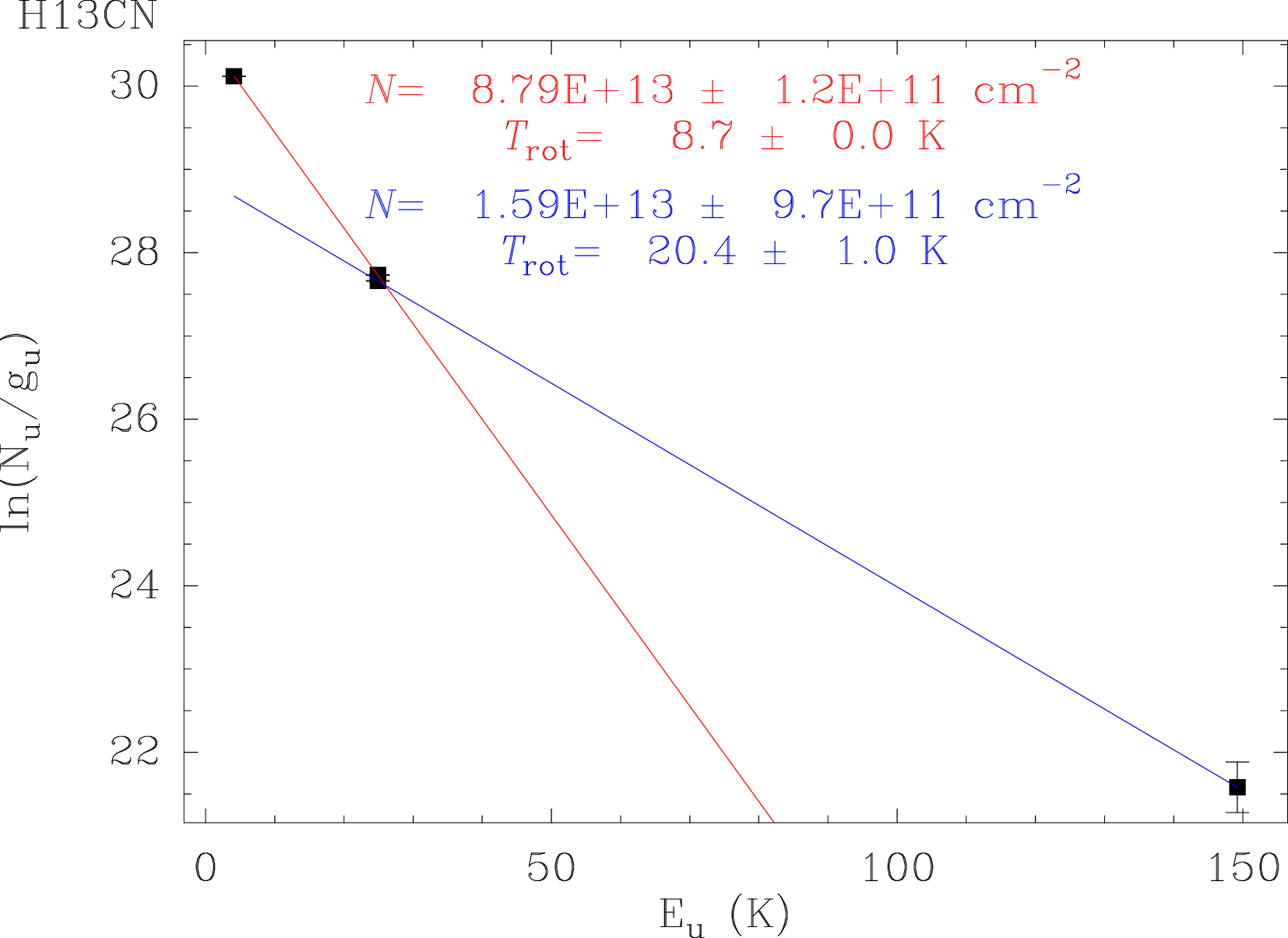}
   \includegraphics[width=8cm]{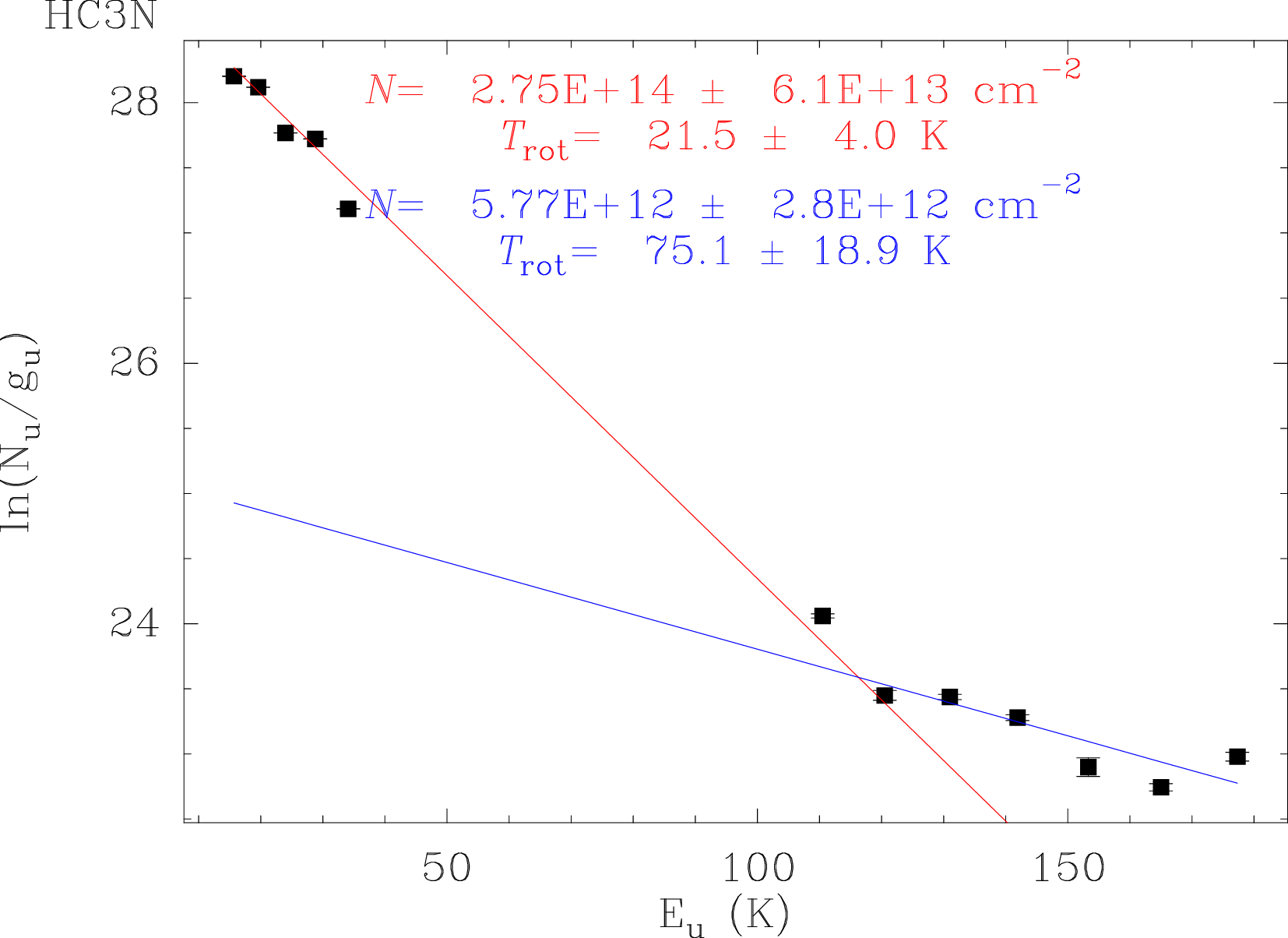}
   \includegraphics[width=8cm]{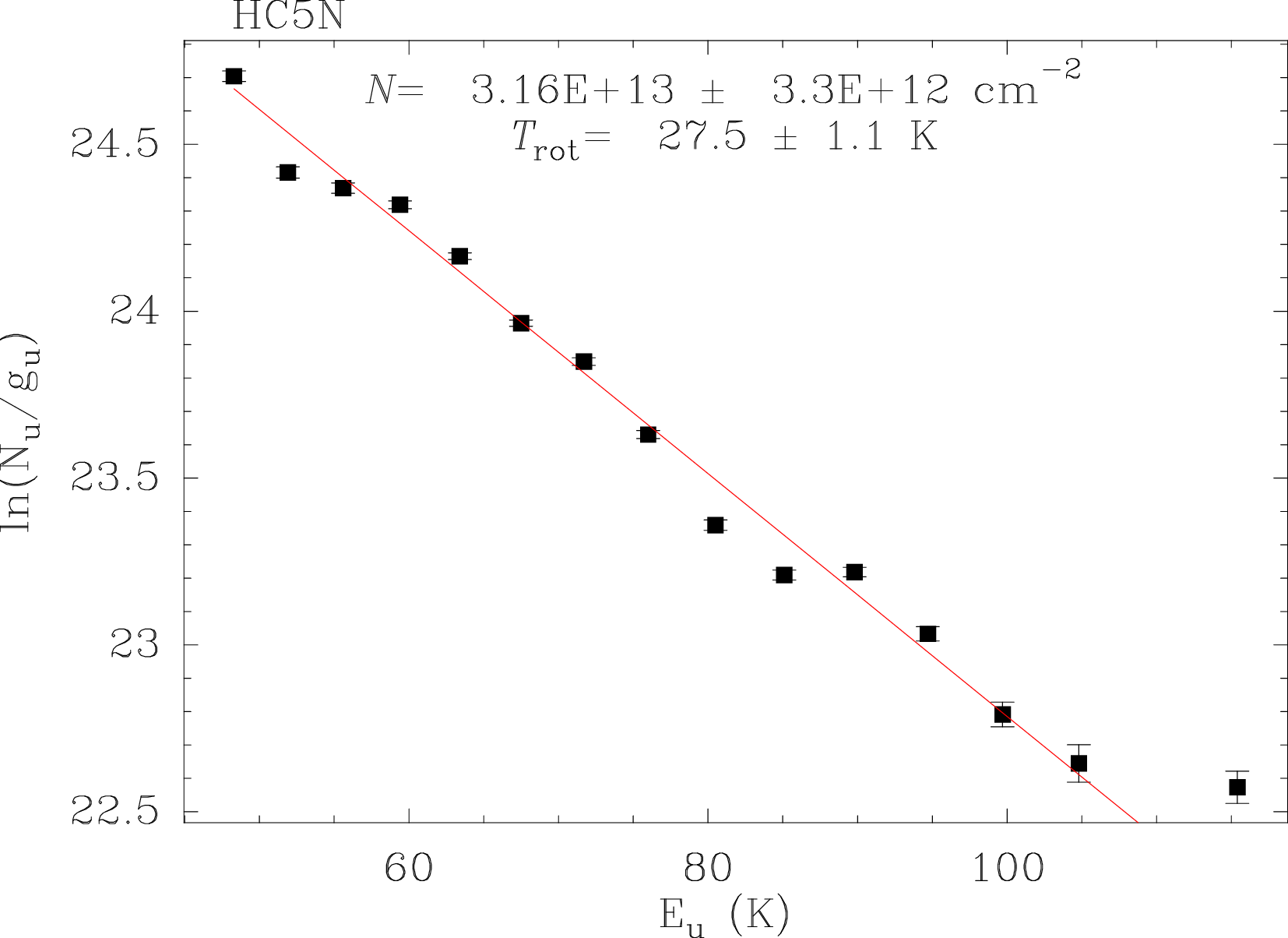}
   \includegraphics[width=8cm]{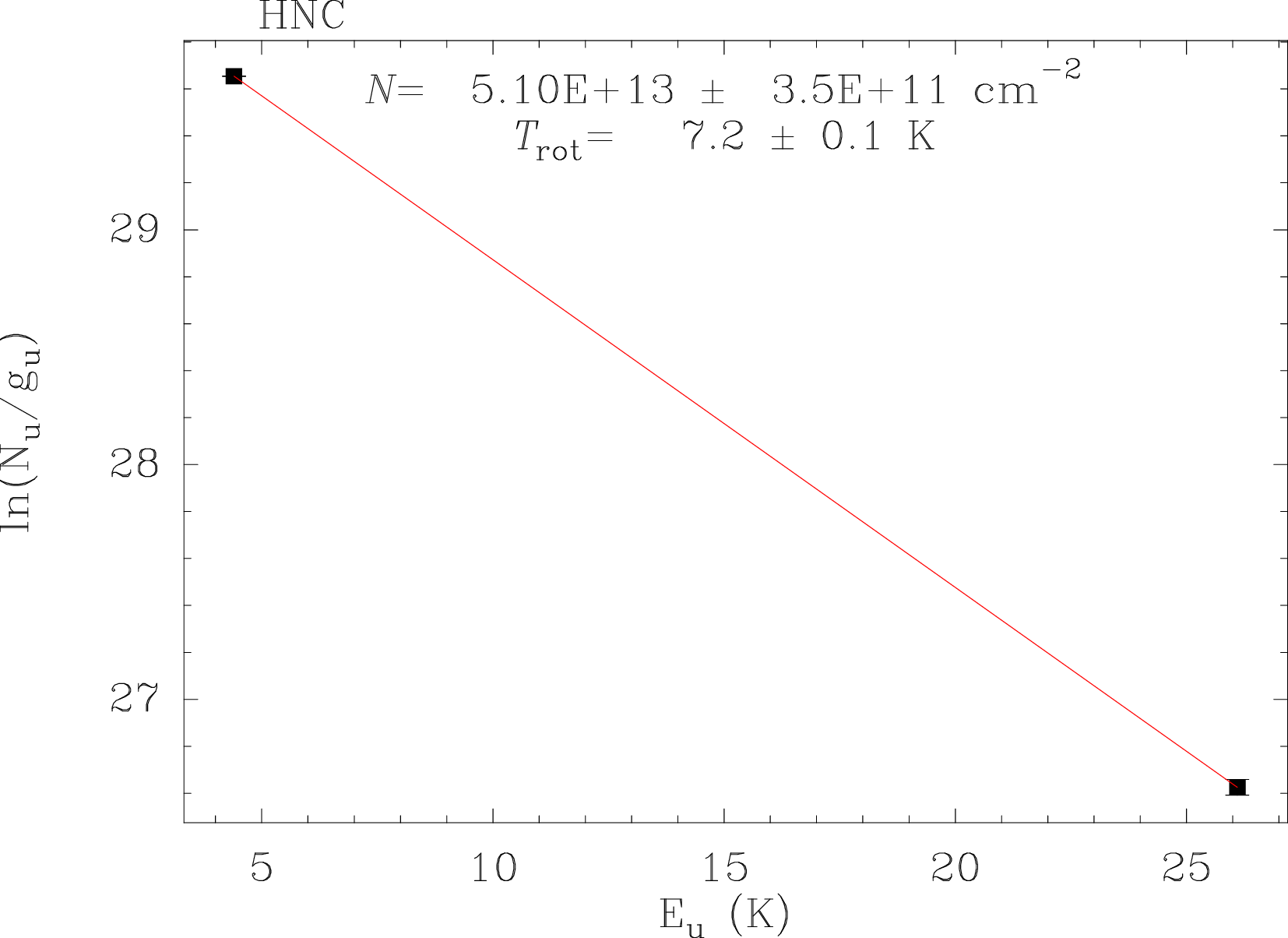}
      \caption{Results of the rotational diagram fitting II.}
         \label{rot2}
   \end{figure}
   
      \begin{figure}
   \centering
   \includegraphics[width=8cm]{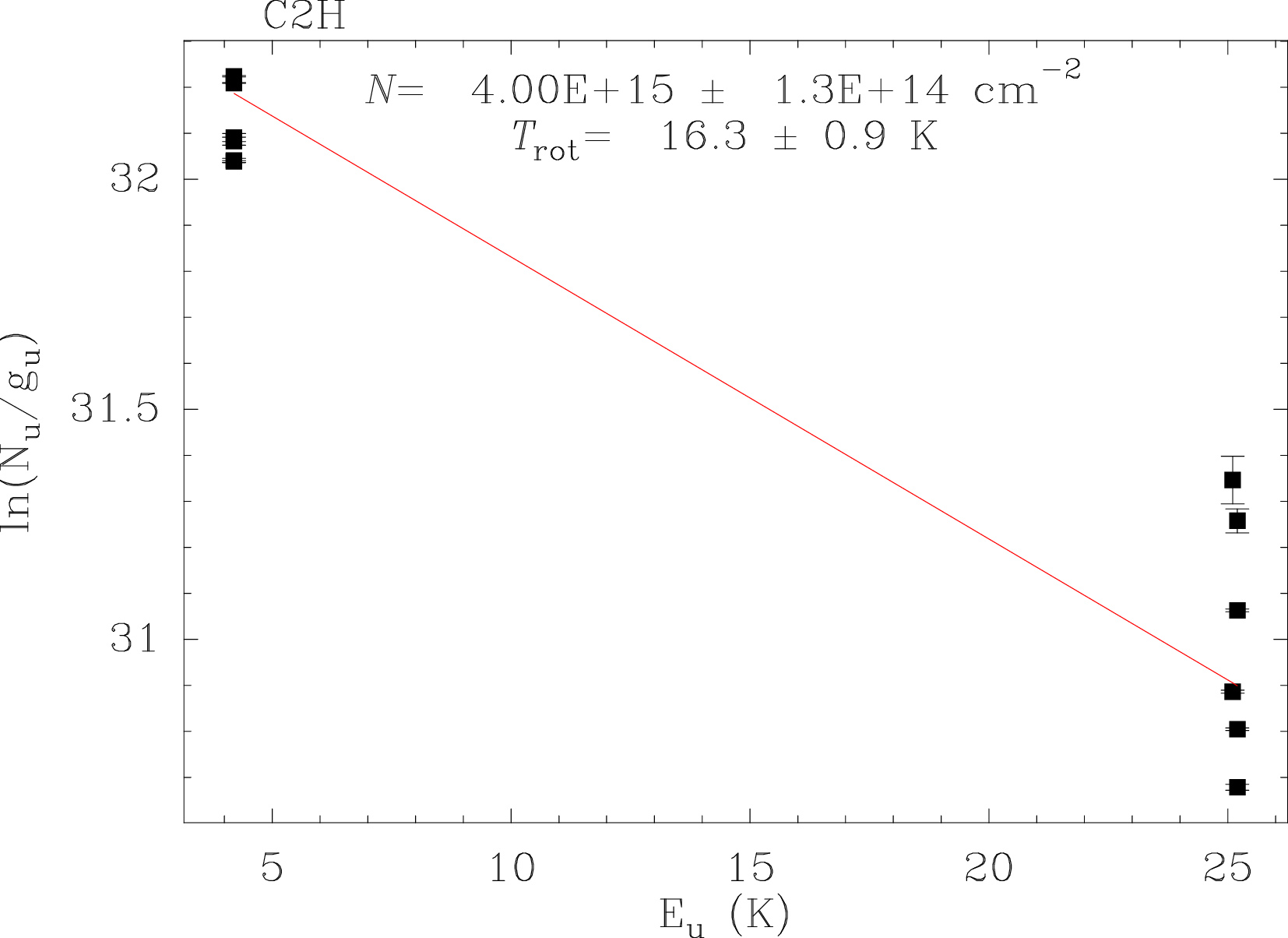}
   \includegraphics[width=8cm]{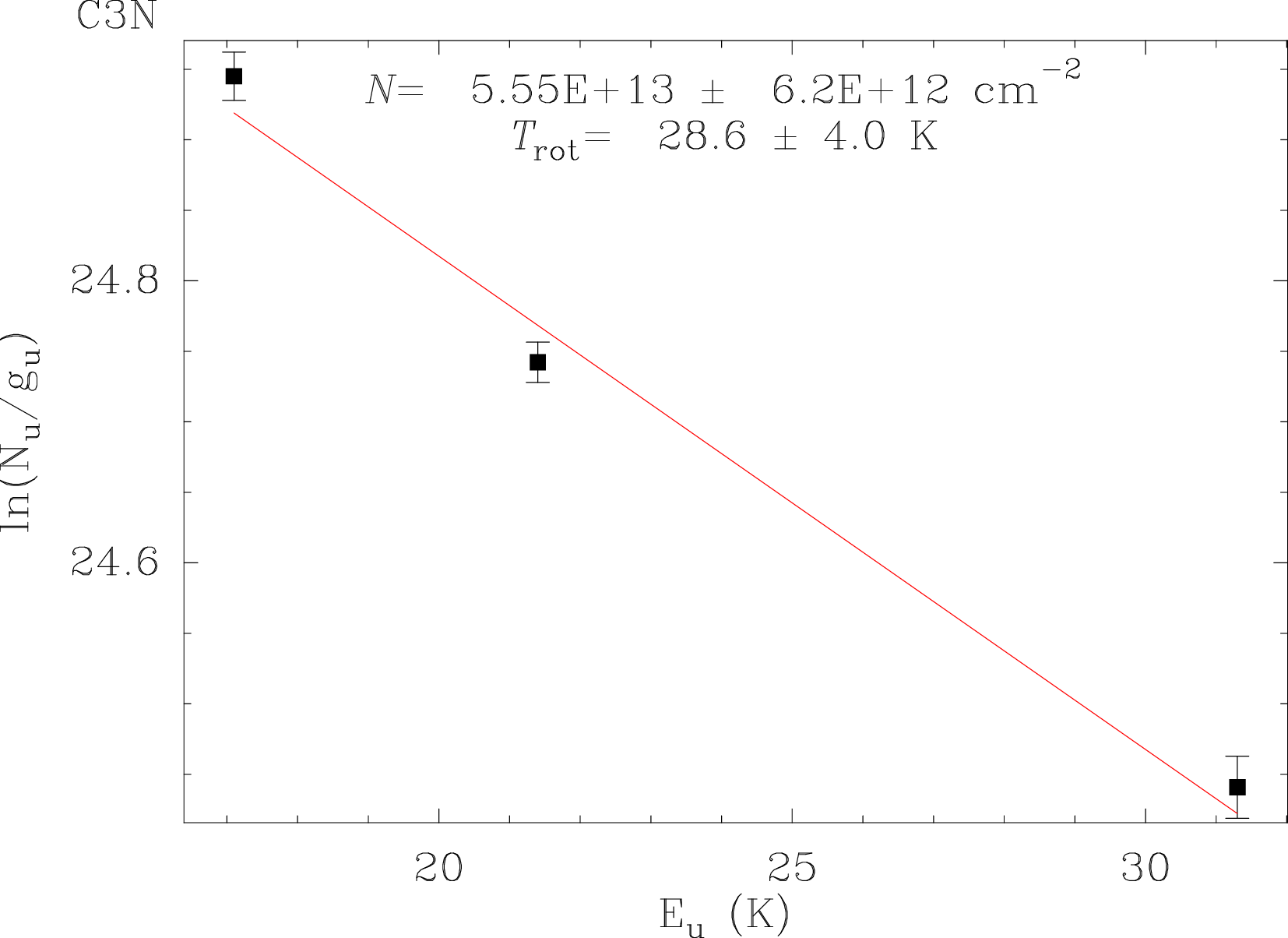}
   \includegraphics[width=8cm]{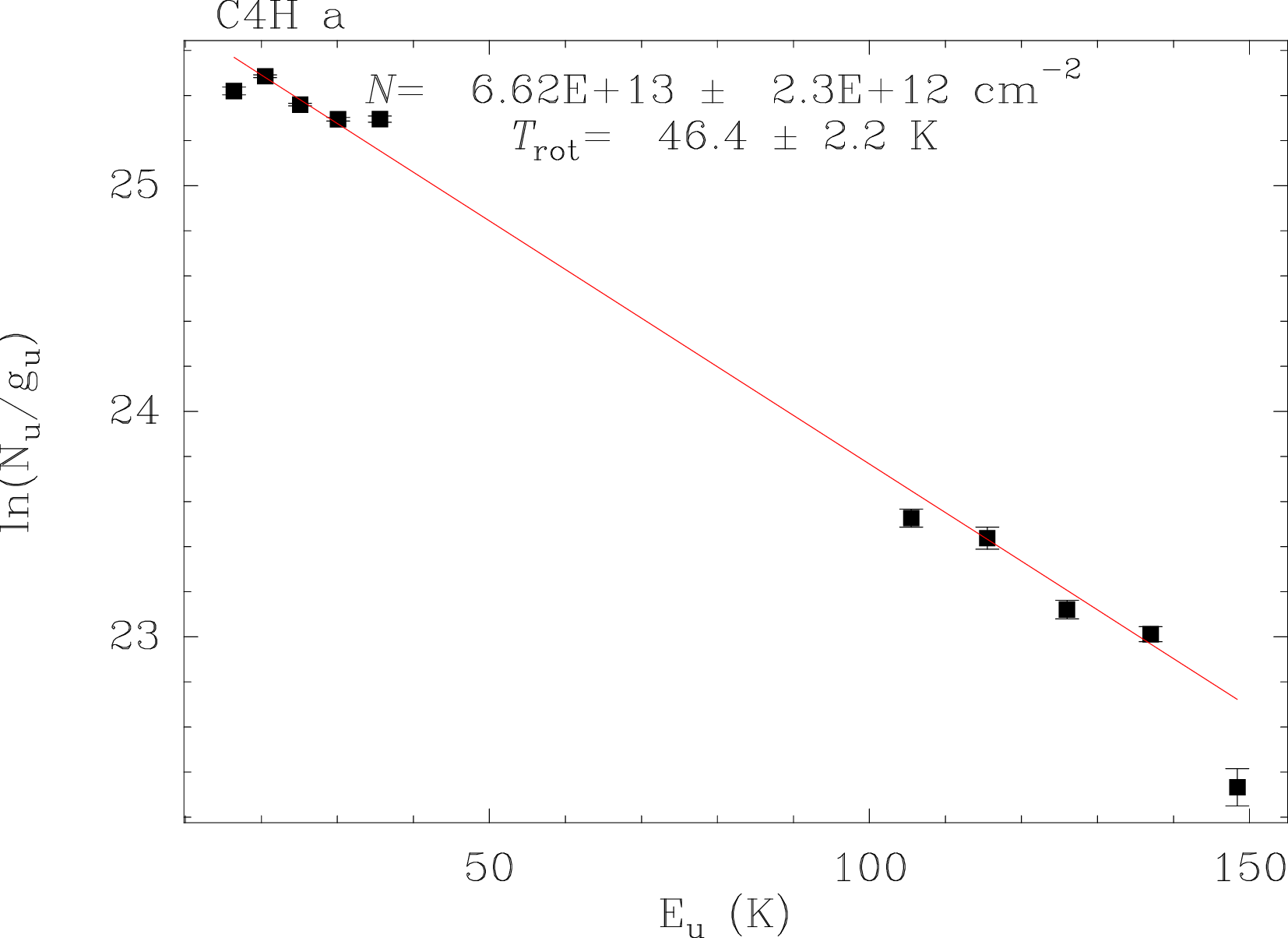}
   \includegraphics[width=8cm]{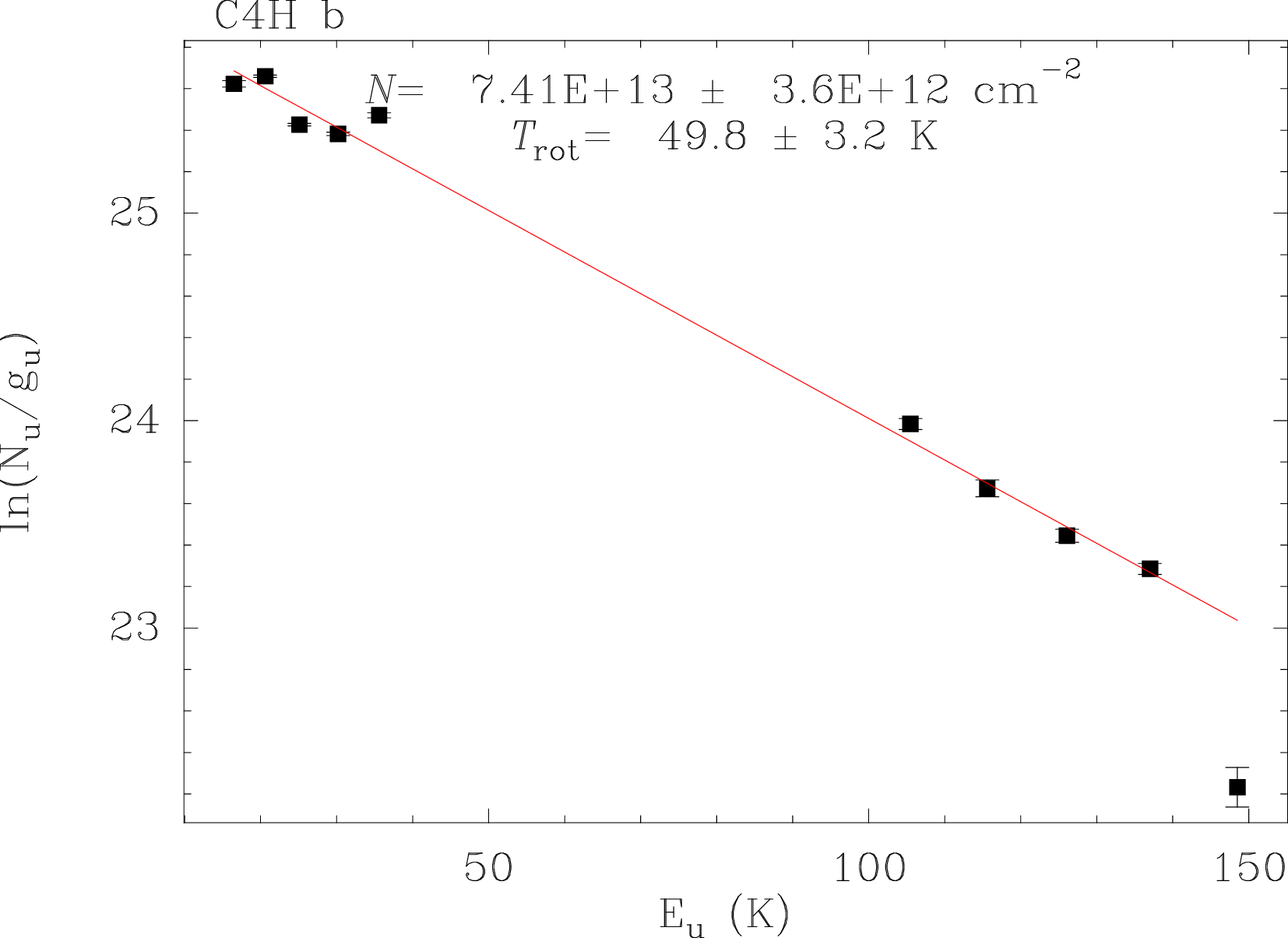}
   \includegraphics[width=8cm]{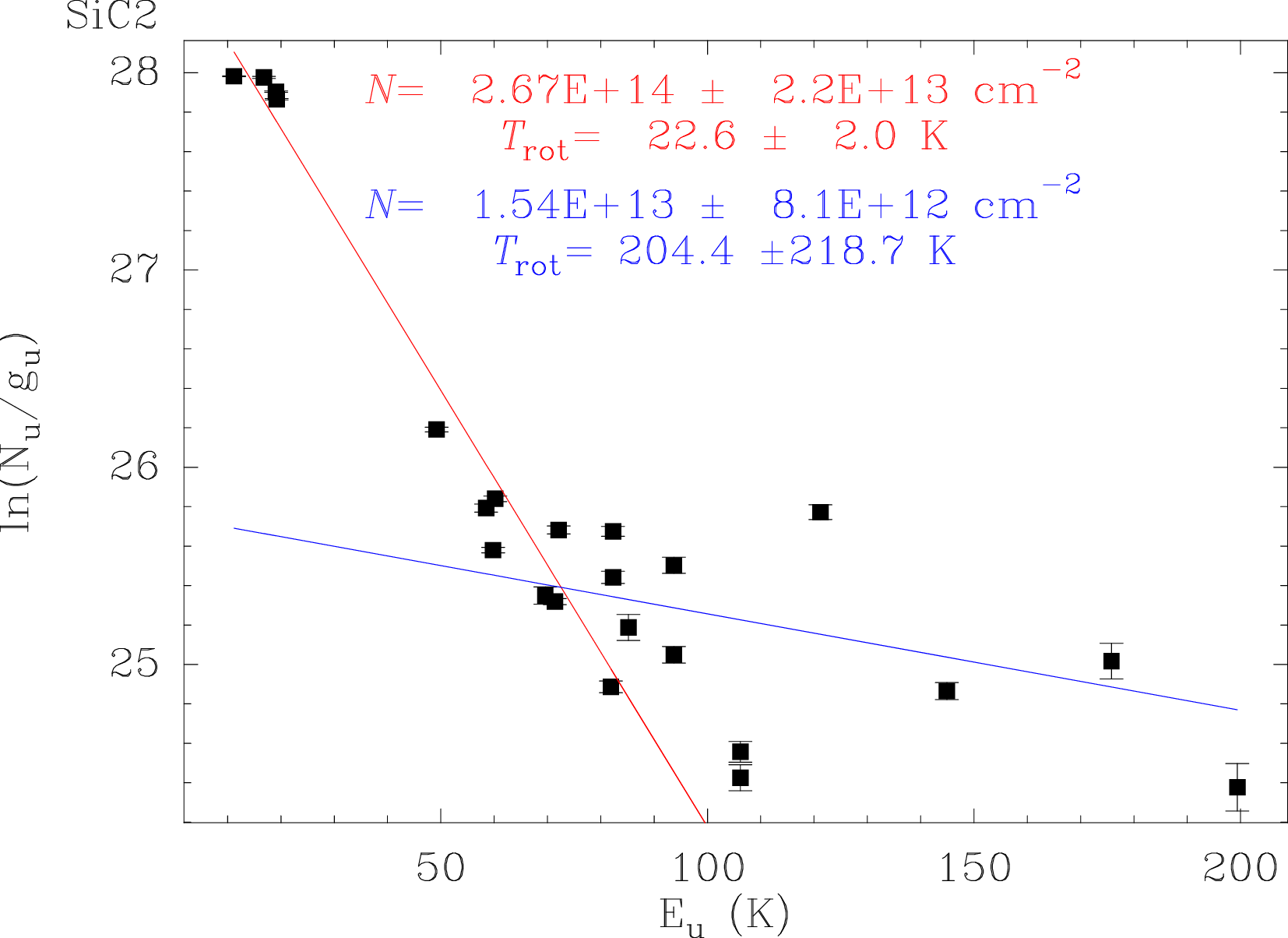}
      \caption{Results of the rotational diagram fitting III.}
         \label{rot3}
   \end{figure}

\section{Molecular line maps}
\label{sect:maps}

         \begin{figure}
   \centering
   \includegraphics[width=16cm]{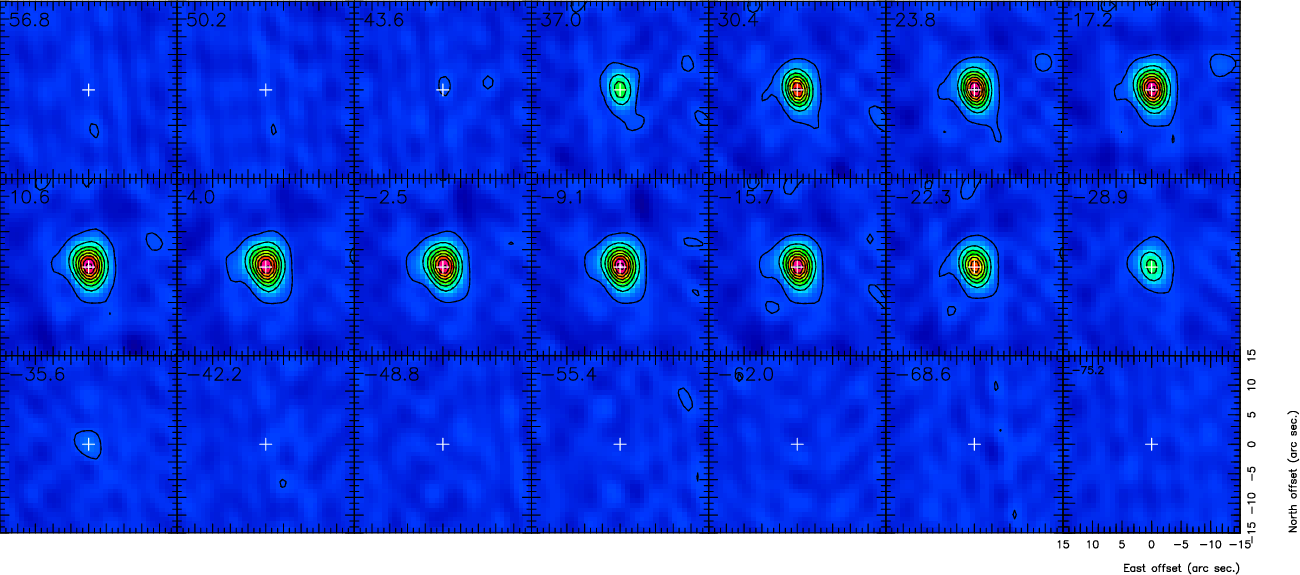}
      \caption{Molecular channel map of SiO 2--1. HPBW = $4\secp35\times2\secp84$. Level step is 20$\sigma$, being $\sigma = 2.39$\,mJy/beam and the first contour at 3$\sigma$. }
         \label{SiO}
   \end{figure}

         \begin{figure}
   \centering
   \includegraphics[width=16cm]{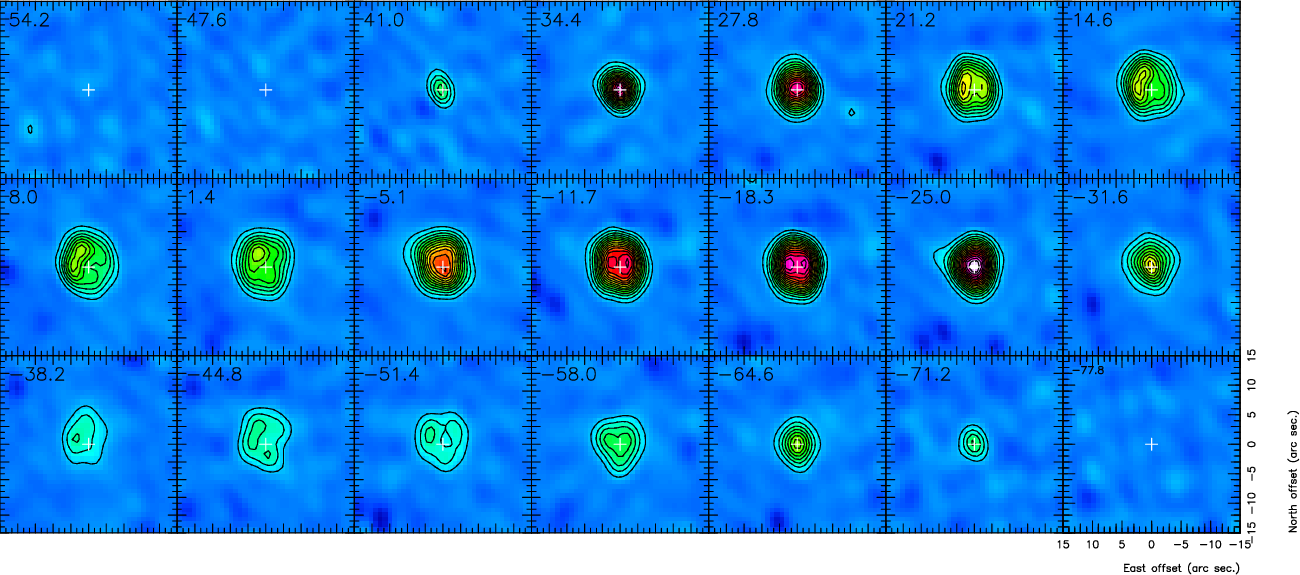}
      \caption{Molecular channel map of C$_2$H. HPBW = $4\secp32\times2\secp82$. Level step is 3$\sigma$, being $\sigma = 1.85$\,mJy/beam and the first contour at 3$\sigma$.               }
         \label{}
   \end{figure}
         \begin{figure}
   \centering
   \includegraphics[width=16cm]{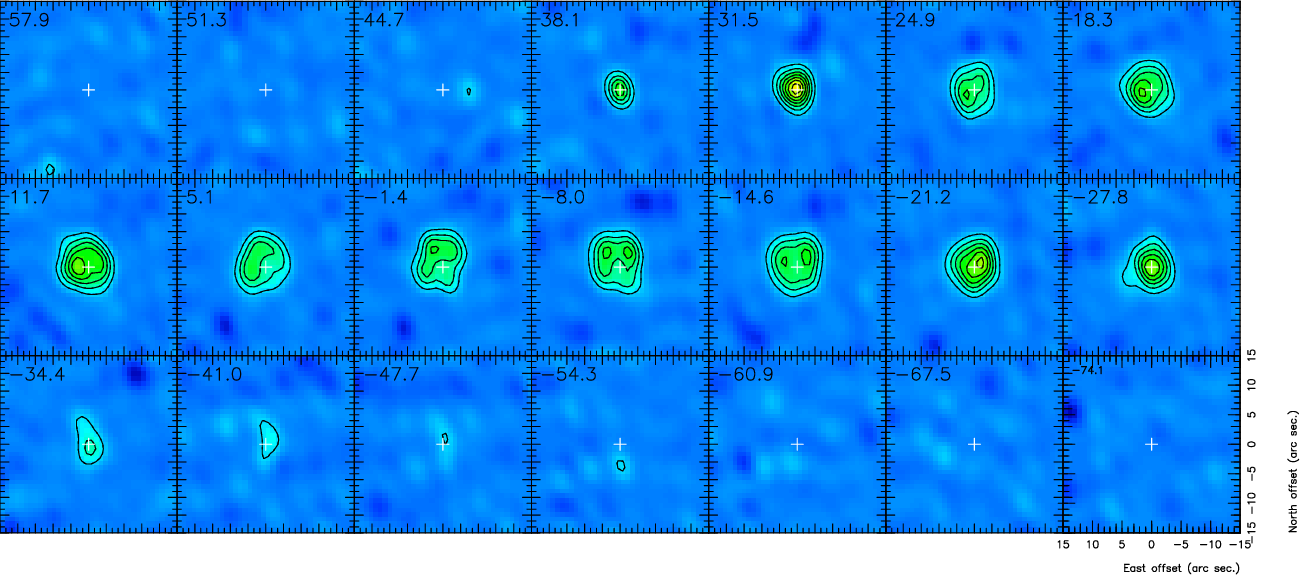}
      \caption{Molecular channel map of C$_2$H. HPBW = $4\secp31\times2\secp82$. Level step is 3$\sigma$, being $\sigma = 2.22$\,mJy/beam and the first contour at 3$\sigma$. }
         \label{}
   \end{figure}
         \begin{figure}
   \centering
   \includegraphics[width=16cm]{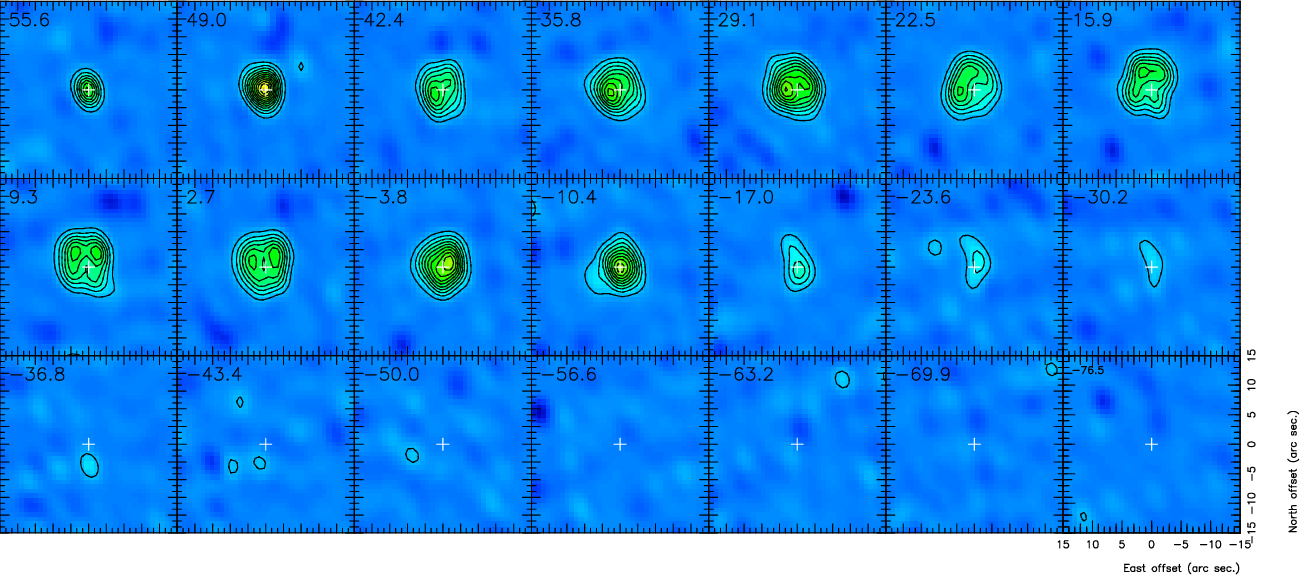}
      \caption{Molecular channel map of C$_2$H. HPBW = $4\secp36\times2\secp83$. Level step is 3$\sigma$, being $\sigma = 1.43$\,mJy/beam and the first contour at 3$\sigma$. }
         \label{}
   \end{figure}
         \begin{figure}
   \centering
   \includegraphics[width=16cm]{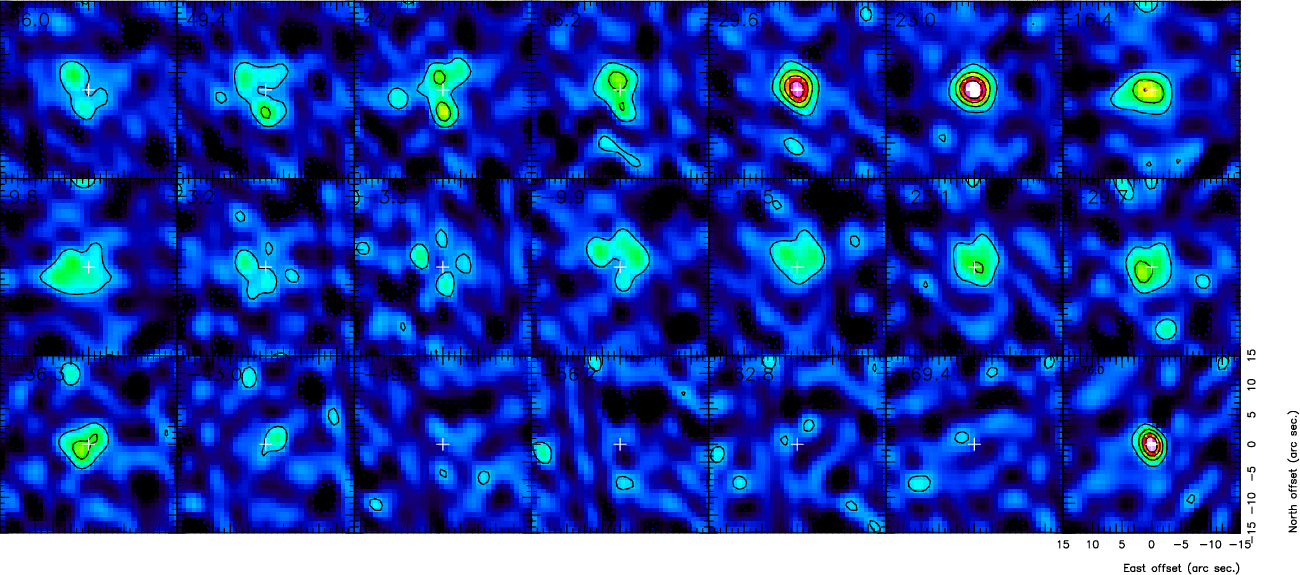}
      \caption{Molecular channel map of C$_3$N. HPBW = $4\secp25\times2\secp78$. Level step is 3$\sigma$, being $\sigma = 1.48$\,mJy/beam and the first contour at 3$\sigma$. }
         \label{}
   \end{figure}
         \begin{figure}
   \centering
   \includegraphics[width=16cm]{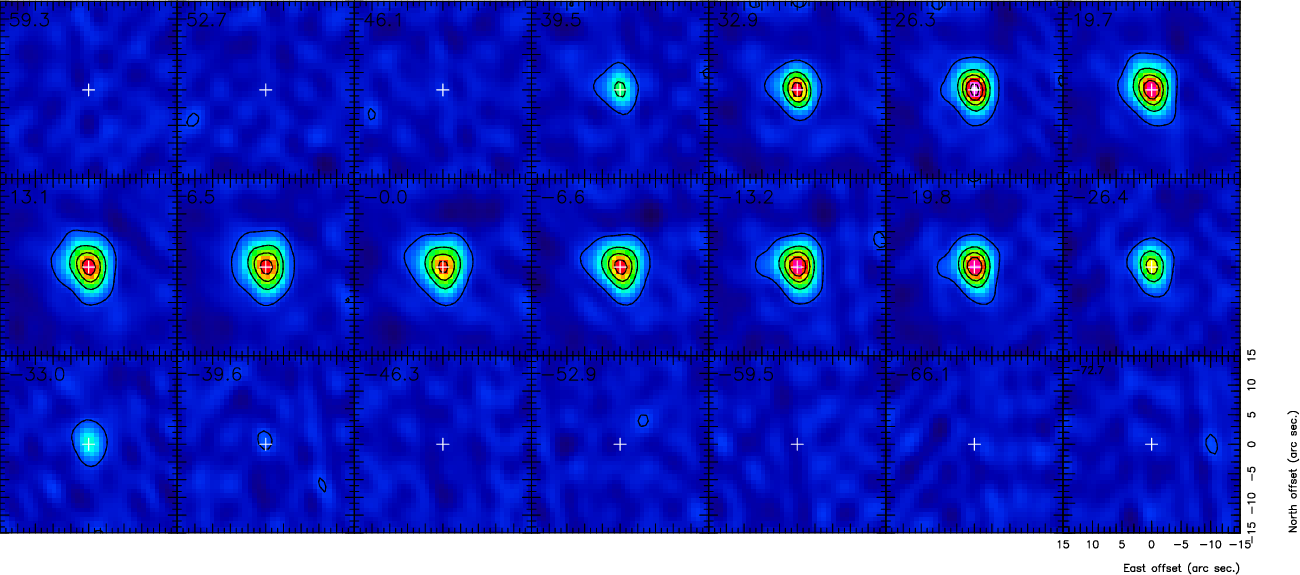}
      \caption{Molecular channel map of H$^{13}$CN. HPBW = $4\secp37\times2\secp86$. Level step is 20$\sigma$, being $\sigma = 2.13$\,mJy/beam and the first contour at 3$\sigma$.  }
         \label{}
   \end{figure}

         \begin{figure}
   \centering
   \includegraphics[width=16cm]{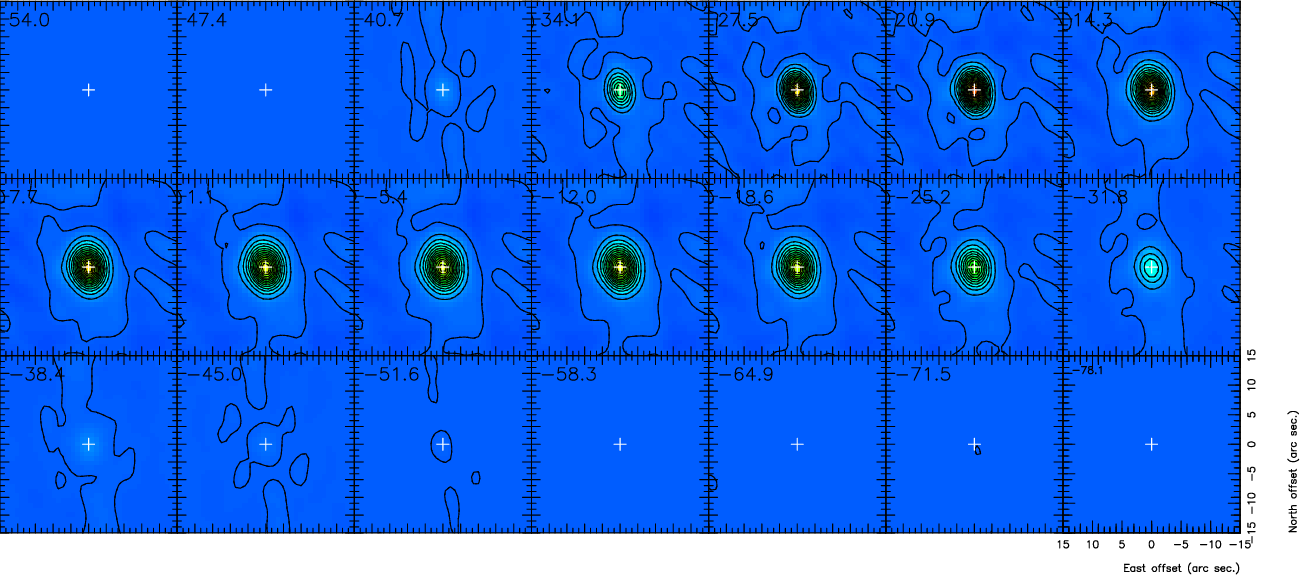}
      \caption{Molecular channel map of HCN. HPBW = $4\secp27\times2\secp79$. Level step is 100$\sigma$, being $\sigma = 2.22$\,mJy/beam and the first contour at 3$\sigma$.              }
         \label{}
   \end{figure}
         \begin{figure}
   \centering
   \includegraphics[width=16cm]{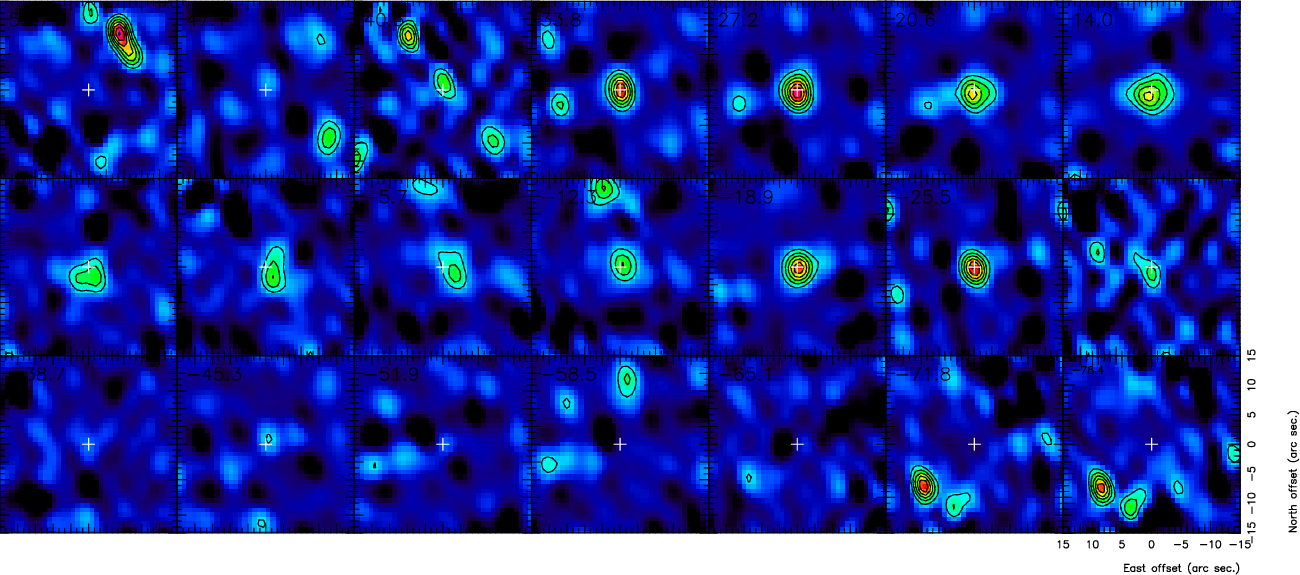}
      \caption{Molecular channel map of HC$_5$N. HPBW = $4\secp31\times2\secp82$. Level step is 2$\sigma$, being $\sigma = 1.41$\,mJy/beam and the first contour at 3$\sigma$.               }
         \label{HC5N}
   \end{figure}

\section{Spectral line survey}
\label{surveyimages}

\begin{figure*}[h!]
\centering
\includegraphics[angle=0,width=12cm]{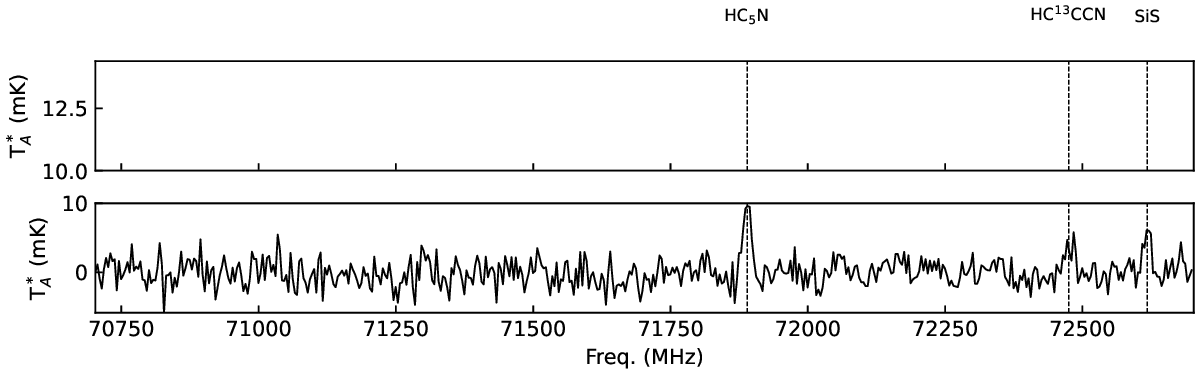}
\includegraphics[angle=0,width=12cm]{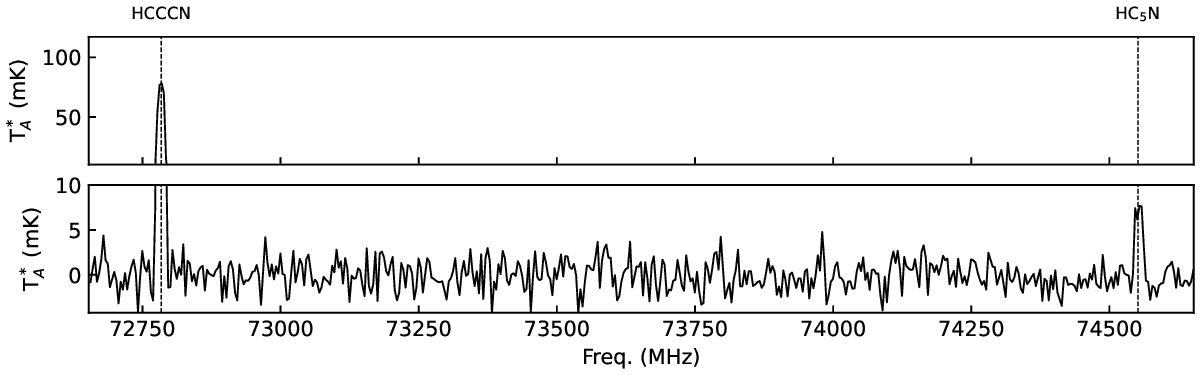}
\includegraphics[angle=0,width=12cm]{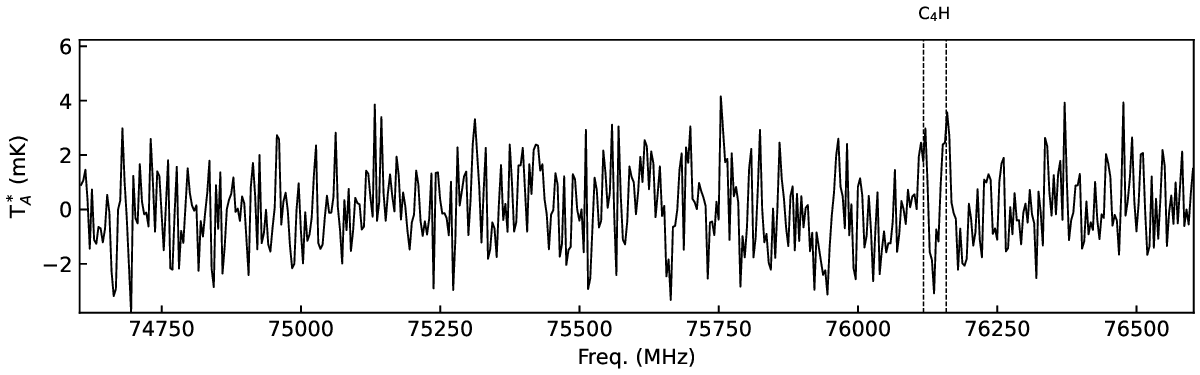}
\includegraphics[angle=0,width=12cm]{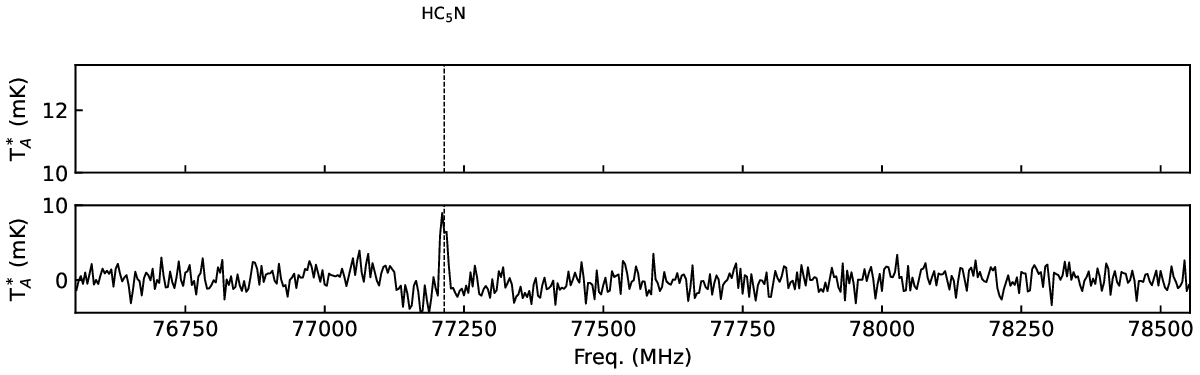}
\caption{Spectral survey obtained with the IRAM 30m telescope at the atmospheric window of 3\,mm. }
\label{Fig3mm}%
\end{figure*}
\begin{figure*}[h!]
\centering
\ContinuedFloat
\includegraphics[angle=0,width=12cm]{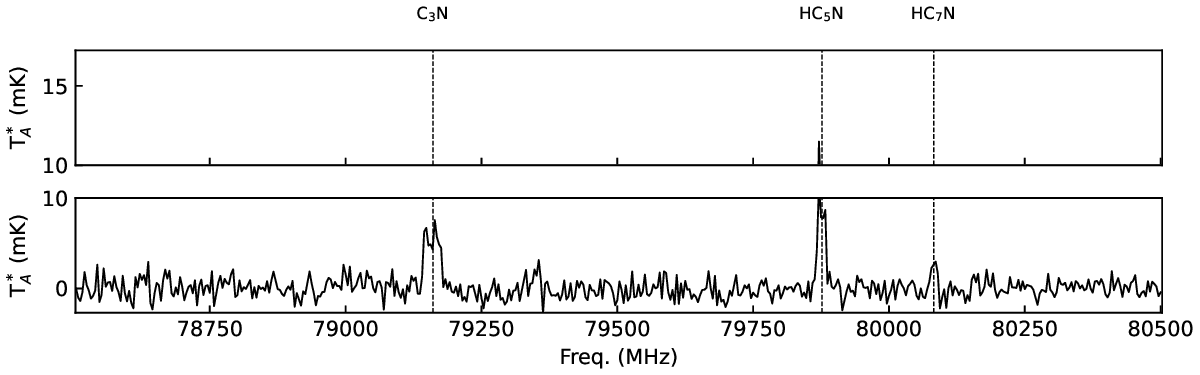}
\includegraphics[angle=0,width=12cm]{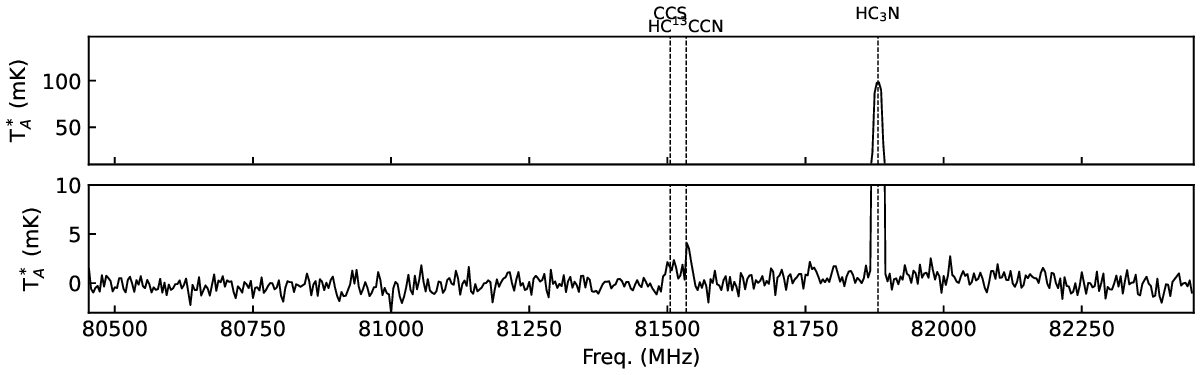}
\includegraphics[angle=0,width=12cm]{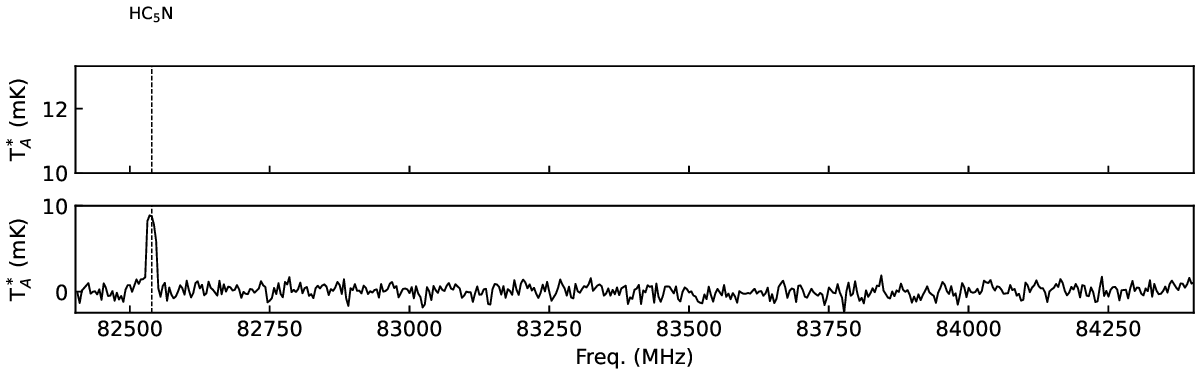}
\includegraphics[angle=0,width=12cm]{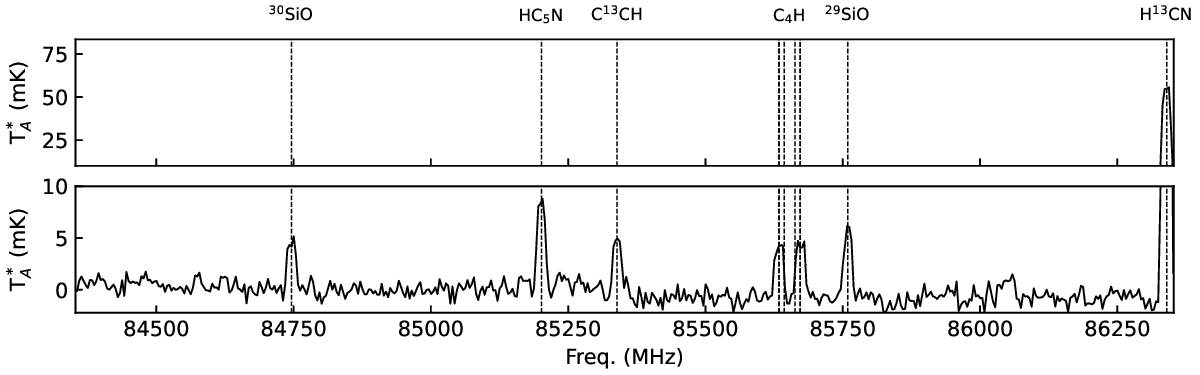}
\caption{Spectral survey obtained with the IRAM 30m telescope at the atmospheric window of 3\,mm. (continued) }
\label{Fig3mm}%
\end{figure*}
\begin{figure*}[h!]
\centering
\ContinuedFloat
\includegraphics[angle=0,width=12cm]{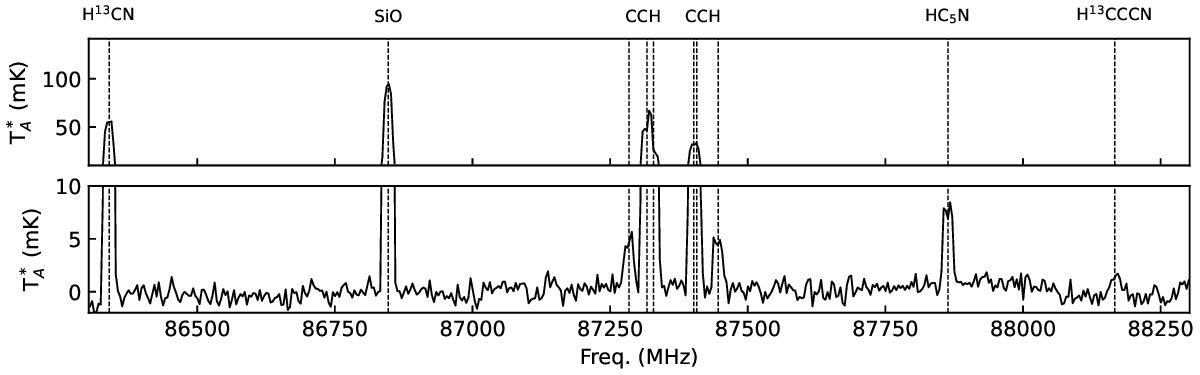}
\includegraphics[angle=0,width=12cm]{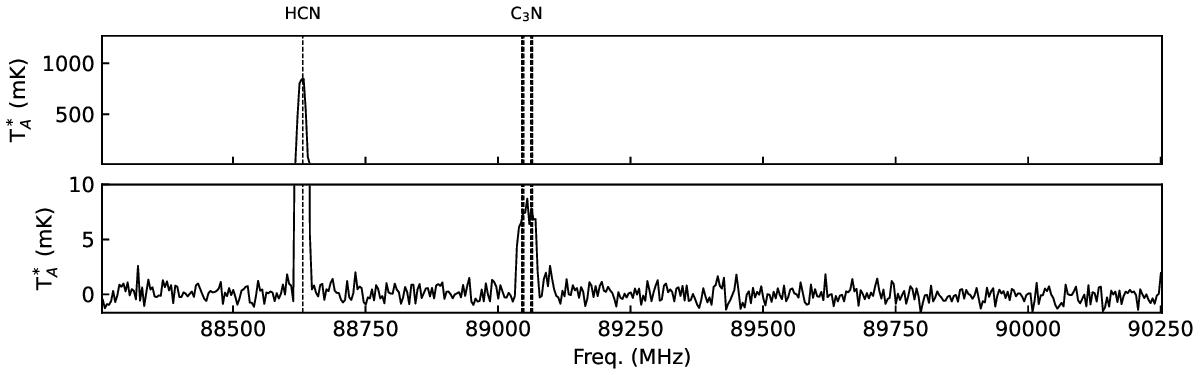}
\includegraphics[angle=0,width=12cm]{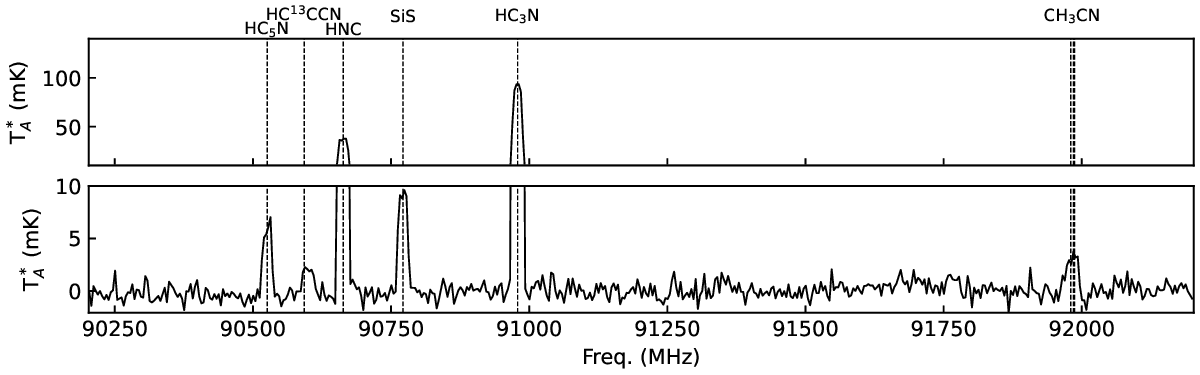}
\includegraphics[angle=0,width=12cm]{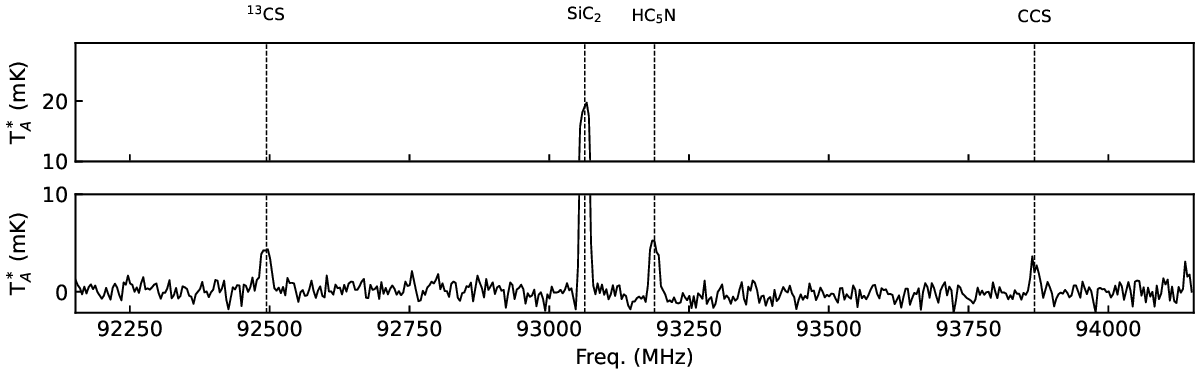}
\caption{Spectral survey obtained with the IRAM 30m telescope at the atmospheric window of 3\,mm. (continued) }
\label{Fig3mm}%
\end{figure*}
\begin{figure*}[h!]
\centering
\ContinuedFloat
\includegraphics[angle=0,width=12cm]{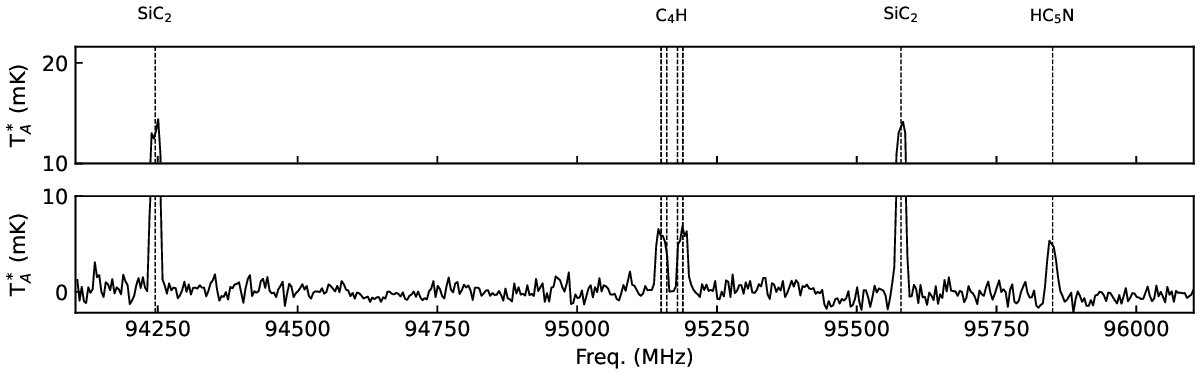}
\includegraphics[angle=0,width=12cm]{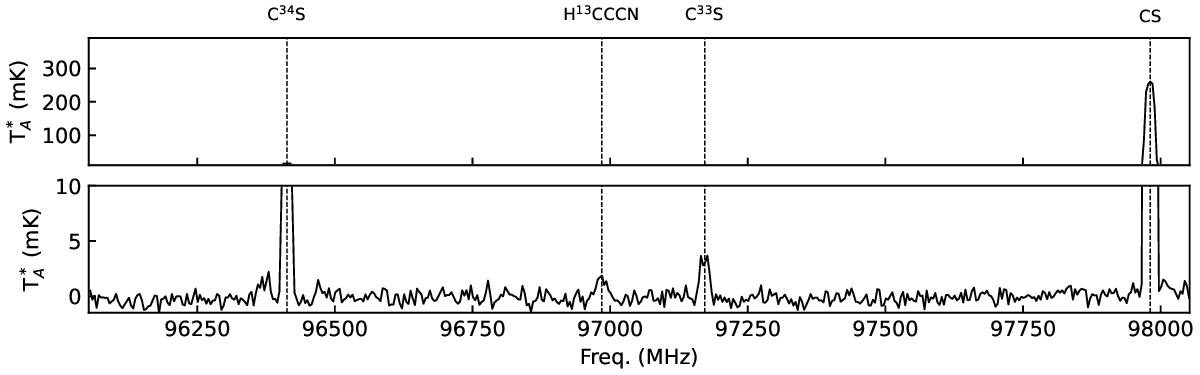}
\includegraphics[angle=0,width=12cm]{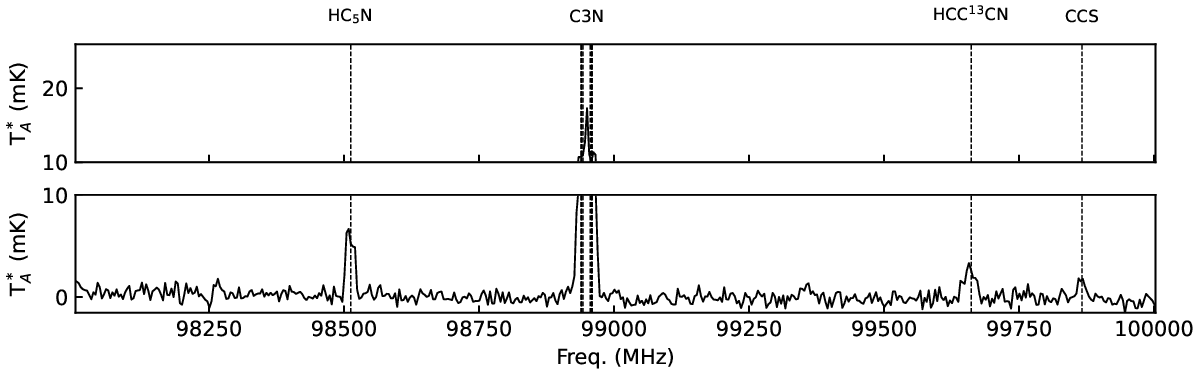}
\includegraphics[angle=0,width=12cm]{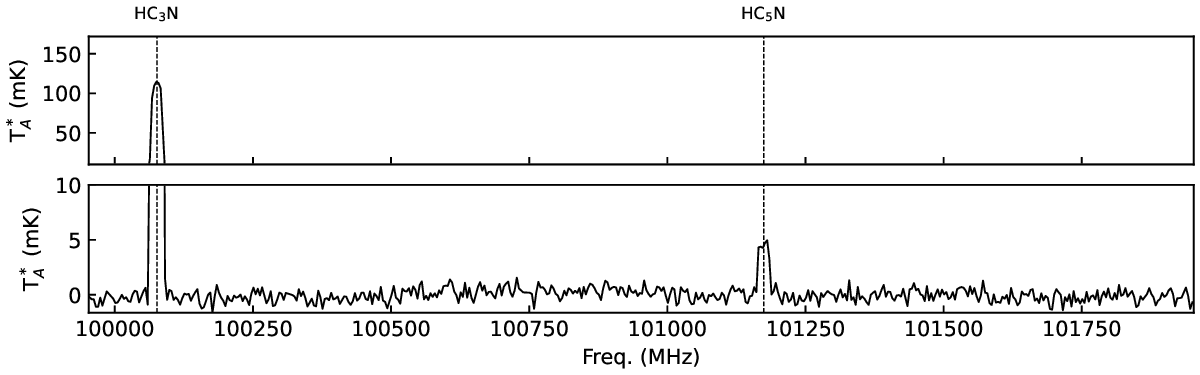}
\caption{Spectral survey obtained with the IRAM 30m telescope at the atmospheric window of 3\,mm. (continued) }
\label{Fig3mm}%
\end{figure*}
\begin{figure*}[h!]
\centering
\ContinuedFloat
\includegraphics[angle=0,width=12cm]{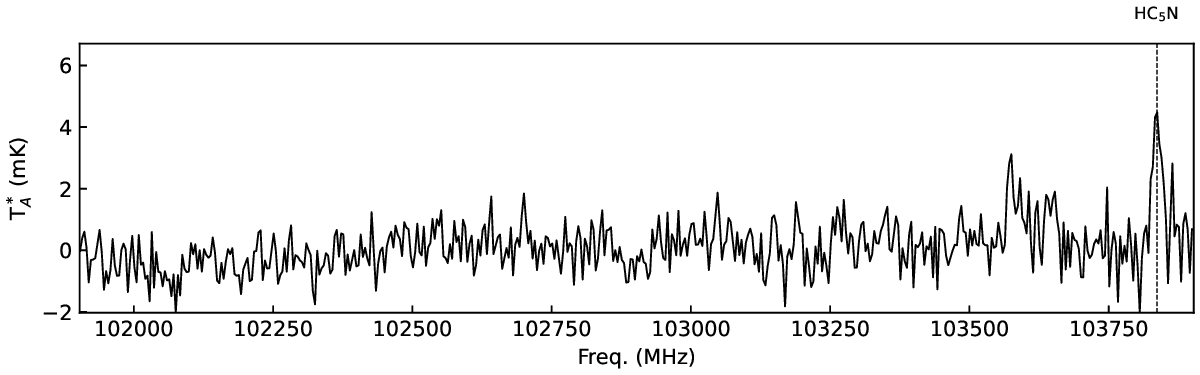}
\includegraphics[angle=0,width=12cm]{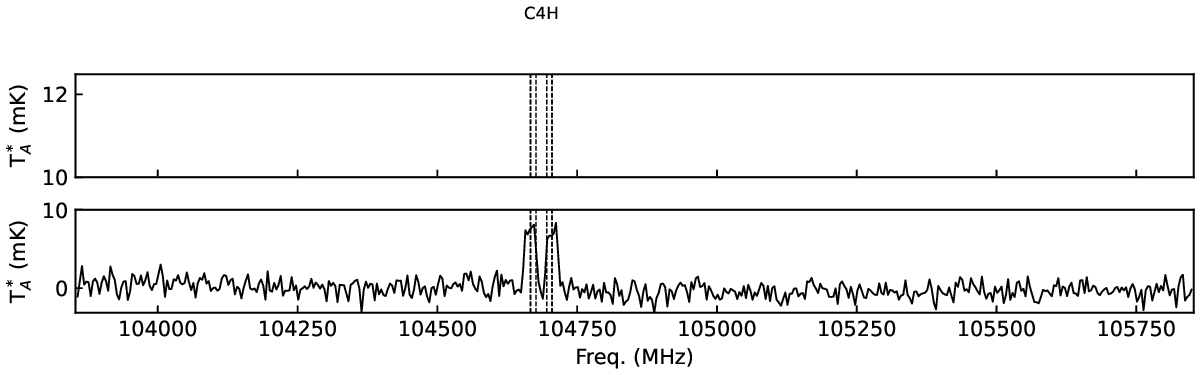}
\includegraphics[angle=0,width=12cm]{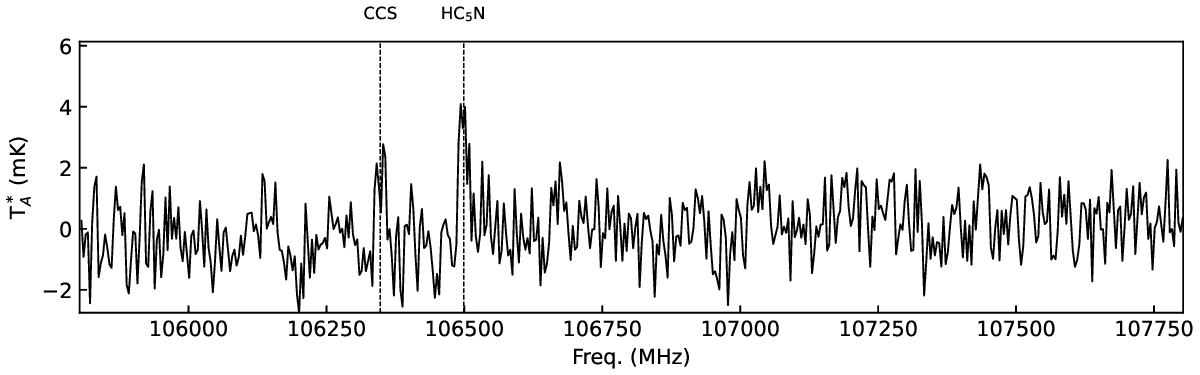}
\includegraphics[angle=0,width=12cm]{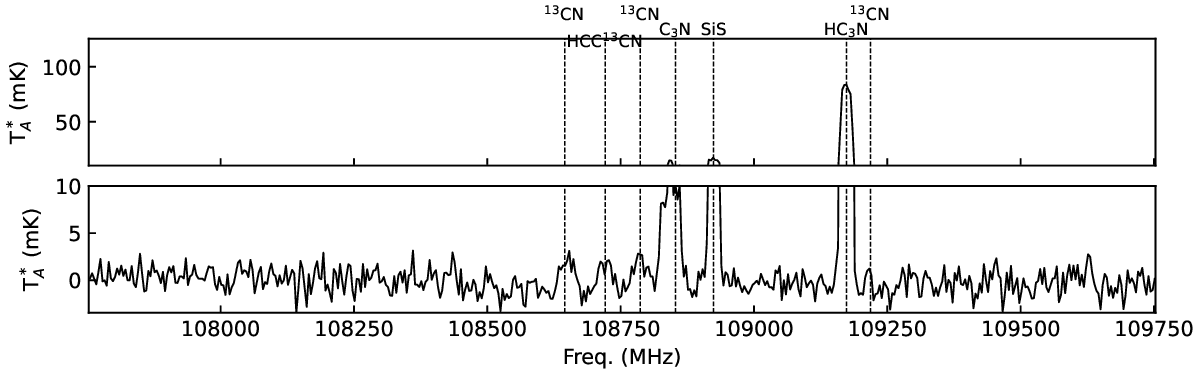}
\caption{Spectral survey obtained with the IRAM 30m telescope at the atmospheric window of 3\,mm. (continued) }
\label{Fig3mm}%
\end{figure*}
\begin{figure*}[h!]
\centering
\ContinuedFloat
\includegraphics[angle=0,width=12cm]{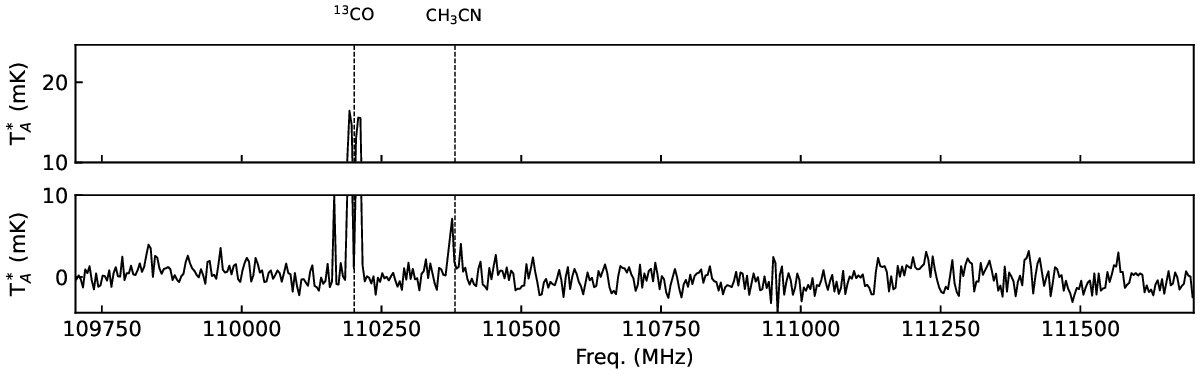}
\includegraphics[angle=0,width=12cm]{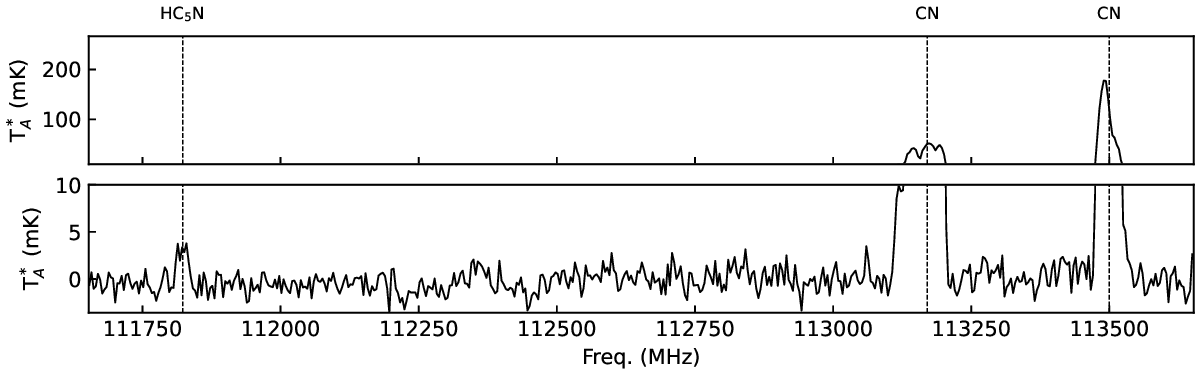}
\includegraphics[angle=0,width=12cm]{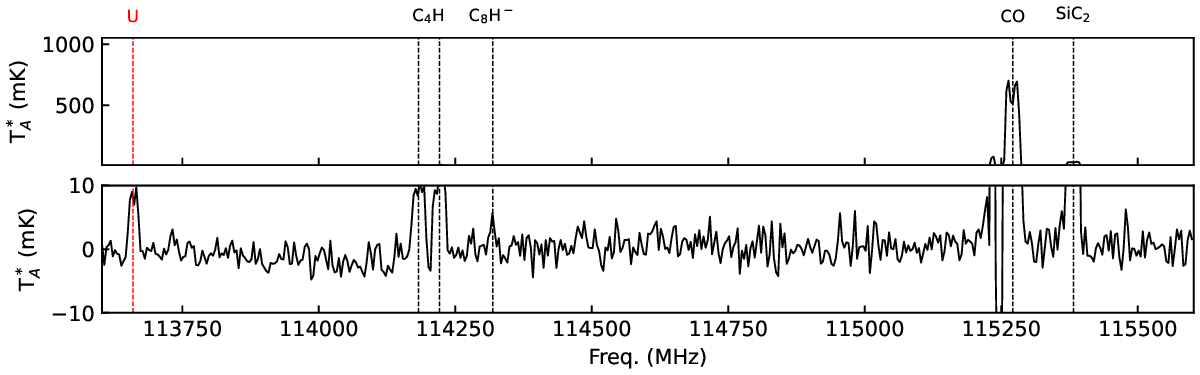}
\caption{Spectral survey obtained with the IRAM 30m telescope at the atmospheric window of 3\,mm. (continued) }
\label{Fig3mm}%
\end{figure*}
\begin{figure*}[h!]
\centering
\ContinuedFloat
\includegraphics[angle=0,width=12cm]{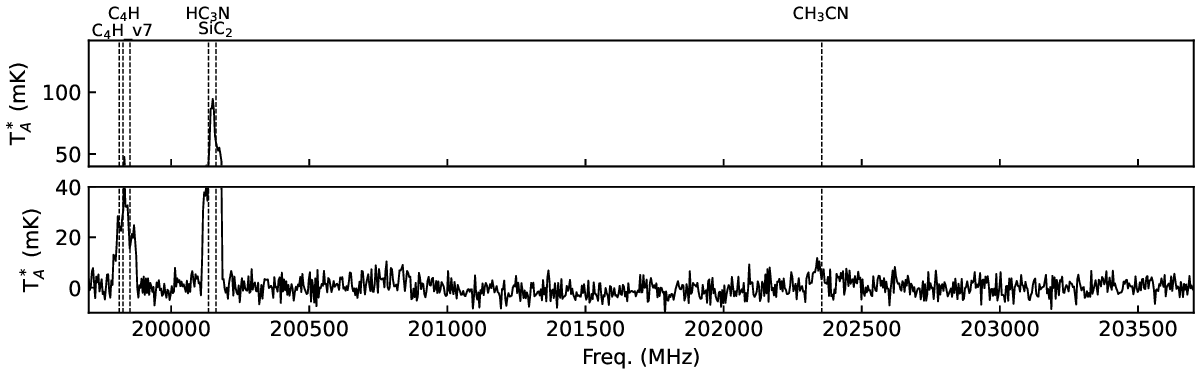}
\includegraphics[angle=0,width=12cm]{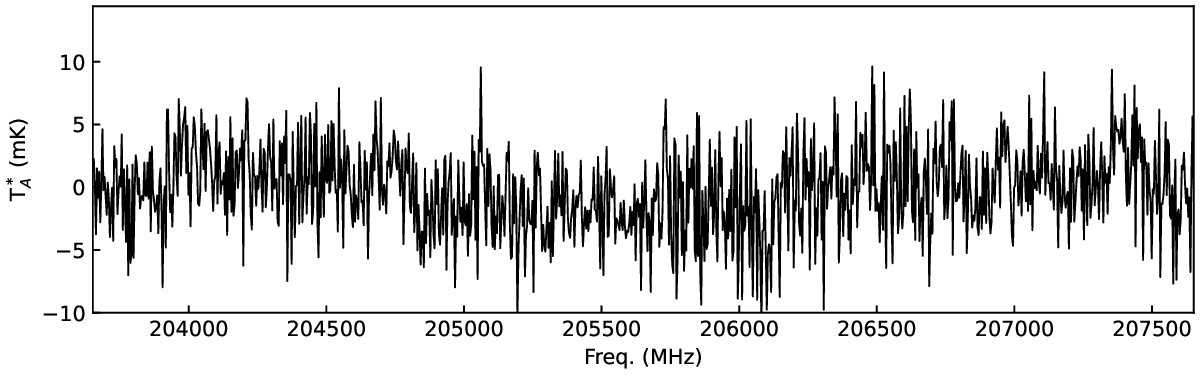}
\includegraphics[angle=0,width=12cm]{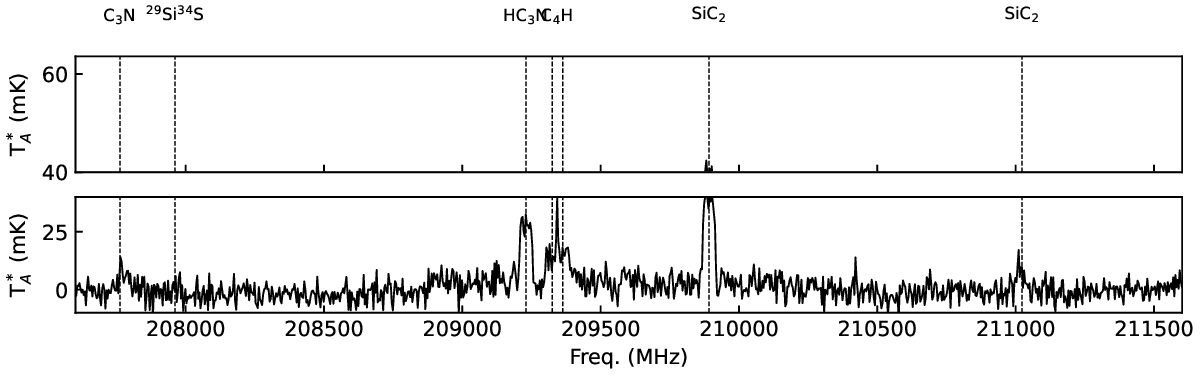}
\includegraphics[angle=0,width=12cm]{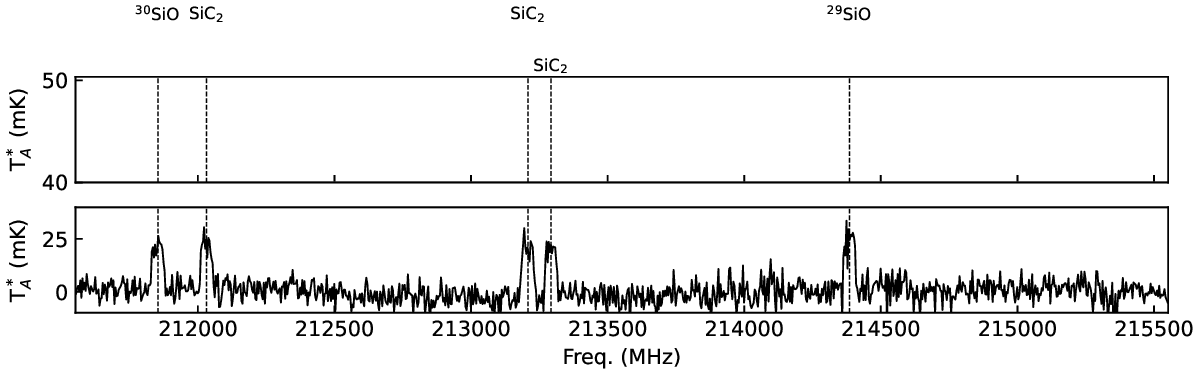}
\caption{Spectral survey obtained with the IRAM 30m telescope at the atmospheric window of 1\,mm.  }
\label{Fig1mm}%
\end{figure*}
\begin{figure*}[h!]
\centering
\ContinuedFloat
\includegraphics[angle=0,width=12cm]{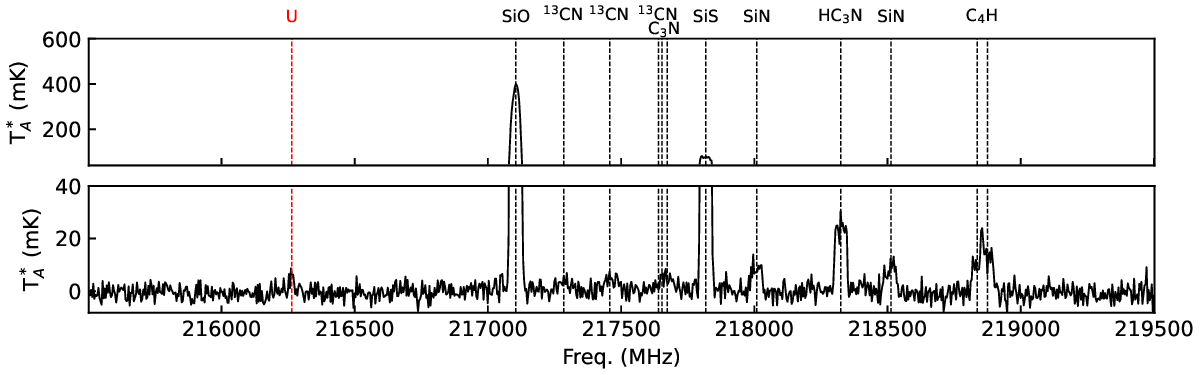}
\includegraphics[angle=0,width=12cm]{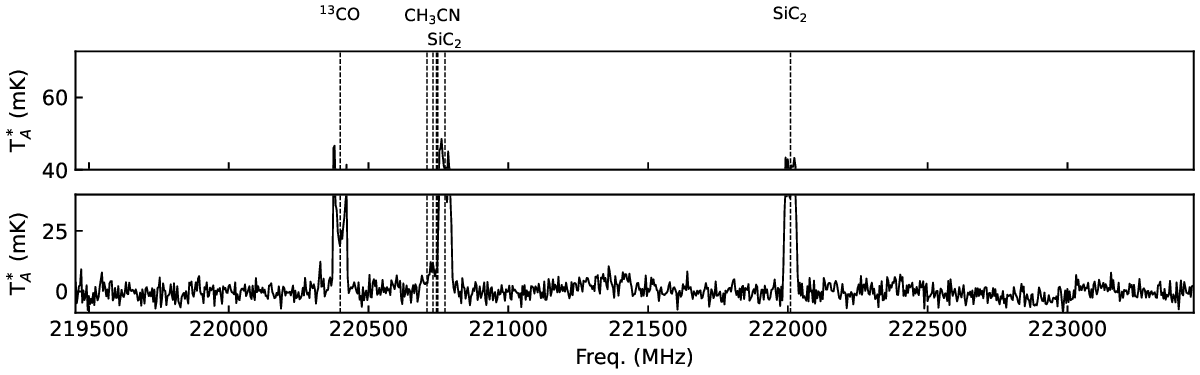}
\includegraphics[angle=0,width=12cm]{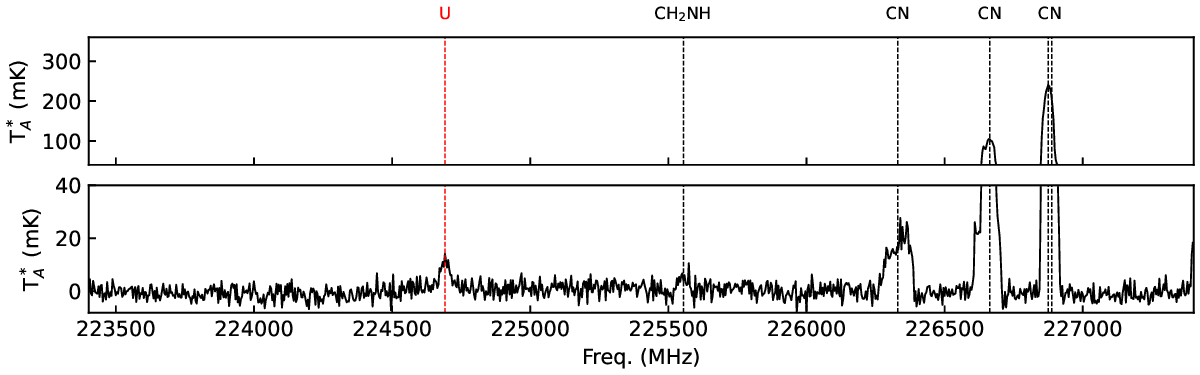}
\includegraphics[angle=0,width=12cm]{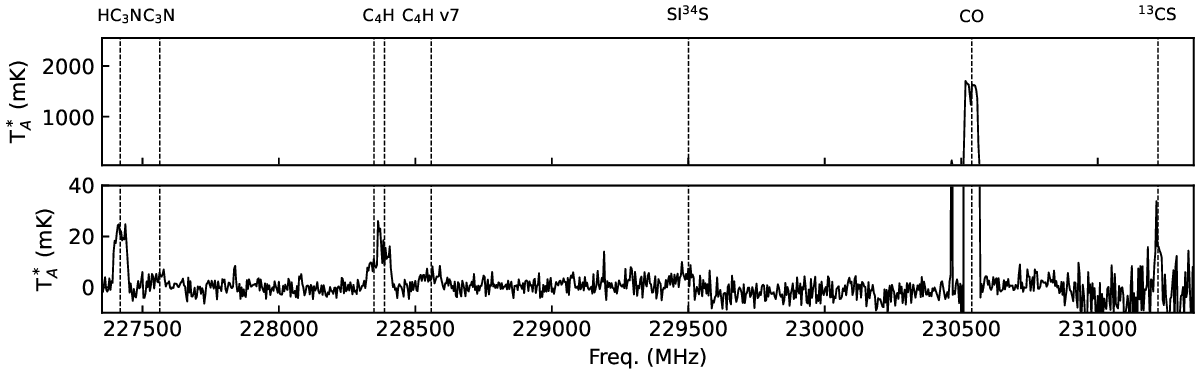}
\caption{Spectral survey obtained with the IRAM 30m telescope at the atmospheric window of 1\,mm. (continued) }
\label{Fig1mm}%
\end{figure*}
\begin{figure*}[h!]
\centering
\ContinuedFloat
\includegraphics[angle=0,width=12cm]{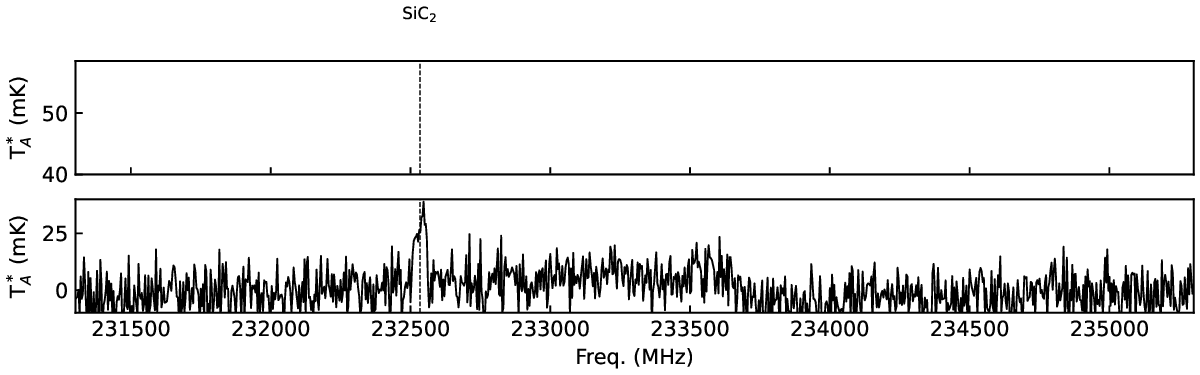}
\includegraphics[angle=0,width=12cm]{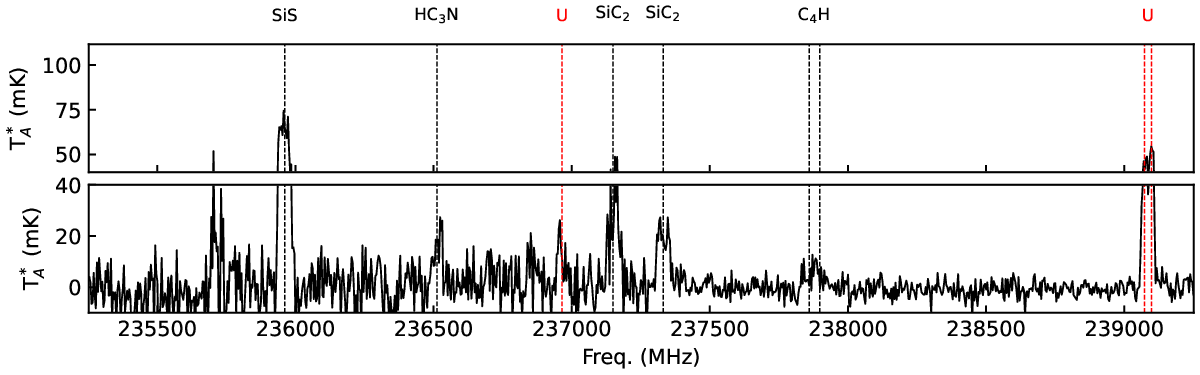}
\includegraphics[angle=0,width=12cm]{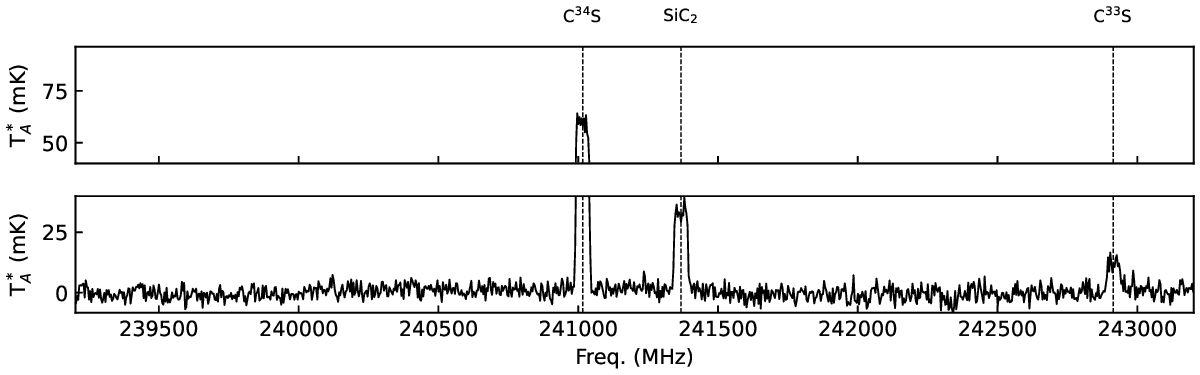}
\includegraphics[angle=0,width=12cm]{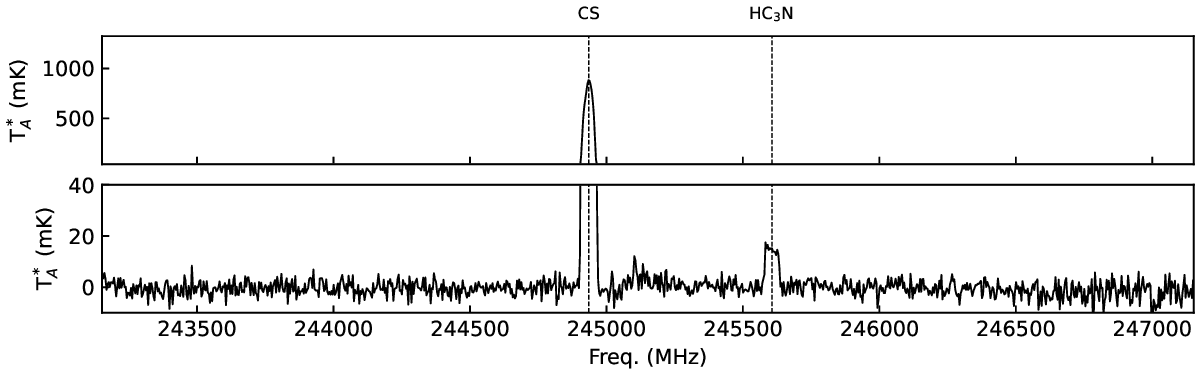}
\caption{Spectral survey obtained with the IRAM 30m telescope at the atmospheric window of 1\,mm. (continued) }
\label{Fig1mm}%
\end{figure*}
\begin{figure*}[h!]
\centering
\ContinuedFloat
\includegraphics[angle=0,width=12cm]{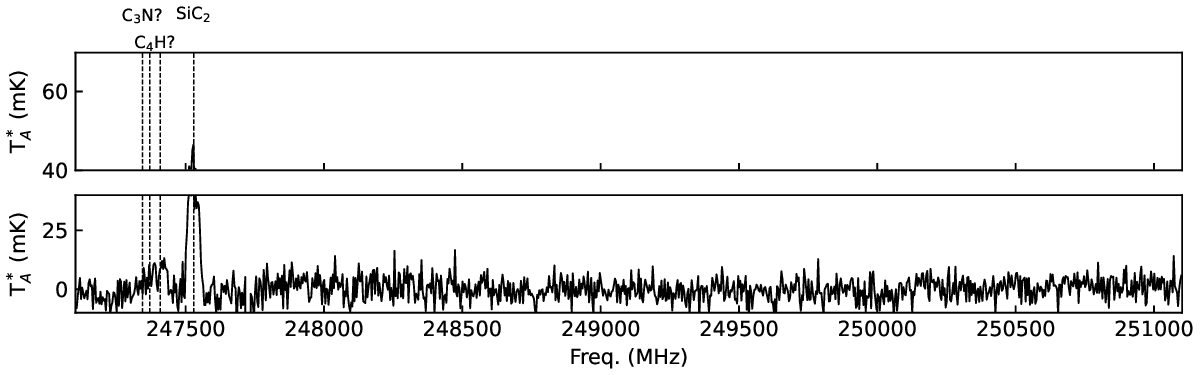}
\includegraphics[angle=0,width=12cm]{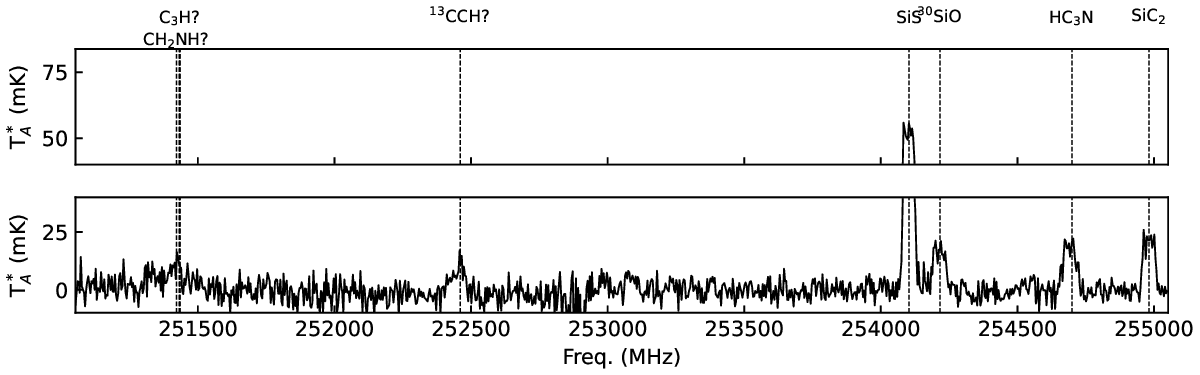}
\includegraphics[angle=0,width=12cm]{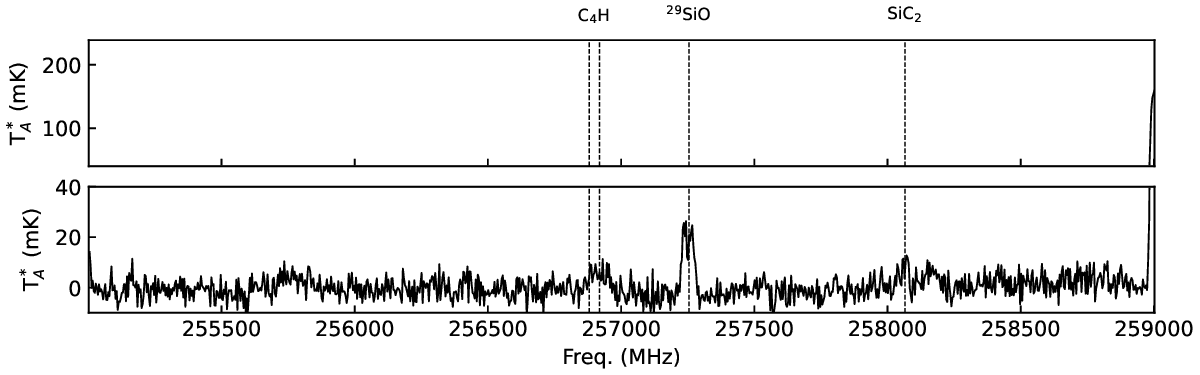}
\includegraphics[angle=0,width=12cm]{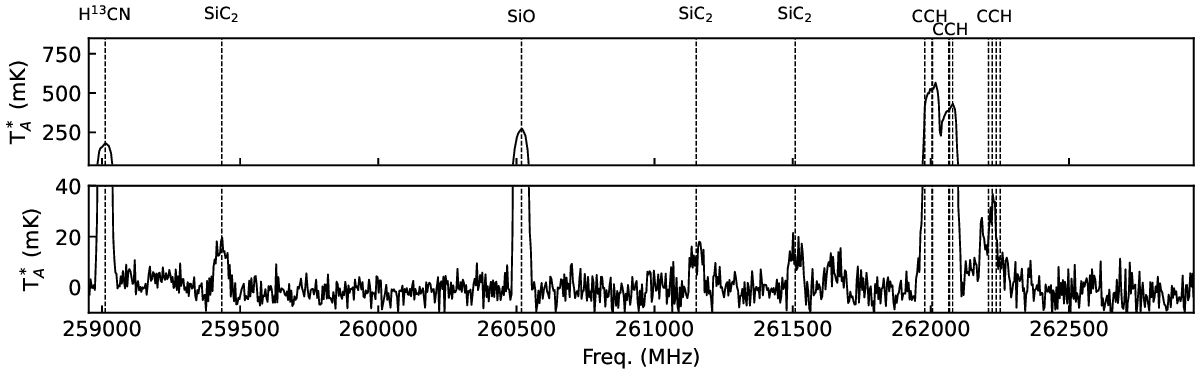}
\caption{Spectral survey obtained with the IRAM 30m telescope at the atmospheric window of 1\,mm. (continued) }
\label{Fig1mm}%
\end{figure*}
\begin{figure*}[h!]
\centering
\ContinuedFloat
\includegraphics[angle=0,width=12cm]{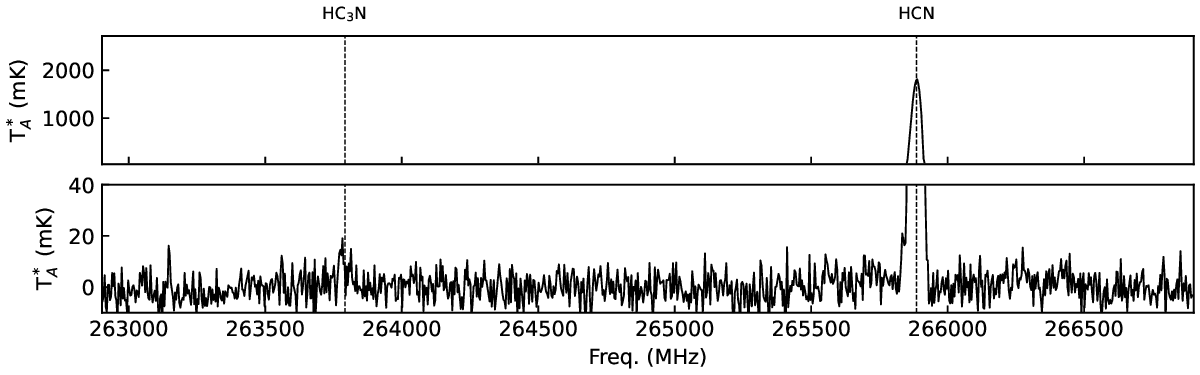}
\includegraphics[angle=0,width=12cm]{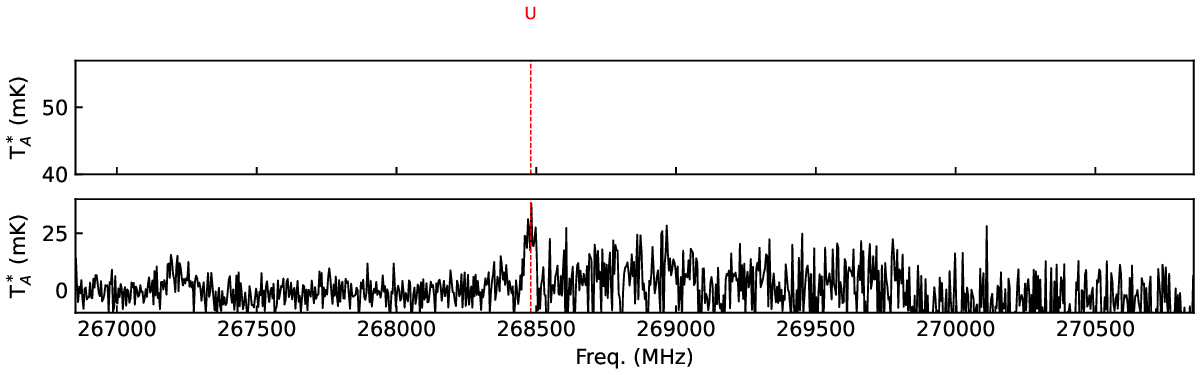}
\includegraphics[angle=0,width=12cm]{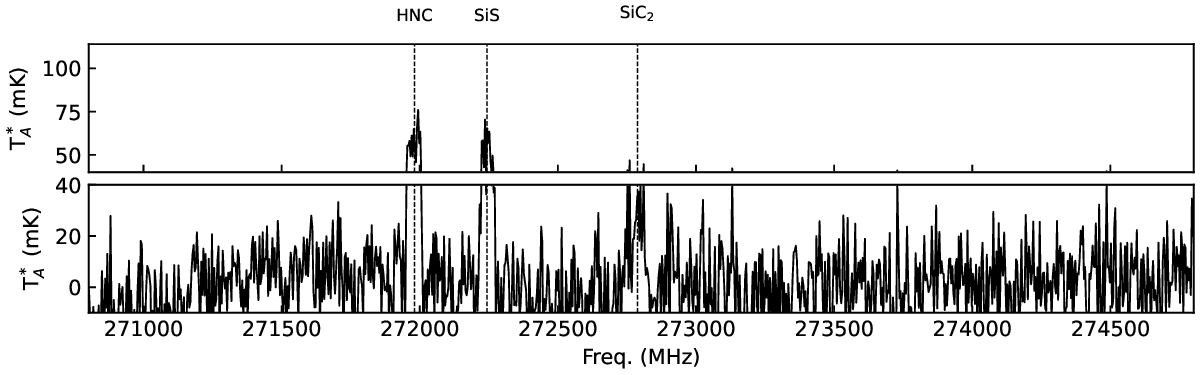}
\caption{Spectral survey obtained with the IRAM 30m telescope at the atmospheric window of 1\,mm. (continued) }
\label{Fig1mm}%
\end{figure*}  
   
\end{appendix}

\end{document}